%% file: main.tex
\documentclass[acmsmall,screen,nonacm]{acmart}

\usepackage[utf8]{inputenc}
\usepackage[english]{babel}
\usepackage{graphicx}
\usepackage{tabularx}
\usepackage{listings}
\usepackage{caption}
\usepackage{subcaption}
\usepackage{float}
\usepackage[frozencache,cachedir=_minted-main]{minted}
\usepackage{tikz}
\usetikzlibrary{arrows.meta,positioning}
\usepackage{calc}
\usepackage{mdframed}
\usepackage{cleveref}
\usepackage{todonotes}
\usepackage{xcolor}
\usepackage{fvextra}
\usepackage{xspace}
\usepackage{stmaryrd} % fatsemi

\input{macros.tex}

\usepackage{mathpartir}
\usepackage{mathtools}
\usepackage[mathscr]{euscript} % Eg. \mathscr{E}

\theoremstyle{definition}

\definecolor{codebg}{rgb}{0.98,0.96,0.92}
\newmintinline[javacode]{java}{
  bgcolor=codebg,
  fontsize=\small,
}
\newmintinline[choralcode]{'choral.py:ChoralLexer -x'}{
  bgcolor=codebg,
  fontsize=\small,
}

\usepackage[inline]{enumitem} % shortens spacing ;)
\setlist[enumerate]{noitemsep}
\setlist[enumerate]{leftmargin=*}
\newcommand{\code}[1]{\javacode{#1}}
\newcommand{\choral}[1]{\choralcode{#1}}

\definecolor{foo1}{RGB}{253,238,229}
\definecolor{foo2}{RGB}{254,226,231}
\definecolor{foo3}{RGB}{254,212,236}
\definecolor{foo4}{RGB}{254,200,240}

\AtBeginDocument{%
  }

\setcopyright{acmlicensed}
\copyrightyear{2026}
\acmYear{2026}
\acmDOI{XXXXXXX.XXXXXXX}

\begin{document}

\title{Accompanist: A Runtime for Resilient Choreographic Programming}
%\title{One-Click Decentralisation: Automated Synthesis of Microservice Choreographies}

\author{Viktor Strate Kløvedal}
\email{viktorstrate@imada.sdu.dk}
\affiliation{%
  \institution{University of Southern Denmark}
  \city{Odense}
  \country{Denmark}
}

\author{Dan Plyukhin}
\email{dplyukhin@imada.sdu.dk}
\affiliation{%
  \institution{University of Southern Denmark}
  \city{Odense}
  \country{Denmark}
}

\author{Marco Peressotti}
\email{peressotti@imada.sdu.dk}
\affiliation{%
  \institution{University of Southern Denmark}
  \city{Odense}
  \country{Denmark}
}

\author{Fabrizio Montesi}
\email{fmontesi@imada.sdu.dk}
\affiliation{%
  \institution{University of Southern Denmark}
  \city{Odense}
  \country{Denmark}
}

\begin{abstract}
In service-oriented architecture, services coordinate in one of two ways: directly, using point-to-point communication, or indirectly, through an intermediary called the \emph{orchestrator}. Orchestrators tend to be more popular because their local state is a `single source of truth' for the status of ongoing workflows, which simplifies fault recovery and rollback for distributed transactions that use the `saga' pattern. But orchestration is not always an option because of hardware constraints and security policies. Without a central orchestrator, resilient saga transactions are hard to implement correctly.

A natural idea is to use \emph{choreographic programming}, a paradigm that brings the `global view' of orchestrators to a decentralised setting. Unfortunately, choreographic programming relies on strong assumptions about network reliability and service uptime that often do not hold. Recent work weakens some of these assumptions with `failure-aware' language features, but these features make programs more complex. We propose a complementary approach: to co-design the programming interface with a customizable \emph{runtime} that can replay computation to mask faults. Our approach keeps programs simple, does not require modifying the compiler, and lends itself to a clean separation of concerns in formal proofs.

We present \Accompanist{}, a resilient runtime for the Choral choreographic programming language. With \Accompanist{}, programmers can implement decentralised saga transactions as choreographic programs and deploy the compiled code to `sidecars' that run alongside services in a pre-existing codebase. Our key assumptions are that choreographic programs should be deterministic, transactions within a saga should be idempotent, and messages should be written to a durable message queue. Based on these assumptions, we present a formal model and prove that target code is correct-by-construction: choreographic sagas are deadlock-free, terminate under bounded restarts, and upon termination their participants have all succeeded or all invoked compensating transactions to undo the workflow.
\end{abstract}

\keywords{orchestration, choreographic programming, workflows, saga transactions}

\settopmatter{printfolios=true}
\maketitle

\input{introduction.tex}

\input{motivation.tex}

\input{choral.tex}

\input{accompanist.tex}

\input{fault-tolerance.tex}

\input{evaluation.tex}

\input{conclusion.tex}

\bibliographystyle{ACM-Reference-Format}
\bibliography{main}

\end{document}

%% file: macros.tex
\newcommand{\Accompanist}{\textsc{Accompanist}\xspace}
\newcommand{\manualTheorem}[2]{
  \noindent\\{\bfseries\upshape #1} #2
}

\newcommand{\p}{\mathsf{p}}
\newcommand{\q}{\mathsf{q}}
\renewcommand{\r}{\mathsf{r}}
\newcommand{\PName}{\mathbb P}
\newcommand{\epp}[1]{\llbracket #1 \rrbracket}
\newcommand{\conf}[1]{\langle #1 \rangle}

\newcommand{\trans}[3]{\conf{#1} \xrightarrow{#2} \conf{#3}}

\newcommand{\restrict}[2]{#1 \!\!\upharpoonright_{#2}}

\DeclareMathOperator{\pn}{pn}
\DeclareMathOperator{\cfg}{cfg}

%% file: introduction.tex
\section{Introduction}%
\label{sec:intro}

There are many reasons to split a single `monolithic' application into distinct services~\cite{DGLMMMS17,quigley2009ros}.
Services can be developed, deployed, and scaled independently.
Services can also be written in different programming languages, giving organizations access to a larger software ecosystem and talent pool.
By making services simple and small (\emph{micro}services), programmers can make their applications more resilient to hardware and software faults~\cite{newman2021}. However, writing workflows to coordinate services is not always an easy feat.

%However, service-oriented architectures pose a challenge for performance and maintenance. Splitting a monolith into services adds a \emph{network latency tax} ranging from microseconds to hundreds of milliseconds~\cite{seemakhupt2023}. Oftentimes the latency tax can be kept low by carefully provisioning and co-locating tightly coupled services, but service-level objectives, data sovereignty, and hardware requirements sometimes make this impossible~\cite{park2024,oracleMulticloud,seemakhupt2023,quigley2009ros}. In these scenarios, network costs are high and programmers must take pains to eliminate unnecessary network trips. But programmers also need to keep applications maintainable, and sometimes this means sacrificing precious latency.

\begin{figure*}
  \centering
  \includegraphics[width=\textwidth]{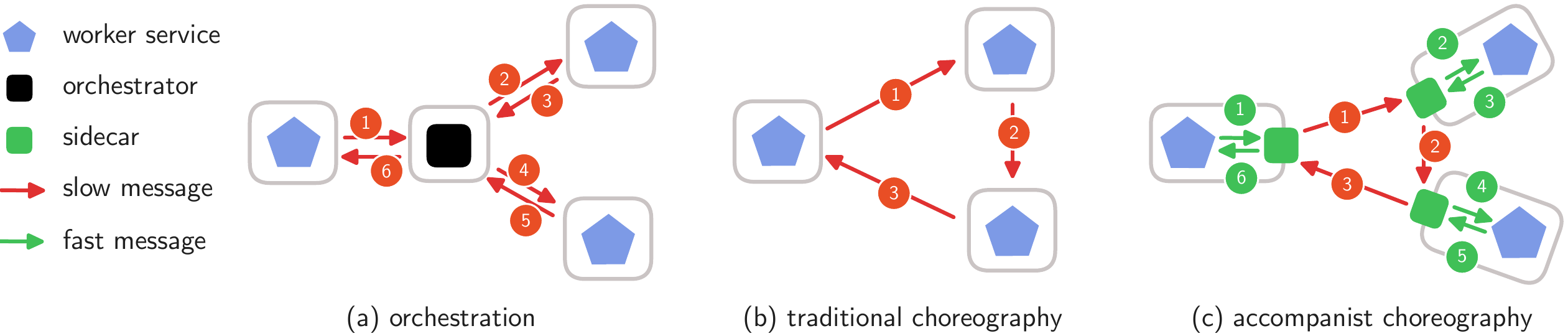}
  \caption{A workflow with three services, implemented in three different styles.}%
  \label{fig:orchestration-choreography-accompanist}
\end{figure*}

There are two ways to coordinate services: orchestration and choreography~\cite{camundaOrchestrationVsChoreography}. In orchestration, a central `orchestrator' service makes remote procedure calls (RPCs) to `worker' services that handle domain logic (c.f.\ \Cref{fig:orchestration-choreography-accompanist}a). Many organizations cite that orchestration makes complex workflows easy to monitor and maintain~\cite{netflixtechnologyblog2016,awsStepFunctionsWhy}. The key advantages are:
\begin{enumerate}[label=A\arabic*., ref=A\arabic*]
  \item \label{adv:global-view} \emph{(Global view)} The entire workflow lives at one service---perhaps even one procedure.
  \item \label{adv:one-language}\emph{(One language)} Workers may use different languages, but the orchestrator uses one language.
  \item \label{adv:safety}\emph{(Safety)} Deadlock-freedom, determinism, and message type safety are easy to check statically.
  \item \label{adv:separate-deployment}\emph{(Separate deployment)} New workflows can be deployed without redeploying the workers.
  \item \label{adv:single-source-of-truth}\emph{(Single source of truth)} In case of faults, workflows can be resumed by querying the orchestrator's local database.
\end{enumerate}

However, making all interactions go through a single service can incur unnecessary network hops, with each hop ranging from microseconds to hundreds of milliseconds~\cite{seemakhupt2023,lai2020}. The impact of unnecessary hops on end-to-end latency is what we call the \emph{orchestration tax}. %Moreover, the performance of an orchestrated system is highly sensitive to the placement of the orchestrator relative to other services~\cite{wang2021}.
Organizations can try to dodge the orchestration tax with optimizations like RPC chains and distributed futures~\cite{DBLP:conf/nsdi/SongAKM09,wang2021} or by dynamically merging orchestrators with worker services like in ServiceWeaver~\cite{ghemawat2023}. But sometimes this is not enough: optimizations do not give the programmer complete control over data movement, and some services cannot be co-located because they are pinned to specific hardware~\cite{quigley2009ros}. Security policies and data sovereignty requirements can also add tight constraints on how services coordinate and preclude the use of orchestrators altogether~\cite{DBLP:conf/csfw/AcayGRM24,park2024,oracleMulticloud,seemakhupt2023}.

%Researchers have developed numerous tools to improve responsiveness in service-oriented architectures without sacrificing maintainability. Most notable is ServiceWeaver~\cite{ghemawat2023}, which allowed programmers to write microservice applications as logical monoliths in Go and merged tightly-coupled components at runtime; however, ServiceWeaver's approach assumes the entire application can be rewritten in one language, which is infeasible for large codebases~\cite{serviceweaverteam2024}.
% In Spark, the throughput for geodistributed workloads was improved by moving computation closer to data and adopting a push-based communication model~\cite{lai2020} -- but it is not clear how those optimizations could be applied to general-purpose microservice architectures.
% What programmers need is a microservice framework that offers control over how data is moved and where computation is executed, without significant rewrites to existing applications.

Without orchestration, worker services need to use choreography: direct, point-to-point communication with no inherent center of control (c.f.\ \Cref{fig:orchestration-choreography-accompanist}b). Traditionally, choreographies split a workflow's implementation across several different services, sacrificing advantages~\ref{adv:global-view}--\ref{adv:single-source-of-truth}.
\emph{Choreographic programming}~\cite{montesi2014,montesi2023} helps recover some of these advantages. The key idea is to extend traditional languages with a notion of `roles' (the location of a value or computation) and `communications' (a way to bring values from one location to another). The resulting source code, which we will call \emph{choreographic programs}, can be compiled to produce \emph{choreographies} as target code.
% In contrast to other distributed programming paradigms~\cite{weisenburger2020,giallorenzo2021,wael2015,charles2005,casadei2023}, choreographic programming languages offer fine-grained control over data movement, are deterministic and deadlock-free by default, and generate readable target code without runtime overhead~\cite{giallorenzo2024}.
Researchers have developed choreographic programming languages based on Haskell~\cite{shen2023}, Rust~\cite{bates2025}, Elixir~\cite{chorex}, Clojure~\cite{klor}, and more; we will focus on Choral~\cite{giallorenzo2024}, a choreographic programming language based on Java.

Unfortunately, real service-oriented applications have a laundry list of requirements that are not satisfied by any choreographic programming language:

%\crefname{enumi}{requirement}{requirements}
%\Crefname{enumi}{Requirement}{Requirements}
\begin{enumerate}[label=R\arabic*., ref=R\arabic*]
  \item \label{req:concurrency} \emph{(Concurrency)} Services should not need to finish one workflow before joining another.
  \item \label{req:agreement} \emph{(Agreement)} At the end of a workflow, all participants succeed or all participants fail.
  \item \label{req:durability} \emph{(Durability)} Workflows can make progress despite crashes and dropped messages.
  \item \label{req:reactivity} \emph{(Reactivity)} Once a workflow starts, services can add new participants at runtime.
\end{enumerate}
Requirements \ref{req:concurrency}, \ref{req:agreement}, and \ref{req:durability} are properties of \emph{saga} transactions~\cite{sagas}. Sagas are distributed transactions that sacrifice the consistency and isolation requirements of classical ACID transactions, improving performance in service-oriented applications.
%Decentralized workflows also inherently lack advantage \ref{adv:single-source-of-truth}.
Our key challenge is to design an architecture that satisfies requirements \ref{req:concurrency}--\ref{req:reactivity} and advantages \ref{adv:global-view}--\ref{adv:separate-deployment}, despite the lack of \ref{adv:single-source-of-truth}.

% For example, consider a user checkout workflow in an online shopping application. R1 reflects the fact that a service in the workflow might crash because of a software error, or network partitions could cause packets to be dropped. R2 ensures that...
Researchers have aimed to satisfy requirements \ref{req:concurrency}--\ref{req:reactivity} by making choreographic programming more expressive. Graversen et al proposed frames to recover from dropped messages~\cite{graversen2025}, Plyukhin et al proposed a new semantics to increase concurrency~\cite{plyukhin2024}, Lugovi\'c and Montesi proposed an event-driven pattern to make choreographies reactive~\cite{lugovic2024}, and Wiesdorf and Greenman proposed synchronized checkpoints to achieve all-or-nothing semantics~\cite{chorex}. Each of these extensions only addresses a subset of our requirements, and some require nontrivial changes to the compiler. Some language extensions also make programs more complex. To combine them all in one place would weaken the elegant correspondence between orchestrators and choreographic programs.
% The paradigm also assumes the application will be written in just one language, so choreographic code can invoke local code. Hence, although choreographic programming meets criteria \ref{adv:global-view} to \ref{adv:safety}, it does not provide \ref{adv:separate-deployment} or \ref{adv:single-source-of-truth}.

Our solution is \Accompanist: a language runtime for resilient choreographic programs. Programmers use \Accompanist{} by writing ordinary Choral programs and deploying the compiled code to \emph{sidecars}~\cite{sidecar,richardson_microservices_2019} that accompany worker services, as shown in \Cref{fig:orchestration-choreography-accompanist}c. Sidecars communicate point-to-point as dictated by the Choral program (the three red arrows in \Cref{fig:orchestration-choreography-accompanist}c) and they communicate with worker services using RPCs (the six green arrows in \Cref{fig:orchestration-choreography-accompanist}c). In contrast to recent work that assumes the entire application is written in a choreographic language~\cite{lugovic2024,plyukhin2024,shen2023,bates2025,chorex,giallorenzo2024}, our approach can be gracefully integrated with pre-existing polyglot applications. Indeed, from the perspective of worker services, RPCs from sidecars are the same as RPCs from an orchestrator---although we expect lower latency in the former because sidecars can be deployed to the same locations as worker services.

In its default mode, \Accompanist{} provides basic necessities like monitoring, reactivity, resource multiplexing, and at-most-once semantics. We show the default mode is practical by porting orchestrators from two open-source microservice applications and we observe improved end-to-end latency when services are deployed across zones or regions. \Accompanist{} also has a fault-tolerant mode that re-runs compiled Choral programs until they signal completion or failure. Our initial findings show that the overhead of running choreographic saga transactions in fault-tolerant mode is not too high---in fact, due to architectural differences, we observe a 5.9$\times$ improvement in median end-to-end latency compared to a mature saga orchestration framework used in industry.

While language runtimes and replay-based recovery are certainly not new, our work is the first to apply them to choreographic programming---and our evaluation shows that the approach still achieves good performance. We complement these empirical results with a novel theoretical result: the first formalization of decentralized correct-by-construction saga transactions. Our approach offers a clean separation of concerns between (1) proving deadlock-freedom and saga safety at the level of choreographic programs, (2) proving a standard bisimulation result between choreographic programs and compiled choreographies, and (3) proving that every execution with restarts is equivalent to a fault-free execution at the level of networks.

In the following sections, we begin with a gentle introduction to service-oriented architectures, choreographic programming, and their limitations. We proceed to give an overview of the \Accompanist{} runtime in \Cref{sec:accompanist}, which motivates the constructs of our formal model in \Cref{sec:fault-tolerance}. We conclude with a performance evaluation in \Cref{sec:evaluation}, related work in \Cref{sec:related-work}, and we reflect on future work in \Cref{sec:conclusion}.

%%% Local Variables:
%%% mode: latex
%%% TeX-master: "main"
%%% End:

%% file: motivation.tex
\section{Motivation: Two Case Studies}%
\label{sec:case-study}

\subsection{Online Boutique}%
\label{sec:case-study-webshop}

\begin{figure*}
  \centering
  \includegraphics[width=\textwidth]{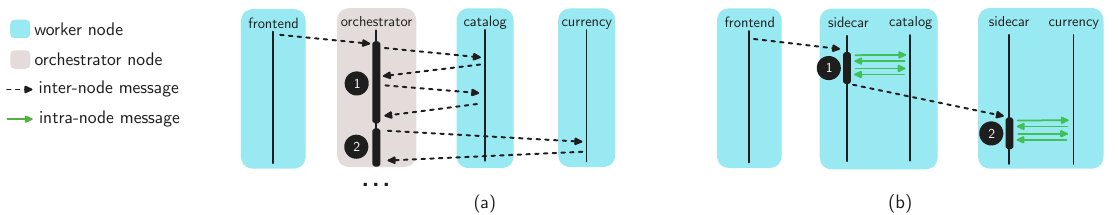}
  \Description{Diagram showing a sequence diagram for orchestration versus Accompanist.}
  \caption{A user checkout workflow implemented with orchestration (a) versus \Accompanist (b).}%
  \label{fig:baseline-vs-choral-diagram}
\end{figure*}

\emph{Online Boutique}~\cite{microservices-demo} is an open-source web app developed by Google to showcase modern microservice architecture. Services in the application are written in C\#, Java, Go, Python, and JavaScript, and all communicate using gRPC~\cite{grpc}. We use the Online Boutique codebase as a case study to identify some of the best practices and pathologies of orchestrated microservices.

\subsubsection{Orchestrators}
Worker services that handle domain logic, like product catalog or billing services, are implemented as RPC servers. These servers handle RPCs locally and do not interact with one another, minimizing coupling.

Most user requests are handled in the application frontend by making a few RPCs to worker services; for example, viewing the user's shopping cart is implemented with an RPC to the shopping cart service. For complex interactions with many services, the functionality is factored into a dedicated orchestrator service. For example, when a user makes a checkout request, the frontend delegates to an orchestrator called the \emph{checkout service}.

There are several advantages to handling complex workflows with a dedicated orchestrator, instead of making RPCs directly from the frontend:
\begin{enumerate}
  \item Programming errors in the orchestration will not affect other API endpoints in the frontend.
  \item Operators can deploy orchestrators close to the workers they communicate with.
  \item Orchestrators and the frontend can be maintained by different teams and can scale independently.
\end{enumerate}
However, orchestration scales poorly with network latency and we shall see next how it requires worker services to develop specialized APIs for their clients to mitigate that latency.

\subsubsection{Unnecessary Sequencing} Perhaps surprisingly, all services in the Online Boutique use blocking RPCs instead of nonblocking ones. Blocking RPCs are easy to use because they closely resemble ordinary procedure calls in popular imperative programming languages, whereas nonblocking RPCs have a more complex API that can introduce unwanted nondeterminism and distributed concurrency bugs~\cite{LLLG16}. The gRPC library, in particular, does not guarantee nonblocking calls will be handled in program order. It should therefore come as no surprise that developers prefer blocking RPCs for maintainability.

\subsubsection{Non-Local Computation}\label{sec:non-local-computation}
Not only does the checkout service use blocking RPCs -- it also executes them in a loop. As shown in \Cref{fig:baseline-vs-choral-diagram}a, the checkout service (1) asks the product catalog to get the price of each item in the shopping cart, and then (2) asks the currency service to convert each price into the user's local currency. It would of course be simpler and faster for the checkout service to make a single RPC with a list of all items in the user's cart, but -- crucially -- the product catalog only offers methods for getting the price of a single item.

This simple example illustrates a fundamental problem: good performance requires specializing worker services for their orchestrators, but the developers of worker services cannot predict all the ways an orchestrator might use them. In the Online Boutique, the checkout team could reasonably ask the product team to add a method for lists of items. But in large organizations, the product team may take a long time to respond; if the product team is in a separate partner organization, the request may never be met at all; and even if the request is handled promptly, one would expect the product catalog API to gradually become bloated with specialized methods for various clients. As a result, performance concerns introduce \emph{API coupling} between orchestrators and worker services.

\subsection{DeathStar Hotel Reservation}

As another case study, we consider \emph{DeathStar}~\cite{gan2019}: a suite of polyglot microservice applications using a variety of technologies. We focus specifically on the \emph{Hotel Reservation} application, because it presents a realistic scenario where multiple organizations collaborate: a booking service making requests to remote worker services, each of which manages some finite resource like hotels and flights and may require expensive trips over WAN to access. We found that Hotel Reservation has similar features to the Online Boutique: queries to remote services are managed by an orchestrator using blocking RPCs, and although the search orchestrator does not execute RPCs in a loop, it does demonstrate API coupling: worker services have specialized endpoints to handle queries about multiple hotels at once. %In \Cref{sec:evaluation-hotel} we extend the application with a basic flight service located in a different data center, and show how the overhead from repeated RPCs over WAN yields terrible end-to-end latency.

With orchestration, programmers can add sophisticated workflows to a microservice application without modifying any worker services. However, in high-latency environments, the centralized nature of orchestration can introduce significant overhead. In situations where orchestration does not scale, we would like an alternative that offers the same level of modularity but a greater level of control over data locality.

%%% Local Variables:
%%% mode: latex
%%% TeX-master: "main"
%%% End:

%% file: choral.tex
\section{Choral and Choreographic Programming}\label{sec:choral}

\newcommand{\blackcircle}[1]{%
  \tikz[baseline=(char.base)]{
    \node[shape=circle, fill=black, text=white, inner sep=1pt] (char) {#1};
  }%
}

\begin{figure*}
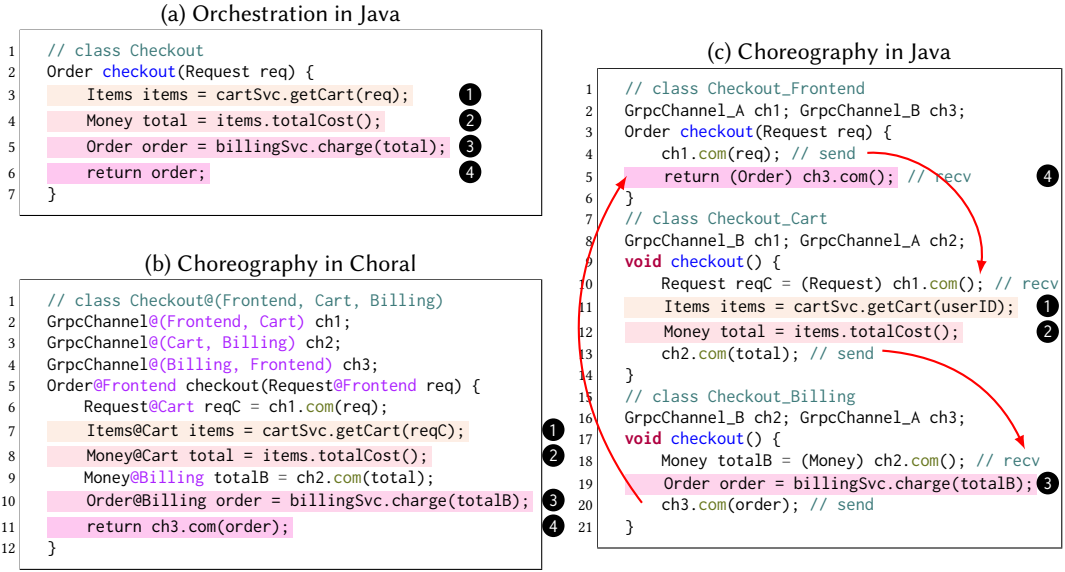

  \begin{minipage}[b]{0.5\textwidth}

    \begin{subfigure}{\textwidth}
      \caption{Orchestration in Java}
      \label{fig:orchestrated}

      \centering
      \scriptsize
      \begin{mdframed}
        \begin{minted}[linenos, escapeinside=!!]{java}
// class Checkout
Order checkout(Request req) {
!\colorbox{foo1}{    Items items = cartSvc.getCart(req);}!     !\blackcircle{1}!
!\colorbox{foo2}{    Money total = items.totalCost();}!        !\blackcircle{2}!
!\colorbox{foo3}{    Order order = billingSvc.charge(total);}! !\blackcircle{3}!
!\colorbox{foo4}{    return order;}!                           !\blackcircle{4}!
}
        \end{minted}
      \end{mdframed}
    \end{subfigure}

    %\vspace{1cm}
    \vspace{2mm}

    \begin{subfigure}{\textwidth}
      \caption{Choreography in Choral}
      \label{fig:choreographic}
      \centering
      \scriptsize
      \begin{mdframed}
\begin{minted}[linenos, escapeinside=!!]{'choral.py:ChoralLexer -x'}
// class Checkout@(Frontend, Cart, Billing)
GrpcChannel@(Frontend, Cart) ch1;    !\label{ch1}!
GrpcChannel@(Cart, Billing) ch2;     !\label{ch2}!
GrpcChannel@(Billing, Frontend) ch3; !\label{ch3}!
Order@Frontend checkout(Request@Frontend req) { !\label{chHeader}!
    Request@Cart reqC = ch1.com(req); !\label{chReqToCart}!
!\colorbox{foo1}{    Items@Cart items = cartSvc.getCart(reqC);}!        !\blackcircle{1}!!\label{chGetCart}!
!\colorbox{foo2}{    Money@Cart total = items.totalCost();}!            !\blackcircle{2}! !\label{chTotalCost}!
    Money@Billing totalB = ch2.com(total); !\label{chTotalToBilling}!
!\colorbox{foo3}{    Order@Billing order = billingSvc.charge(totalB);}! !\blackcircle{3}!!\label{chCharge}!
!\colorbox{foo4}{    return ch3.com(order);}!                           !\blackcircle{4}!!\label{chOrderToFrontend}!
}
\end{minted}
      \end{mdframed}
    \end{subfigure}

  \end{minipage}
  \hspace{0.04\textwidth}
  \begin{subfigure}[b]{0.445\textwidth}
    \caption{Choreography in Java}
    \label{fig:projection}
    \centering
    \scriptsize
    \begin{mdframed}
\begin{minted}[linenos, escapeinside=!!]{java}
// class Checkout_Frontend
GrpcChannel_A ch1; GrpcChannel_B ch3; !\label{chFrontend}!
Order checkout(Request req) { !\label{proHeaderFrontend}!
    ch1.com(req); // send !\label{proReqToCart}!
!\colorbox{foo4}{    return (Order) ch3.com();}! // recv       !\blackcircle{4}!!\label{proOrderFromFrontend}!
}
// class Checkout_Cart
GrpcChannel_B ch1; GrpcChannel_A ch2; !\label{chCart}!
void checkout() { !\label{proHeaderCart}!
    Request reqC = (Request) ch1.com(); // recv !\label{proReqFromCart}!
!\colorbox{foo1}{    Items items = cartSvc.getCart(userID);}!  !\blackcircle{1}!
!\colorbox{foo2}{    Money total = items.totalCost();}!        !\blackcircle{2}!
    ch2.com(total); // send !\label{proTotalToBilling}!
}
// class Checkout_Billing
GrpcChannel_B ch2; GrpcChannel_A ch3; !\label{chBilling}!
void checkout() { !\label{proHeaderBilling}!
    Money totalB = (Money) ch2.com(); // recv !\label{proTotalFromBilling}!
!\colorbox{foo3}{    Order order = billingSvc.charge(totalB);}!!\blackcircle{3}!
    ch3.com(order); // send !\label{proOrderToFrontend}!
}
\end{minted}
    \end{mdframed}

    \tikz[overlay, remember picture] {
      \draw [red,thick,-latex] (0.5, 5.3) to [bend left=50] (2.0, 3.7);
      \draw [red,thick,-latex] (0.7, 2.7) to [bend left=30] (2.6, 1.4);
      \draw [red,thick,-latex] (-2.5, 0.7) to [bend left=35] (-2.7, 5);
    }
  \end{subfigure}

  \caption{A simple but inefficient orchestration (a); a more efficient choreographic version in Choral (b); and the equivalent Java choreography (c). Manually converting \Cref{fig:orchestrated} into \Cref{fig:projection} would require `shredding' the code, destroying program structure and safety guarantees.}%
  \label{fig:code-comparison}
\end{figure*}

In this section we introduce choreographic programming as a natural solution to the problems in \Cref{sec:case-study}. As a running example, we will use the simple Java orchestration in \Cref{fig:orchestrated} based on the Online Boutique case study. The orchestrator begins by making an RPC to the cart service to get the user's shopping cart (line 3), then computes the total cost of the items (line 4), makes an RPC to the billing service (line 5) and returns the result.

When network latency is high, the orchestrated solution is inefficient: the cart service sends the items to the orchestrator, only for the orchestrator to add up the prices and forward the total to the billing service. What if our programming language could tell the cart service to compute the cost \emph{locally}, and to send the total cost \emph{directly} to the billing service? Such a language would need a way to specify `where' a computation should happen, and `how' data should be moved. Choreographic programming is a simple, low-level abstraction that provides precisely this feature.

\subsection{Migrating to Choral}\label{sec:migrating-to-choral}

The \emph{Choral}~\cite{giallorenzo2024} compiler adds choreographic programming to Java. We can make the \code{Checkout} class from \Cref{fig:orchestrated} into a Choral choreography by annotating the class name with a list of \emph{locations} (also called \emph{roles}): \choral{Checkout@(A,B,C)} is a Choral class with locations \code{A}, \code{B}, and \code{C}. Locations are arbitrary names, and the Choral compiler will translate the annotated class into three Java classes named \code{Checkout_A}, \code{Checkout_B}, and \code{Checkout_C}:

\vspace{\abovedisplayskip}
{\noindent\hfill%
\begin{minipage}{0.25\columnwidth}
\begin{minted}[fontsize=\footnotesize]{'choral.py:ChoralLexer -x'}
// Choral choreography
class Checkout@(A,B,C) {
  ...
}
\end{minted}
\end{minipage}\hfill%
{\large $\xrightarrow{\text{Choral Compiler}}$}%
\hfill%
\begin{minipage}{0.25\columnwidth}
\begin{minted}[fontsize=\footnotesize]{java}
// Java choreography
class Checkout_A {...}
class Checkout_B {...}
class Checkout_C {...}
\end{minted}
\end{minipage}\hfill%}
\vspace{\belowdisplayskip}

Inside a Choral class, we specify a location for every piece of data by annotating its type. \code{Request@Frontend req} is a \code{Request} object at location \code{Frontend}; in the compiled code, this line will appear as \code{Request req} in \code{Checkout_Frontend} and it will not appear in any of the other classes. Annotating our simple orchestrator from \Cref{fig:orchestrated}, we get the following:

\begin{minted}[fontsize=\footnotesize,escapeinside=!!]{'choral.py:ChoralLexer -x'}
// class Checkout@(Frontend,Cart,Billing)
Order@Frontend checkout(Request@Frontend req) {
    Items@Cart items = cartSvc.getCart(req);
    Money@Cart total = items.totalCost();
    Order@Billing order = billingSvc.charge(total);
    return order;
}
\end{minted}

\noindent Notice we specified that \code{total} has type \code{Money@Cart}, meaning the cart service will compute the total cost locally.

%Notice that \code{items} is located at \code{Cart}, so the expression \code{items.totalCost()} will only appear in \code{Checkout\_Cart}.

Although the code above is syntactically correct, Choral's type checker will reject it: the variable \code{req} is located at \code{Frontend}, but \code{cartSvc} is located at \code{Cart}. In high-level distributed programming languages like X10~\cite{charles2005}, communication would be managed by an opaque runtime layer. Choral choreographies are a \emph{low-level} abstraction because they add no runtime overhead, and all communication is done explicitly by invoking methods on user-defined objects.

To make our code typecheck, we create a \emph{channel} by implementing a pair of classes extending the Java interfaces \code{DiChannel_A} and \code{DiChannel_B}. For \Accompanist, we implemented a gRPC channel in less than 100 lines of plain Java.

\vspace{1em}

\noindent
\begin{minipage}{0.45\columnwidth}
\begin{minted}[fontsize=\footnotesize]{java}
class GrpcChannel_A<T>
implements DiChannel_A<T> {
// send method
<M extends T> Unit com(M m) {
    ...
}}
\end{minted}
\end{minipage}\hfill%
\begin{minipage}{0.45\columnwidth}
\begin{minted}[fontsize=\footnotesize]{java}
class GrpcChannel_B<T>
implements DiChannel_B<T> {
// recv method
<M extends T> M com() {
    ...
}}
\end{minted}
\end{minipage}

\vspace{1em}

\noindent Choral understands \code{GrpcChannel_A} and \code{GrpcChannel_B} as a single class \choral{GrpcChannel@(A,B)}. The unified class provides a method \code{com} that turns objects of type \code{M@A} into objects of type \code{M@B} by invoking the methods on our custom channel.

Using our custom channel, the full Choral version of our orchestrator appears in \Cref{fig:choreographic}.  On line \ref{chReqToCart}, the frontend sends the request to the cart service. On line \ref{chGetCart}, the cart service gets the user's cart. On lines \ref{chTotalCost} and \ref{chTotalToBilling}, the cart service computes the total cost locally and sends the result to the billing service. On lines \ref{chCharge} and \ref{chOrderToFrontend}, the billing service charges the user's card and sends the result back to the frontend. Although the Choral choreography is more verbose than the orchestration, it allows us to be more precise about where data is located and how it moves.

The Java code generated by the Choral compiler appears in \Cref{fig:projection}. It is important to notice that Choral adds no runtime overhead aside from the code we explicitly wrote in \Cref{fig:choreographic}. This is not because the Choral compiler performs complex optimizations; the transformation it performs is fairly simple, and in our experience it is very easy to predict how the source and target code correspond. For example:
\begin{enumerate}
  \item The method header on line \ref{chHeader} of \Cref{fig:choreographic} is translated into the method headers on lines \ref{proHeaderFrontend}, \ref{proHeaderCart}, and \ref{proHeaderBilling} of \Cref{fig:projection}. Notice the \code{Request} parameter and \code{Order} return type both appear in the \code{Checkout_Frontend} class, and do not appear in the other two classes.
  \item The communication on line \ref{chReqToCart} of \Cref{fig:choreographic} is translated into a `send' on line \ref{proReqToCart} of \Cref{fig:projection} and a `receive' on line \ref{proReqFromCart} of \Cref{fig:projection}, corresponding to the methods of our hand-written \code{GrpcChannel} classes.
\end{enumerate}
We will discuss the limitations of this translation in \Cref{sec:choral-limitations}. But first, we summarize the advantages of this new programming model.

\subsection{The Choreographic Programming Model}

The Java code in \Cref{fig:projection} could just as well have been written by hand. There are three key reasons why Choral is a more convenient frontend:

\paragraph{Structure}
Choral's choreographic code matches the structure of the orchestration: migrating from \Cref{fig:orchestrated} to \Cref{fig:choreographic} is straightforward, whereas migrating from \Cref{fig:orchestrated} to \Cref{fig:projection} less so. Notice in \Cref{fig:projection}, control flow jumps from line \ref{proReqFromCart} to line \ref{proReqToCart} when the frontend sends a message to the cart service, and there are similar jumps from line \ref{proTotalToBilling} to line \ref{proTotalFromBilling} and from line \ref{proOrderToFrontend} to line \ref{proOrderFromFrontend}. In contrast, the control flow of the Choral code stays inside a single procedure and follows a simple straight-line pattern. Choral also supports conditional control flow and recursion, so it can mirror the structure of an orchestrator even in complex workflows.

\paragraph{Safety}
If a programmer translated \Cref{fig:orchestrated} into \Cref{fig:projection} manually, it would be easy to make mistakes -- like forgetting to send the message on line \ref{proOrderToFrontend} or casting the message on line \ref{proReqToCart} to the wrong type. Such mistakes are easy to catch in our pedagogical example, but not so easy to catch when there is conditional control flow with complex input preconditions~\cite{LLLG16}. The Choral compiler guarantees generated code is deadlock-free, type-safe, and free from communication mismatches~\cite{giallorenzo2024}.

\paragraph{Deterministic Implicit Parallelism}
Choreographic programming provides an \emph{implicitly parallel} concurrency model, where any two expressions can execute in parallel as long as (1) they have different locations, and (2) neither expression depends on the other~\cite{montesi2023}. The concurrency model is also \emph{deterministic}, in that all executions have the same effects and return the same result, as long as services don't share state. As a result, converting an orchestrator to a Choral choreography will preserve the original behavior and possibly introduce safe, additional parallelism that wasn't in the original orchestration. There are cases where a nondeterministic concurrency model provides even greater parallelism; the programmer can opt in to handle these cases~\cite{plyukhin2024}.

\subsection{Limitations}\label{sec:choral-limitations}

Choreographic programming (CP) is a convenient programming model, but existing CP proposals have significant limitations that make it impractical in real systems:

\crefname{enumi}{disadvantage}{disadvantages}
\Crefname{enumi}{Disadvantage}{Disadvantages}
\begin{enumerate}[label=D\arabic*., ref=D\arabic*]
  \item \label{d4} CP requires services to block so they cannot participate in multiple choreographies at once;
  \item \label{d2} CP assumes a reliable network and non-faulty services;
  \item \label{d1} CP does not let services join a running choreography;
  \item \label{d3} CP requires rewriting the application from scratch.
\end{enumerate}
%Our key contribution in the following sections will be to bridge the gap using \Accompanist.

For example, on line \ref{proReqFromCart} of \Cref{fig:projection}, the cart service is expected to block until it receives the request; hence a slow checkout for one user will have knock-on effects for others in the pipeline. This blocking call also assumes communication is reliable; dropped or duplicated messages can cause data corruption~\cite{plyukhin2024}. CP languages are also not reactive: in \Cref{fig:projection} the frontend, cart service, and billing service must all know in advance to invoke their respective \code{checkout} methods. Previous attempts to make CP languages more reactive require modifying the choreography, either `shredding' it into event handlers~\cite{lugovic2024} or sacrificing determinism~\cite{plyukhin2024}.

In the next section, we show how these limitations can be addressed at the \emph{runtime level} instead of the language level, allowing developers to use Choral's programming model in real systems.

%Choreographic programming can also be added to conventional programming languages as a library~\cite{kashiwa2023,shen2023}. The advantage of Choral is (1) it extends a well-known language; (2) it supports advanced features like inheritance; and (3) we can manually inspect the generated code, which is written in plain Java.

\begin{figure}
  \begin{subfigure}{0.45\textwidth}
    \centering
    \includegraphics[width=\textwidth]{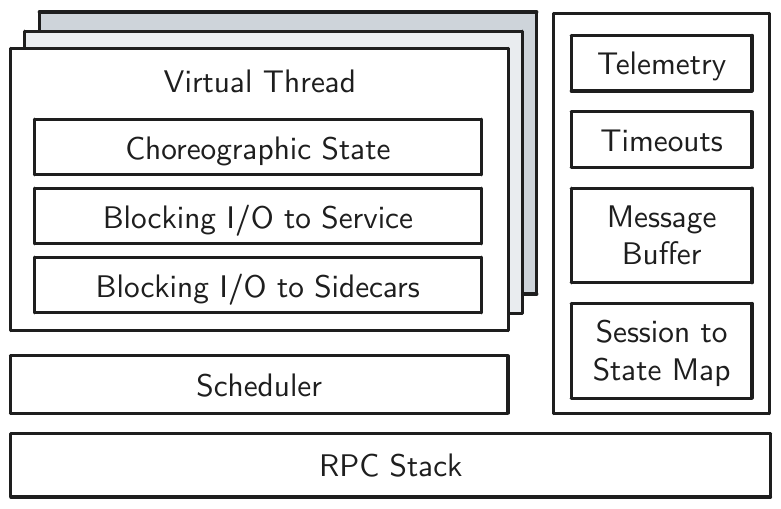}
    \Description{Accompanist consists of a scheduler that schedules choreographies in virtual threads}
    \caption{Overview of an \Accompanist sidecar.}%
    \label{fig:accompanist-sidecar-overview}
  \end{subfigure}
  \hfill
  \begin{subfigure}{0.45\textwidth}
    \centering
    \includegraphics[width=\textwidth]{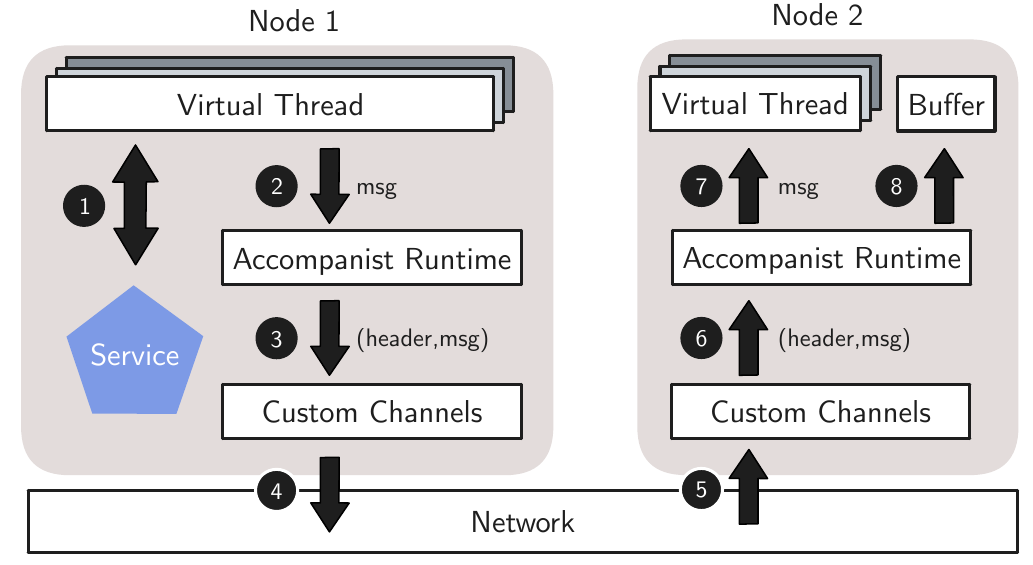}
    \Description{Accompanist sidecar sends messages through the runtime which appends headers to the messages before they are sent over the network to another sidecar.}
    \caption{Communication between  sidecars.}%
    \label{fig:accompanist-communication-overview}
  \end{subfigure}
  \caption{\Accompanist sidecars}
\end{figure}

\begin{table}
  \centering
  \begin{tabularx}{\columnwidth}{l X}
    \hline
    \textbf{Header} & \textbf{Description} \\
    \hline
    \emph{Session ID} & Uniquely identifies the choreography invocation. \\
    \emph{Choreography ID} & Uniquely identifies the Choral class that implements the choreography. \\
    \emph{Sender ID} & Uniquely identifies the sidecar that sent the message.  \\
    \emph{Seqnum} & A sequence number that counts messages sent from this sender to this receiver within this session.  \\
    \emph{Telemetry} & Metadata used to correlate OpenTelemetry spans with their session. \\
    \hline
  \end{tabularx}
  \caption{Headers attached to messages.}%
  \label{table:accompanist-headers}
\end{table}

\begin{figure*}
  \centering
  \includegraphics[width=0.8\textwidth]{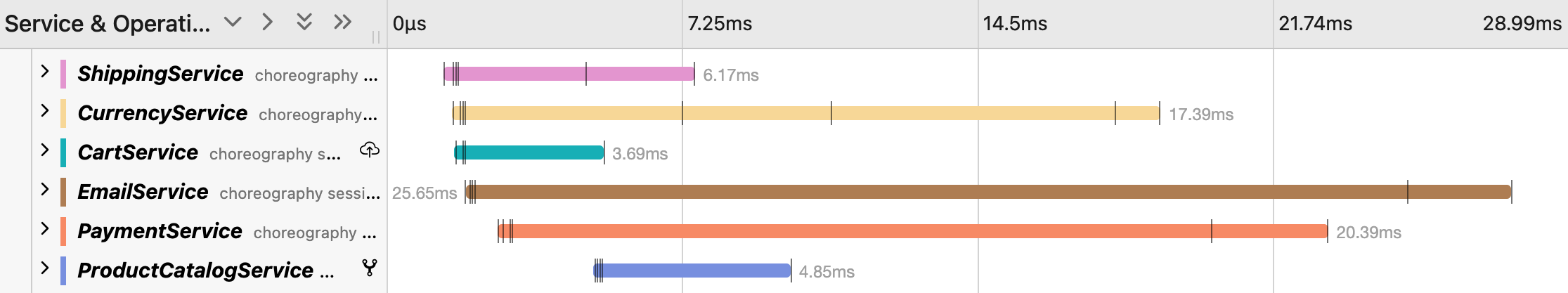}
  \Description{Screenshot of a Jaeger trace showing how each sidecar interacts with each other.}
  \caption{Telemetry visualization for the \Accompanist Online Boutique.}%
  \label{fig:jaeger}
\end{figure*}

%% file: accompanist.tex
\section{Decentralize Your Orchestrator with \Accompanist}\label{sec:accompanist}

To make Choral practical for service-oriented applications, we developed \Accompanist. The runtime enables incremental replacement of high-latency orchestrators in an existing codebase with more efficient choreographies---all without modifying any of the underlying services.
% The runtime dynamically instantiates compiled Choral code on demand and multiplexes resources, so services can participate in multiple choreographies at once.

\subsection{Overview}

\Cref{fig:accompanist-sidecar-overview} shows an overview of the runtime environment. Every invocation of a Choral choreography is mapped to a lightweight virtual thread; the thread's stack encapsulates any local state the sidecar needs while the choreography is executing, and the thread is descheduled when making blocking RPCs to worker services or when it is waiting for a message from another sidecar. Each choreography invocation is called a \emph{session}, and the runtime environment maintains a mapping from session IDs to virtual threads. The runtime also maintains telemetry information for each session, timeouts to handle failures, and a message buffer to handle out-of-order messages. Communication between sidecars is handled by the network stack, which can be managed by the programmer with Choral's channel API (\Cref{sec:migrating-to-choral}).

\Accompanist multiplexes the network by adding headers to outgoing messages, summarized in \Cref{table:accompanist-headers}. When a choreography is first invoked, the initiator generates a unique session ID.
The choreography ID is used for services to dynamically join a running choreography, as explained below. The sender ID and seqnum headers are used to handle faults and out-of-order delivery, which we also explain below.
Telemetry headers allow Choral choreographies to integrate with distributed tracing and logging tools, as shown in \Cref{fig:jaeger}.

\subsection{Communication}

\Cref{fig:accompanist-communication-overview} breaks down a choreographic interaction between two nodes in our framework. In step \blackcircle{1}, target code generated by Choral runs in a virtual thread and interacts with its associated worker service via blocking RPCs. These RPCs are efficient because sidecars and worker services are always deployed to the same node, and other virtual threads can execute while the original thread is blocked. Notably, the sidecar uses RPCs instead of IPCs so the worker service need not be modified; thus \Accompanist choreographies can coexist with traditional orchestrators.

In step \blackcircle{2} of \Cref{fig:accompanist-communication-overview}, the sidecar on Node 1 sends a message to the sidecar on Node 2. The message is intercepted by the runtime environment, which adds the headers shown in \Cref{table:accompanist-headers} and forwards the message to the user-defined communication channel (step \blackcircle{3}). If the message arrives at its destination in steps \blackcircle{4}--\blackcircle{6}, the recipient's runtime inspects the message headers. If the recipient has not seen the session ID before, a new virtual thread is instantiated using the choreography ID to determine which choreography to run; if the recipient has seen the session ID before and the seqnum and sender ID match their expected values, the message is delivered to the virtual thread (step \blackcircle{7}). If the seqnum or sender ID do not match their expected value, the message is buffered because it may have been delivered out of order (step \blackcircle{8}). If the choreography timed out before the message was received (\Cref{sec:termination}) the runtime drops the message.

%The extra concurrency means that whereas a single choreography is deterministic, the interleaving of two choreographies might not be. This is the same problem faced by orchestrated code: applications must make sure their workflows are serializable.

\subsection{Termination}%
\label{sec:termination}

When a virtual thread finishes a choreography (or takes too long to finish) the session is killed and its resources reclaimed. Any subsequent messages pertaining to a killed session are ignored.
Virtual threads also handle dropped and reordered messages by checking sequence numbers attached to each communication; messages that arrive out of sequence are buffered until they can be processed in order or the session is killed.
This gives us an \emph{at most once} execution model, where faults may prevent the choreography from completing, but resources are always reclaimed.

\subsection{Fault Tolerance}%
\label{sec:accompanist-fault-tolerance}

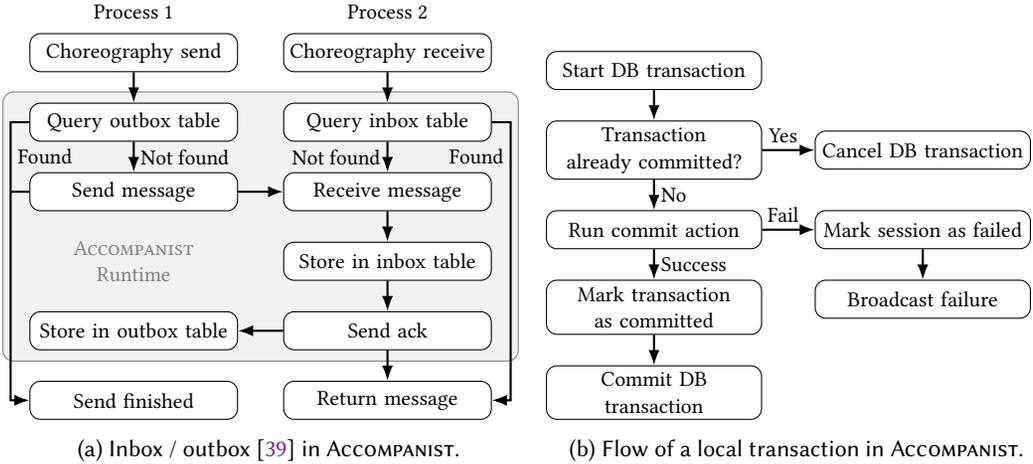
\begin{figure*}
  \begin{subfigure}{0.51\textwidth}
    \input{graphics/inbox-outbox-diagram.tex}
    \caption{Inbox / outbox~\cite{richardson_microservices_2019} in \Accompanist.}
    \label{fig:inbox-outbox-diagram}
  \end{subfigure}
  \hfill
  \begin{subfigure}{0.48\textwidth}
    \input{graphics/accompanist-transaction-flow.tex}
    \caption{Flow of a local transaction in \Accompanist.}
    \label{fig:transaction-flowchart}
  \end{subfigure}
  \caption{The durable communication flows (a), and transaction flow (b), with fault tolerance.}
\end{figure*}

\Accompanist{} also provides a fault-tolerant mode with \emph{at least once} semantics, assuming all operations in the choreography are deterministic\footnote{Given the same initial state, every execution results in an identical trace of messages at each process where message content must be equivalent in terms of application logic.} and idempotent. The extension uses process-local databases to store metadata for each choreography invocation: the session ID, the choreography name, and its \emph{status}---which may be ``started'', ``failed'', or ``completed''. For simplicity, all faults are handled by restarting the session. Adding customizable retry policies would be straightforward, for example, by adding an optional parameter to the \texttt{com} method on line 8 of \Cref{fig:accompanist-saga-choreography}.

When a sidecar restarts, the database is checked to see if any sessions need to be restored. \Accompanist{} restores sessions by \emph{replaying} them from their initial state. This requires a durable message queue, which we implement using the inbox/outbox pattern~\cite{richardson_microservices_2019} as shown in \Cref{fig:inbox-outbox-diagram}. When a Choral program sends or receives a message, it will interact with the \Accompanist runtime (shaded in gray). When the runtime receives a message, the contents are stored in a persistent `inbox' table, acknowledged, and returned to the Choral program. When the runtime sends a message, an asynchronous listener is installed to await the acknowledgment and control immediately returns to the Choral program. When a restarted process needs to receive a message, the runtime will first query its inbox to see if the message was already received. Likewise, the runtime will drop any sent messages already found in the outbox. If a message or acknowledgment is not received within a specified timeout, the session is restarted. Timeouts ensure messages are re-sent until successful, and message identifiers ensure duplicate messages are ignored.

\begin{figure*}
  \begin{subfigure}{0.49\textwidth}
    \begin{mdframed}[innerleftmargin=1mm]
    \begin{minted}[fontsize=\scriptsize]{java}
class ProductTransaction implements Transaction {

String transactionName() {
  return "updateProductStock";
}

boolean commit(int sessionID, SQLTransaction trans) {
  int stock = trans.exec("SELECT stock FROM product");
  if (stock == 0) { return false; }
  trans.exec("UPDATE product SET stock = stock - 1");
  return true;
}

void compensate(int sessionID, SQLTransaction trans) {
  trans.exec("UPDATE product SET stock = stock + 1");
}}
    \end{minted}
    \end{mdframed}
    \caption{Implementing a local transaction in Java.}
    \label{fig:transaction-impl}
  \end{subfigure}
  \hfill
  \begin{subfigure}{0.43\textwidth}
    \begin{mdframed}[innerleftmargin=1mm]
    \begin{minted}[fontsize=\scriptsize,linenos]{'choral.py:ChoralLexer -x'}
class ProductWorkflow@(Product, Payment) {
  void run(
    Order@Product order,
    FaultSessionContext@Product ctx_prd,
    FaultSessionContext@Payment order_pmt,
  ) {
    ctx_prd.transaction(
      new ProductTransaction@Product(order));
    Order@Payment order_pmt = ch.com(order);
    order_pmt.transaction(
      new PaymentTransaction@Payment(order));
  }
}
    \end{minted}
    \end{mdframed}
    \caption{Example of a saga transaction in Choral using the \textit{fault tolerant extension} of \Accompanist.}
    \label{fig:accompanist-saga-choreography}
  \end{subfigure}
  \caption{Transactions in \Accompanist with fault tolerance.}
\end{figure*}

% Motivation
%The \textit{fault tolerance extension} presented in \Cref{sec:accompanist-fault-tolerance}, ensures an \textit{at-least-once} execution model of the choreography. This model, however, falls short in two aspects. First, if an action must only happen once, the programmer must make the action idempotent, for instance by also storing the fact that the action took place when committing the action.

%Secondly, and perhaps more importantly, many practical workflows allow for failures.

\subsection{Saga Transactions}
\label{sec:accompanist-transactions}

As an application of our framework, we show how \emph{saga transactions}~\cite{sagas} can be implemented in a choreographic programming language. To motivate the problem, consider a workflow for buying a product. First the \texttt{product} service performs a local transaction that reserves the product. Next the \texttt{payment} service performs a local transaction that processes the payment. Either transaction could fail: the product might be out of stock and the user might have insufficient funds. If the first transaction fails, we do not want to charge the user's card; if the second transaction fails, we want to \emph{roll back} the product reservation. Sagas are distributed transactions that consist of several local transactions and provide the following guarantee: either (a) every local transaction eventually succeeds, or (b) some transaction fails and every local transaction is rolled back.

% Solution
With \Accompanist{}, programmers implement local transactions by extending the \texttt{Transaction} interface (\Cref{fig:transaction-impl}). Local transactions can be combined into choreographic sagas as shown in \Cref{fig:accompanist-saga-choreography}. When a process tries to commit a transaction, it follows the flow in \Cref{fig:transaction-flowchart}. First, it starts a new database transaction. Then it checks if the transaction has already been committed (due to a restart of the session). If so, the new transaction is canceled and the process proceeds; this ensures transactions are idempotent. If the transaction has not already been committed, the process runs the commit action. If committing succeeds, the transaction is marked complete. If committing fails, the entire session is marked as failed and all previous transactions for this session rolled back. The process then broadcasts the session failure to the other participants, ensuring they will abort the choreography and roll back their transactions.

%\subsection{Development and Deployment}%
%\label{sec:deployment}

%We used \Accompanist to rewrite the Online Boutique checkout orchestrator and the Hotel Reservation search orchestrator (\Cref{sec:case-study}) as choreographies. Each choreography is implemented as a single Choral method that interacts with services via auto-generated gRPC client stubs.\footnote{Currently, the Choral compiler requires manually writing headers to use Java code; a future version could make this boilerplate unnecessary.} Compiled Choral code is deployed to sidecars alongside each service, with gRPC connections between the sidecars, as depicted in \Cref{fig:orchestration-choreography-accompanist}c.

%Conceptually, our approach replaces the `monolithic' orchestrator service with a decentralized collection of sidecars. Sidecars still interact with services via RPC like a normal orchestrator, but the calls are fast because they come from within the same Kubernetes pod.

%%% Local Variables:
%%% mode: latex
%%% TeX-master: "main"
%%% End:

%% file: graphics/inbox-outbox-diagram.tex
\scriptsize
\usetikzlibrary{shapes.geometric, arrows.meta, positioning, fit, backgrounds}
\begin{tikzpicture}[
    node distance=0.4cm and 0.6cm,
    every node/.style={font=\footnotesize},
    process/.style={rectangle, draw, text width=2.6cm, align=center, rounded corners, minimum height=0.5cm, fill=white},
    arrow/.style={-{Latex[length=2mm]}, thick}
  ]

  % Nodes
  \node[process] (B1) {Choreography receive};
  \node[process, below=of B1] (B2) {Query inbox table};
  \node[process, below=of B2] (B3) {Receive message};
  \node[process, below=of B3] (B4) {Store in inbox table};
  \node[process, below=of B4] (B5) {Send ack};
  \node[process, below=of B5] (B6) {Return message};

  \node[process, left=of B1] (A1) {Choreography send};
  \node[process, below=of A1] (A2) {Query outbox table};
  \node[process, left=of B3] (A3) {Send message};
  \node[process, left=of B5] (A4) {Store in outbox table};
  \node[process, below=of A4] (A5) {Send finished};

  \node[above=1mm of B1] (B) {Process 2};
  \node[above=1mm of A1] (A) {Process 1};

  % Arrows
  \draw[arrow] (A1) -- (A2);
  \draw[arrow] (A2) -- node[right]{Not found} (A3);
  \draw[arrow] (A3) -- (B3);
  \draw[arrow] (B5) -- (A4);
  % \draw[arrow] (A4) -- (A5);

  \draw[arrow] (B1) -- (B2);
  \draw[arrow] (B2) -- node[left]{Not found} (B3);
  \draw[arrow] (B3) -- (B4);
  \draw[arrow] (B4) -- (B5);
  \draw[arrow] (B5) -- (B6);

  \draw[arrow] (B2.east) -- ++(2.5mm,0) |- node[pos=0.061, left] {Found} (B6.east);

  \draw[arrow] (A2.west) -- ++(-2.5mm,0) |- node[pos=0.061, right] {Found} (A5.west);
  \draw[style=thick] (A3.west) -- ++(-2.5mm,0);

  \node[align=center,below=of A3, yshift=1mm, text=gray]{\Accompanist \\ Runtime};

  \begin{scope}[on background layer]
    \node[
      fill=gray!10,
      draw=gray,
      % dashed,
      inner xsep=10pt,
      inner ysep=4pt,
      rounded corners,
      fit=(A2) (B2) (A4) (B5)
    ] (box) {};
  \end{scope}

\end{tikzpicture}

%% file: graphics/accompanist-transaction-flow.tex
\scriptsize
\usetikzlibrary{shapes.geometric, arrows.meta, positioning}
\begin{tikzpicture}[
    node distance=0.4cm and 0.7cm,
    every node/.style={font=\footnotesize},
    process/.style={rectangle, draw, text width=2.7cm, align=center, rounded corners, minimum height=0.5cm},
    arrow/.style={-{Latex[length=2mm]}, thick}
  ]

  % Nodes
  \node[process] (A) {Start DB transaction};
  \node[process, below=of A] (B) {Transaction \\already committed?};
  \node[process, below=of B] (C) {Run commit action};
  \node[process, right=of B] (D) {Cancel DB transaction};
  \node[process, below=of C] (E) {Mark transaction as committed};
  \node[process, below=of E] (G) {Commit DB transaction};
  \node[process, right=of C] (F) {Mark session as failed};
  \node[process, below=of F] (H) {Broadcast failure};

  % Arrows
  \draw[arrow] (A) -- (B);
  \draw[arrow] (B) -- node[right]{No} (C);
  \draw[arrow] (B) -- node[pos=0.4, above]{Yes} (D);
  \draw[arrow] (C) -- node[right]{Success} (E);
  \draw[arrow] (C) -- node[pos=0.4, above]{Fail} (F);
  \draw[arrow] (E) -- (G);
  \draw[arrow] (F) -- (H);

\end{tikzpicture}

%% file: fault-tolerance.tex
\section{Formal Model}\label{sec:fault-tolerance}

We will now use ideas from the previous sections to build a formal model for choreographic saga transactions that are correct-by-construction, where processes may crash and restart back to their initial state at any time. At first blush, the task seems monumental! But a clean separation of concerns allows us to prove an end-to-end correctness result with three simple pieces:

\begin{enumerate}
  \item First, we extend an ordinary choreographic programming model with transactions that can succeed, fail, or be compensated. At this level, it is easy to prove that transactions are deadlock-free and that \emph{either} (a) all processes succeed, \emph{or} (b) all processes invoke compensating transactions.
  \item Second, we introduce the network model and establish a textbook bisimulation result that relates choreographic programs to compiled choreographies at the network level---assuming networks are fault-free, i.e., processes never restart.
  \item Third, we design a novel \emph{efficiency relation} $(\preceq)$ that relates fault-free executions to fault-prone executions. This relation relies on the existence of a durable message queue and the assumption that processes are deterministic.
\end{enumerate}
By composing the pieces together, properties about choreographic programs extend down to the level of fault-prone networks. We focus on the high-level ideas here; our precise formalization is in the Supplemental Material.

\subsection{Definitions}

The set of participating process names in a network is denoted $\PName$.
A \textit{messaging state}, $K: \PName \times \PName \times \mathbb{N} \rightarrow \mathbf{Value} + \bot$,
is a function that maps a sender, a receiver, and a sequence number to a value or $\bot$ if no value is present.

A \textit{sequence number state}, $S: \PName \times \PName \times \{ \textsc{send, receive} \} \rightarrow \mathbb{N}$, is a function that maps a local process, a remote process, and the labels send and receive to a sequence number.
It is used to keep track of the next sequence numbers to send and receive between processes.

The operation $\textit{incSend}(S,\p,\q)$ produces a sequence number state where $\p$'s send count to $\q$ has been incremented.
Likewise, the operation $\textit{incRecv}(S,\p,\q)$ produces a sequence number state where $\p$'s receive count to $\q$ has been incremented.
The operation $restart(S,\p)$ produces a sequence number state where $\p$'s send and receive counts have been reset to zero.

A \textit{choreographic store}, $\Sigma : \PName \times \textit{Var} \rightarrow \textit{Value}$, denotes the local state of each process in a choreography or network.

A \textit{transaction state}, $T : \PName \rightarrow [\mathbf{Trans}]$, is a mapping from process names to sequences of transactions,
where a transaction $\mathbf{Trans}$ can either be a commit action $t(v) = v'$, which reads ``committing transaction $t$ with input $v$ produces result $v'$'', or a compensation action, $c(t(v) = v')$.

\subsection{Choreography Model}

\begin{figure*}
  \centering
  \footnotesize

  \begin{align*}
    C & \coloneq I ; C \mid \mathbf{0} \\
    I & \coloneq \p.e \rightarrow \q.x \mid \p \rightsquigarrow \q.x \mid \p \rightarrow \q[\textsc{l}] \mid \p \rightsquigarrow \q[\textsc{l}] \mid \p.x \coloneq e \mid \mathrm{if}\ \p.e\ \mathrm{then}\ C_1\ \mathrm{else}\ C_2 \mid \p.x \coloneq t(e) \\
    e & \coloneq v \mid x \mid f(\vec{e})
  \end{align*}

  \begin{align*}
    \inferrule* [Right=\textsc{C-local}]
    {
      \Sigma(\p) \vdash e \downarrow v \\
      \p \in A
    }
    {
      \trans
      {\p.x \coloneq e ; C, \Sigma, K, S, T, A}
      { \tau @ \p }
      {C, \Sigma[\p.x \mapsto v], K, S, T, A}
    }
  \end{align*}
  \begin{align*}
    \inferrule* [Right=\textsc{C-delay}]
    {
      \trans{C, \Sigma, K, S, T, A}{\mu}{C', \Sigma', K', S', T', A'} \\
      \p \neq \pn(\mu)
    }
    {
      \trans
      {I ; C, \Sigma, K, S, T, A}
      { \mu }
      {I ; C', \Sigma', K', S', T', A'}
    }
  \end{align*}
  \begin{align*}
    \inferrule* [Right=\textsc{C-send-val}]
    {
      i = S(\p, \q, \textsc{send}) \\
      \Sigma(\p) \vdash e \downarrow v \\
      \p \in A
    }
    {
      \trans
      {\p.e \rightarrow \q.x ; C, \Sigma, K, S, T, A}
      {\p.v \rightarrow \q !}
      {\p.e \rightsquigarrow \q.x ; C, \Sigma, K[\p, \q, i \mapsto v], \mathit{incSend}(S, \p, \q), T, A}
    }
  \end{align*}
  \begin{align*}
    \inferrule* [Right=\textsc{C-recv-val}]
    {
      i = S(\p, \q, \textsc{receive}) \\
      K(\p, \q, i) = v \\
      v \ne \bot \\
      \q \in A
    }
    {
      \trans
      {\p \rightsquigarrow \q.x ; C, \Sigma, K, S, T, A}
      {\p.v \rightarrow \q ?}
      {C, \Sigma[\q.x \mapsto v], K, \mathit{incRecv}(S, \q, \p), T, A}
    }
  \end{align*}
  \begin{align*}
    \inferrule* [Right=\textsc{C-success}]
    {
      \Sigma(\p) \vdash e \downarrow v \\
      t(v) = v' \\
      \p \in A
    }
    {
      \trans
      {\p.x \coloneq t(e);C, \Sigma, K, S, T, A}
      {\tau @ \p}
      {C, \Sigma[\p.x \mapsto v'], K, S, T[\p \Mapsto t(v)=v'], A}
    }
  \end{align*}
  \begin{align*}
    \inferrule* [Right=\textsc{C-fail}]
    {
      \p \in A \\
      t \notin T(\p)
    }
    {
      \trans
      {\p.x \coloneq t(e);C, \Sigma, K, S, T, A}
      {\textsc{compensate} @ \p}
      {C, \Sigma, K, S, \mathit{comp}(T, \p), A \setminus \{ \p \}}
    }
  \end{align*}
  \begin{align*}
    \inferrule* [Right=\textsc{C-compensate}]
    {
      A \ne A_\textit{start} \\
      \p \in A
    }
    {
      \trans
      {C, \Sigma, K, S, T, A}
      {\textsc{compensate} @ \p}
      {C, \Sigma, K, S, \mathit{comp}(T, \p), A \setminus \{\p\}}
    }
  \end{align*}
  \caption{Selected syntax and operational semantics for choreographies.}%
  \label{fig:formal-fault-choreography}
\end{figure*}

We assume an unbounded set of variables denoted $x$ in the syntax, and process names $\p,\q,\r$. A choreography $C$ is either empty $\mathbf 0$ or a sequence $I;C$ where $I$ is an instruction and $C$ is its continuation. An instruction can be one of the following.
Instruction $\p.e \rightarrow \q.x$ indicates that $\p$ puts a value on the message queue addressed to $\q$, whereas $\p \rightsquigarrow \q.x$ indicates that $\q$ receives the message from the message queue.
The $\p \rightsquigarrow \q.x$ instruction exists only as a runtime term, and is not allowed to be used directly in a choreographic program.
Instructions $\p \rightarrow \q[\textsc{l}]$ and $\p \rightsquigarrow \q[\textsc{l}]$ are the same except for communicating labels used to determine knowledge of choice.
$\p.x \coloneq e$ is a local assignment,
and `$\text{if $\p.e$ then $C_1$ else $C_2$}$' indicates a conditional evaluated at $\p$. Lastly, the transaction instruction $\p.x \coloneq t(e)$ reads ``process $\p$ performs transaction $t$ with the result of evaluating $e$ as input and stores the result of the transaction in variable $x$''. The syntax for fault tolerant choreographies is summarized at the top of \Cref{fig:formal-fault-choreography}.

The configuration for choreographies is a 6-tuple $\conf{ C, \Sigma, K, S, T, A }$,
where $A$ is the set of active processes (processes that have not compensated due to a failure).
The initial configuration is the tuple $\conf{ C_\textit{start}, \Sigma_\textit{start}, K_\textit{start}, S_\textit{start}, T_\textit{start}, A_\textit{start} }$.

\Cref{fig:formal-fault-choreography} also presents selected rules for the operational semantics of choreographies. The \textsc{C-local} rule performs a local assignment by evaluating the expression $e$ using $\p$'s local state. The \textsc{C-delay} rule captures the out-of-order semantics of choreographic programs; it allows process $\q$ to execute an instruction in $C$ without waiting for other processes to execute $I$.

The \textsc{C-send-val} and \textsc{C-recv-val} rules define how processes communicate. When a message is sent, it is atomically appended to the queue $K$. In practice, $K$ might be implemented as a durable message queue on a remote service---in which case, the sending process will repeatedly attempt to add the message to the queue until the queue replies with an acknowledgment. We can safely model this as an atomic action because we assume the sending process will not proceed until it receives an acknowledgment. To model when a process receives a message, the \textsc{C-recv-val} rule reads a value from $K$ and stores it in process $\p$'s local state. The sequence number $S(\q, \p, \textsc{receive})$ is used to index the message in the queue. %Messages are never removed from the queue; this will allow messages to be received again if the process restarts.

The semantics of sagas is captured by $\textsc{C-success}$, $\textsc{C-fail}$,
$\textsc{C-compensate}$. The $\textsc{C-success}$ rule commits a transaction to the transaction store. In the $\textsc{C-fail}$ rule, the transaction fails and $\p$ compensates its previously committed transactions. Once a process has failed, all other processes eventually use the $\textsc{C-compensate}$ rule to also compensate their transactions.

At the choreography level, it is relatively easy to see that choreographies either reduce to $\mathbf{0}$ or all perform compensating transactions. This is formalized by the following two theorems.

\manualTheorem{Lemma 3}{
  (Deadlock-Freedom for Choreographies).
  If $s$ is a terminated choreographic configuration, then either:
  \begin{enumerate}
    \item $s$ has the form $\conf{\mathbf 0,\Sigma,K,S,T,A}$ for some $\Sigma,K,S,T,A$; or
    \item $s$ has the form $\conf{C,\Sigma,K,S,T,\emptyset}$ for some $C,\Sigma,K,S,T$.
  \end{enumerate}
}

\manualTheorem{Theorem 2}{
  (Weak Atomicity for Choreographies).
  Let $\sigma$ be a terminated choreographic execution and $s_{\textit{final}}$ the final configuration. Then either:
  \begin{enumerate}
    \item $s_{\textit{final}}$ has the form $\langle \mathbf{0}, \Sigma, K, S, T, \PName \rangle$
      where $T$ has no compensating transactions; or
    \item $s_{\textit{final}}$ has the form $\langle C, \Sigma, K, S, T, \emptyset \rangle$
      where $T(\p)$ has the form $\langle t_1, t_2, \dots, t_i, c(t_i), \dots, \\ c(t_2), c(t_1) \rangle$, for some $i$, for each process $\p \in \PName$.
  \end{enumerate}
}

% TODO(Dan): Are there more relevant theorems to state here?

\noindent\\
Next, we explain how choreographies relate to networks where processes may fail.

\subsection{Network Model}\label{sec:model-networks}

\begin{figure*}
  \centering
  \footnotesize

  \begin{align*}
    P, Q, R &\coloneq I; P \mid \mathbf{0} \\
    I &\coloneq \p!e \mid \p?x \mid \p \oplus \textsc{l}
    \mid \p \& \left\{ (\textsc{l})_i : P_i \right\}_{i \in I} \mid x \coloneq e
    \mid \text{if $e$ then $P$ else $Q$} \mid x \coloneq t(e) \\
    e &\coloneq v \mid x \mid f(\vec{e})
  \end{align*}

  \begin{align*}
    \inferrule* [Right=\textsc{P-local}]
    {
      \Sigma(\p) \vdash e \downarrow v \\
      \p \in A
    }
    {
      \trans
      {\p [ x \coloneq e; P ], \Sigma, K, S, T, A}
      { \tau @ \p }
      {\p [P], \Sigma[\p.x \mapsto v], K, S, T, A}
    }
  \end{align*}
  \begin{align*}
    \inferrule* [Right=\textsc{P-par}]
    {
      \trans{ N, \Sigma, K, S, T, A }{\mu}{ N', \Sigma', K', S', T', A' }
    }
    {
      \trans{ N \mid M, \Sigma, K, S, T, A }{\mu}{ N' \mid M, \Sigma', K', S', T', A' }
    }
  \end{align*}
  \begin{align*}
    \inferrule* [Right=\textsc{P-send-val}]
    {
      i = S(\p, \q, \textsc{send}) \\
      \Sigma(\p) \vdash e \downarrow v \\
      \p \in A
    }
    {
      \trans
      {\p[\q!e ; P], \Sigma, K, S, T, A}
      {\p.v \rightarrow \q !}
      {\p[P], \Sigma, K[\p, \q, i \mapsto v], \mathit{incSend}(S, \p, \q), T, A}
    }
  \end{align*}
  \begin{align*}
    \inferrule* [Right=\textsc{P-recv-val}]
    {
      i = S(\p, \q, \textsc{receive}) \\
      K(\p, \q, i) = v \\
      v \ne \bot \\
      \q \in A
    }
    {
      \trans
      {\q[\p?x ; Q], \Sigma, K, S, T, A}
      {\p.v \rightarrow \q ?}
      {\q[Q], \Sigma[\q.x \mapsto v], K, \mathit{incRecv}(S, \p, \q), T, A}
    }
  \end{align*}
  \begin{align*}
    \inferrule* [Right=\textsc{P-restart}]
    { }
    {
      \trans
      {\p[P], \Sigma, K, S, T, A}
      {\textsc{restart}@\p}
      {\p[N_\textit{start}(\p)], \Sigma(\p \mapsto \Sigma_\textit{start}(\p)), K, \mathit{restart}(S, \p), T, A}
    }
  \end{align*}
  \begin{align*}
    \inferrule* [Right=\textsc{P-success}]
    {
      \Sigma \vdash e \downarrow v \\
      t(v) \downarrow v' \\
      \p \in A
    }
    {
      \trans
      {\p[x \coloneq t(e);P], \Sigma, K, S, T, A}
      {\tau @ \p}
      {\p[P], \Sigma[\p.x \mapsto v'], K, S, T[\p \Mapsto t(v)], A}
    }
  \end{align*}
  \begin{align*}
    \inferrule* [Right=\textsc{P-fail}]
    { \p \in A }
    {
      \trans
      {\p[x \coloneq t(e); P], \Sigma, K, S, T, A}
      {\textsc{compensate} @ \p}
      {\p[P], \Sigma, K, S, \mathit{comp}(T, \p), A \setminus \{ \p \}}
    }
  \end{align*}
  \begin{align*}
    \inferrule* [Right=\textsc{P-compensate}]
    {
      A \subset A_\textit{start} \\
      \p \in A
    }
    {
      \trans
      {\p[P], \Sigma, K, S, T, A}
      {\textsc{compensate} @ \p}
      {\p[P], \Sigma, K, S, \mathit{comp}(T, \p), A \setminus \{\p\}}
    }
  \end{align*}

  \caption{Selected syntax and operational semantics for networks.}%
  \label{fig:formal-fault-process}
\end{figure*}

A process $P$ is either empty $\mathbf 0$ or a sequence $I; P$,
where $I$ is an instruction and $P$ is its continuation. An instruction can be one of the following.
Instruction $\p!e$ denotes sending the result of expression $e$ to the message queue with $\p$ as the recipient.
Similar, $\p?x$ denotes receiving a value from the message queue, where the sender is $\p$, and store it in its local variable $x$.
$\p\oplus\textsc{l}$ sends a label to $\p$, and $\p \& \left\{ (\textsc{l})_i : P_i \right\}_{i \in I}$ receives a label $\textsc{l}_i$ from $\p$, and then continues with the associated program $P_i$.
$x \coloneq e$ is a local assignment, and `$\text{if $e$ then $P$ else $Q$}$' a conditional.
Lastly, the $x \coloneq t(e)$ instruction performs a local transaction at the process.
The expressions $v, x$, and $f(\vec{e})$, respectively denote a value, variable, and a function call.
The syntax for fault tolerant processes is summarized at the top of \Cref{fig:formal-fault-process}.

The configuration for networks is a 6-tuple $\conf{ N, \Sigma, K, S, T, A }$,
where $N$ is a network of processes.
The initial configuration is a tuple $\conf{ N_\textit{start}, \Sigma_\textit{start}, K_\textit{start}, S_\textit{start}, T_\textit{start}, A_\textit{start} }$.

\Cref{fig:formal-fault-process} also presents selected rules for the operational semantics of processes.
Rules \textsc{P-local, P-send-val, P-recv-val, P-success, P-fail}, and \textsc{P-compensate}, are the process versions of the respective choreography rules from \Cref{fig:formal-fault-choreography}.
The \textsc{P-par} rule allows different processes in a network to take steps concurrently.
The special \textsc{restart} rule allows individual processes to restart its execution at any point. The messaging state $K$ persists across restarts.

The termination theorem for networks states that every network will eventually terminate as long, as the number of restart events is unbounded,
since otherwise a network where a process keeps restarting infinitely will never make progress.

\manualTheorem{Theorem 1}{
  (Network Termination). Let $\sigma$ be a finite network execution. Then there exists a terminating execution $\sigma'$ that extends $\sigma$. Moreover, for any $k \in \mathbb N$, there exists $n \in \mathbb N$ such that: any $\sigma'$ that extends $\sigma$ and has at most $k$ \textsc{P-restart} events must terminate in at most $n$ steps. Conversely, if $\sigma$ is an infinite terminated network execution, then $\sigma$ has a finite terminated prefix.
  \\
}

\begin{figure*}
  \centering
  \footnotesize
  \begin{minipage}{0.45\textwidth}
    \begin{align*}
      \epp{ \p.e \rightarrow \q.x ; C }_\mathsf{r} &\triangleq
      \begin{cases}
        \q!e; \epp{C}_\mathsf{r} \qquad& \mathrm{if}\ \mathsf{r = p} \\
        \p?e; \epp{C}_\mathsf{r} \qquad& \mathrm{if}\ \mathsf{r = q} \\
        \epp{C}_\mathsf{r} \qquad& \mathrm{otherwise}
      \end{cases}
      \\
      \epp{ \p \rightsquigarrow \q.x ; C }_\mathsf{r} &\triangleq
      \begin{cases}
        \p?x; \epp{C}_\mathsf{r} \qquad& \mathrm{if}\ \mathsf{r = q} \\
        \epp{C}_\mathsf{r} \qquad& \mathrm{otherwise}
      \end{cases}
    \end{align*}
  \end{minipage}
  \begin{minipage}{0.45\textwidth}
    \begin{align*}
      \epp{ \p.x \coloneq e ; C }_\mathsf{r} &\triangleq
      \begin{cases}
        \p.x \coloneq e ; \epp{C}_\mathsf{r} \qquad& \mathrm{if}\ \mathsf{r = q} \\
        \epp{C}_\mathsf{r} \qquad& \mathrm{otherwise}
      \end{cases}
      \\
      \epp{ \p.x \coloneq t(e); C }_r &\triangleq
      \begin{cases}
        x \coloneq t(e) ; \epp{C}_\mathsf{r} \qquad& \mathrm{if}\ \mathsf{r = q} \\
        \epp{C}_\mathsf{r} \qquad& \mathrm{otherwise}
      \end{cases}
    \end{align*}
  \end{minipage}

  \caption{Selected rules for endpoint projection (EPP).}
  \label{fig:formal-fault-epp}
\end{figure*}

Notice there is no guarantee that all networks are deadlock-free; the network $\p[\q?x] \mid \q[\p?x]$ is terminated, but we say it is \textit{deadlocked} because the processes are not all $\mathbf{0}$. However, we will see that all networks \textit{projected from a choreography} are deadlock-free.

\Cref{fig:formal-fault-epp} shows the endpoint projection function from a choreography to a process in the underlying network.
The projection of the choreography send instruction, $\epp{ \p.e \rightarrow \q.x ; C }$, is a send operation at $\p$ and a receive operation at $\q$.
However, the projection of the choreography receive instruction, $\epp{ \p \rightsquigarrow \q.x ; C }$, is just a receive operation at $\q$, since in this case the message from $\p$ has already been sent.

The relationship between choreographies and their projection is formalized by the following two lemmas that together define the EPP theorem:

\manualTheorem{Lemma 1}{
  (Completeness of Endpoint Projection).
  For all $\conf{ C, \Sigma, K, S, T, A }$, $\mu$, and $\conf{ C', \Sigma', K', S', T', A' }$.
  If $\trans{ C, \Sigma, K, S, T, A }{\mu}{ C', \Sigma', K', S', T', A' }$,
  then $\trans{ \epp{C}, \Sigma, K, S, T, A }{\mu}{ \epp{C'}, \Sigma', K', S', T', A' }$.
}

\manualTheorem{Lemma 2}{
  (Soundness of Endpoint Projection).
  For any $\langle C, \Sigma, K, S, T, A \rangle, N, \mu, N', \Sigma', K',$ and $S'$ such that $\langle C, \Sigma, K, S, T, A \rangle$ is well-formed, $N \sqsupseteq \epp{C}$, and $\mu \neq \textsc{restart}@\p$ for any $\p$,
  $\trans{ N, \Sigma, K, S, T, A }{\mu}{ N', \Sigma', K', S', T', A' }$ implies $\trans{ C, \Sigma, K, S, T, A }{\mu}{ C', \Sigma', K', S', T', A' }$ for some $C'$ \\ such that $N' \sqsupseteq \epp{C'}$.
  \\
}

The soundness direction conspicuously leaves out the \textsc{restart} rule; it is not at all clear what should happen at the choreography level when a process restarts. Instead, we show that executions with restarts are formally related to executions without them; we formalize this next with the efficiency ordering ($\preceq$).

\subsection{Recovery}

\begin{figure*}
  \begin{subfigure}{\textwidth}
  \centering
  \input{graphics/formal-restart-diagram1.tex}
  \caption{A network execution with restarts (top), and its related restart-free execution (bottom). We obtain the bottom execution from the top by pruning away the restart transition and any other transitions needed to bring $\p$ back to its old state $P_{2}$.}%
  \label{fig:execution-equivalence-replay}
\end{subfigure}

\vspace{0.2cm}

  \begin{subfigure}{\textwidth}
    \centering
    \input{graphics/formal-restart-diagram2.tex}
    \caption{A network execution with restarts (top), and its related restart-free execution (bottom). In this example, process $\p$ never reaches its old state $P_{2}$ because it performs the \textsc{compensate} step first.}%
    \label{fig:execution-equivalence-compensate}
  \end{subfigure}

  \vspace{0.2cm}

  \begin{subfigure}{\textwidth}
    \centering
    \input{graphics/formal-restart-diagram3.tex}
    \caption{A network execution with restarts (top), and a related execution with one restart event fewer (bottom). In this example, process $\p$ restarts before it can return to state $P_{2}$.}%
    \label{fig:}
  \end{subfigure}
  \caption{}
\end{figure*}

Next, we show how our process semantics makes failure transparent to the choreography semantics. Namely, we show that any terminated execution is `related' to an execution without restarts. This is because, when a process $\p$ in state $P$ restarts, it replays all steps from $P_{\textit{start}}$ to return to state $P$; these `replay' steps have no effect on the rest of the configuration because (1) transactions are idempotent, and (2) the message queue is durable. Also, since processes are deterministic, $\p$ will either (1) return to state $P$, or (2) fail and perform compensating transactions.

This behavior essentially means that we can reason about executions with restarts as though the restarts never happened,
since we know that the outcome of the executions will be equivalent.
This is shown in \Cref{fig:execution-equivalence-replay}, where the two executions become equivalent if we ignore the restart and replay steps marked in gray.

\begin{figure*}
  \centering
  \begin{minipage}{0.6\textwidth}
    \begin{align*}
      \inferrule* [Right=\textsc{cong-1}]
      {
        \restrict{N'}{A} = \restrict N A\\
        \restrict{\Sigma'}{A} = \restrict \Sigma A\\
        \restrict{S'}{A} = \restrict S A
      }
      {\conf{N',\Sigma',K,S',T,A} \cong \conf{N,\Sigma,K,S,T,A}}
    \end{align*}
  \end{minipage}%
  \begin{minipage}{0.35\textwidth}
    \begin{align*}
      \inferrule* [Right=\textsc{cong-2}]
      {s' \cong s \\ \sigma' \cong \sigma}
      {s' \xrightarrow{\mu} \sigma' \cong s \xrightarrow{\mu} \sigma}
    \end{align*}
  \end{minipage}

  \begin{minipage}{0.25\textwidth}
    \begin{align*}
      \inferrule* [Right=\textsc{prec-1}]
      {
        \p = \pn(\mu)
      }
      {
        s_{1} \preceq_{\p} s_{1} \xrightarrow{\mu} s_{2}
      }
    \end{align*}
  \end{minipage}%
  \begin{minipage}{0.35\textwidth}
    \begin{align*}
      \inferrule* [Right=\textsc{prec-2}]
      {
        \p \neq \pn(\mu)
      }
      {
        s_{1}' \xrightarrow{\mu} s_{2}' \preceq_{\p} s_{1} \xrightarrow{\mu} s_{2}
      }
    \end{align*}
  \end{minipage}%
  \begin{minipage}{0.3\textwidth}
    \begin{align*}
      \inferrule* [Right=\textsc{prec-3}]
      {
        \sigma_{1}' \preceq_{\p} \sigma_{1}\\
        \sigma_{2}' \preceq_{\p} \sigma_{2}\\
      }
      {
        \sigma_{1}'\sigma_{2}' \preceq_{\p} \sigma_{1}\sigma_{2}
      }
    \end{align*}

  \end{minipage}%

  \begin{minipage}{0.4\textwidth}
    \begin{align*}
      \inferrule* [Right=\textsc{prec}]
      {
        \sigma_{1}' = \sigma_{1} \\
        \sigma_{2}' \preceq_{\p} \sigma_{2} \\
        \sigma_{3}' \cong \sigma_{3}
      }
      {
        \sigma_{1}'\sigma_{2}'\sigma_{3}' \preceq \sigma_{1}\sigma_{2}\sigma_{3}
      }
    \end{align*}
  \end{minipage}%
  \begin{minipage}{0.4\textwidth}
    \begin{align*}
      \inferrule* [Right=\textsc{prec-trans}]
      {
        \sigma_1 \preceq \sigma_2\\
        \sigma_2 \preceq \sigma_3
      }
      {
        \sigma_1 \preceq \sigma_3
      }
    \end{align*}
  \end{minipage}%

  \caption{Definition of the efficiency ordering on executions.}
  \label{fig:fault-efficiency-order}
\end{figure*}

A more interesting example appears in \Cref{fig:execution-equivalence-compensate}: in the fault-prone execution, $\p$ restarts and compensates before it can reach its pre-restart state. Nevertheless, we can consider the top and bottom configurations to be `congruent' because their differences are irrelevant for our purposes. The inference rules given in \Cref{fig:fault-efficiency-order} formalize the notion of congruence between network configurations $s\cong s'$ and executions $\sigma \cong \sigma'$. Rule \textsc{cong-1} defines two network configurations to be congruent if their configurations from the perspective of all active processes are the same. By rule \textsc{cong-2}, two executions are congruent if all of their configurations are congruent.

%Concretely, we define a partial order on executions $(\preceq)$. Informally, $\sigma \preceq \sigma'$ means $\sigma$ can be obtained from $\sigma'$ by erasing ``restart'' and ``replay'' events.

To define the $(\preceq)$ relation, we use the notation $\sigma\sigma'$ to denote the concatenation of executions $\sigma$ and $\sigma'$. The concatenation is only defined if the end configuration of $\sigma$ is the same as the start configuration of $\sigma'$. We then say an execution $\sigma$ precedes $\sigma'$ from the perspective of $\p$ (written $\sigma \preceq_\p \sigma'$) if, for all steps $\mu$ in $\sigma$ or $\sigma'$, $\mu$ is either taken by both $\sigma$ and $\sigma'$ (rule \textsc{prec}),
or if $\mu$ is carried out by $\p$, then it is only taken in $\sigma'$ (rule \textsc{prec-1}), or if $\mu$ is carried out by another process, it is only taken in $\sigma$ (rule \textsc{prec-2}).

% TODO: State the recovery theorem. Add an informal proof sketch (~1 paragraph) explaining the proof.

This enables us to state the recovery theorem for processes, to prove that processes will recover safely from restarts.

\manualTheorem{Theorem 3}{
  (Recovery).
  Let $\sigma$ be a finite network execution that terminates in its last step. If $\sigma$ has $k > 0$ restart events, then there is an execution $\sigma'$ with $k - 1$ restart events, such that $\sigma' \preceq \sigma$.
}

% TODO: Explain how we can combine the recovery theorem, EPP theorem, and the choreographic agreement+deadlock-freedom results to get a safety theorem for projected networks.

\noindent \\
By combining the recovery theorem with the EPP theorem and our choreographic safety theorems, we can finally obtain our safety theorem for networks.

\manualTheorem{Theorem 4}{
  (Network Safety).
  Let $\sigma$ be a terminated network execution. Then either:
  \begin{enumerate}
    \item $\cfg(\sigma)$ has the form $\langle \prod_{\p \in \PName} \p[0], \Sigma, K, S, T, \PName \rangle$
      where $T$ has no compensations; or
    \item $\cfg(\sigma)$ has the form $\langle N, \Sigma, K, S, T, \emptyset \rangle$
      where $T(\p)$ has the form $\langle t_1, \dots, t_i, c(t_i), \dots, c(t_1) \rangle$, for some $i$, for each process $\p \in \PName$.
  \end{enumerate}
}

%%% Local Variables:
%%% mode: latex
%%% TeX-master: "main"
%%% End:

%% file: graphics/formal-restart-diagram1.tex
\begin{tikzpicture}[
    x=6.5mm, y=1.5mm,
    >=Stealth,
    every path/.style={line width=0.35pt},
    state/.style={inner sep=1pt, font=\small},
    ell/.style={font=\small, inner sep=0pt},
    lab/.style={font=\small, midway, above},
    map/.style={dashed, blue, -{Stealth[length=1.6mm]}},
    fadednode/.style={inner sep=1pt, font=\small, text=gray},
    node distance=1.4 and 1.4
  ]

  % --------- Top (with restart & replay) ----------
  \node[state] (T0) {$\langle \p[P_2]\ldots\rangle$};
  \node[state, left=0.5 of T0] (T00) {};
  \node[fadednode, right=2.0 of T0] (T1) {$\langle \p[P_0]\ldots\rangle$};
  \node[state, right=1.5 of T1] (T2) {$\langle \p[P_1]\ldots\rangle$};
  \node[fadednode, right=1.5 of T2] (T4) {$\langle \p[P_{1}]\ldots\rangle$};
  \node[fadednode, right=0.8 of T4] (T5) {$\langle \p[P_2]\ldots\rangle$};
  \node[state, right=1.5 of T5] (T6) {$\langle \p[P_{3}]\ldots\rangle$};

  \draw[->] (T00) -- (T0);
  \draw[->, gray] (T0) -- node[lab, text=gray] {$\textsc{restart}@\p$} (T1);
  \draw[->] (T1) -- node[lab] {$\q.v \rightarrow \p !$} (T2);
  \draw[->, gray] (T2) -- node[lab, text=gray] {$\r.v' \rightarrow \p ?$} (T4);
  \draw[->, gray] (T4) -- node[lab, text=gray] {$\tau@\p$} (T5);
  \draw[->] (T5) -- node[lab] {$\q.v \rightarrow \p?$} (T6);

  % --------- Bottom (restart-free) ----------
  \node[state, below=6.5 of T0] (B0) {$\langle \p[P_2]\ldots\rangle$};
  \node[state, left=0.5 of B0] (B00) {};
  \node[state, right=1.5 of B0] (B1) {$\langle \p[P_2]\ldots\rangle$};
  \node[state, right=1.5 of B1] (B4) {$\langle \p[P_{3}]\ldots\rangle$};

  \draw[->] (B00) -- (B0);
  \draw[->] (B0) -- node[lab] {$\q.v \rightarrow \p !$} (B1);
  \draw[->] (B1) -- node[lab] {$\q.v \rightarrow \p?$} (B4);

  % --------- Correspondence (dashed blue) ----------
  \draw[map] (T0.south) .. controls +(0,-5.0) and +(0,5.0) .. (B0.north);
  \draw[map] (T2.south) .. controls +(0,-3.6) and +(0,5.6) .. (B1.north);
  \draw[map] (T6.south) .. controls +(0,-5.0) and +(.0,5.0) .. (B4.north);

\end{tikzpicture}

%% file: graphics/formal-restart-diagram2.tex
\begin{tikzpicture}[
    x=6.5mm, y=1.5mm,
    >=Stealth,
    every path/.style={line width=0.35pt},
    state/.style={inner sep=1pt, font=\small},        % nodes (black)
    fadednode/.style={inner sep=1pt, font=\small, text=gray}, % grey text, no box
    ell/.style={font=\small, inner sep=0pt},
    lab/.style={font=\small, midway, above},           % edge labels
    map/.style={dashed, blue, -{Stealth[length=1.6mm]}},
    node distance=1.4 and 1.4
  ]

  % --------- Top (with restart/replay/failure) ----------
  \node[state] (T0) {$\langle \p[P_2]\ldots\rangle$};
  \node[state, left=0.5 of T0] (T00) {};
  \node[fadednode, right=2.5 of T0] (T1) {$\langle \p[P_0]\ldots\rangle$};
  \node[state, right=2.2 of T1] (T2) {$\langle \p[P_0]\ldots\rangle$};
  \node[fadednode, right=1.0 of T2] (T4) {$\langle \p[P_1]\ldots\rangle$};
  \node[state, right=3.4 of T4] (T5) {$\langle \p[P_1]\ldots\rangle$};

  \draw[->] (T00) -- (T0);
  \draw[->, gray] (T0) -- node[lab, text=gray] {$\textsc{restart}@\p$} (T1);
  \draw[->]       (T1) -- node[lab] {$\textsc{fail}@\q$} (T2);
  \draw[->, gray] (T2) -- node[lab, text=gray] {$\tau@\p$} (T4);
  \draw[->]       (T4) -- node[lab] {$\textsc{compensate}@\p$} (T5);

  % --------- Bottom (restart-free) ----------
  \node[state, below=6.5 of T0] (B0) {$\langle \p[P_2]\ldots\rangle$};
  \node[state, left=0.5 of B0] (B00) {};
  \node[state, right=2.0 of B0] (B1) {$\langle \p[P_2]\ldots\rangle$};
  \node[state, right=3.4 of B1] (B2) {$\langle \p[P_2]\ldots\rangle$};

  \draw[->] (B00) -- (B0);
  \draw[->] (B0) -- node[lab] {$\textsc{fail}@\q$} (B1);
  \draw[->] (B1) -- node[lab] {$\textsc{compensate}@\p$} (B2);

  % --------- Correspondence (dashed blue) ----------
  \draw[map] (T0.south) .. controls +(0,-5.0) and +(0,5.0) .. (B0.north);
  \draw[map] (T2.south) .. controls +(0,-5.6) and +(0,5.6) .. (B1.north);
  \draw[map] (T5.south) .. controls +(0,-5.0) and +(0,5.0) .. (B2.north);

\end{tikzpicture}

%% file: graphics/formal-restart-diagram3.tex
\begin{tikzpicture}[
    x=6.5mm, y=1.5mm,
    >=Stealth,
    every path/.style={line width=0.35pt},
    state/.style={inner sep=1pt, font=\small},        % nodes (black)
    fadednode/.style={inner sep=1pt, font=\small, text=gray}, % grey text, no box
    ell/.style={font=\small, inner sep=0pt},
    lab/.style={font=\small, midway, above},           % edge labels
    map/.style={dashed, blue, -{Stealth[length=1.6mm]}},
    node distance=1.4 and 1.4
  ]

  % --------- Top (with restart/replay/failure) ----------
  \node[state] (T0) {$\langle \p[P_2]\ldots\rangle$};
  \node[state, left=0.5 of T0] (T00) {};
  \node[fadednode, right=2.5 of T0] (T1) {$\langle \p[P_0]\ldots\rangle$};
  \node[state, right=2.2 of T1] (T2) {$\langle \p[P_0]\ldots\rangle$};
  \node[fadednode, right=1.0 of T2] (T4) {$\langle \p[P_1]\ldots\rangle$};
  \node[state, right=3.4 of T4] (T5) {$\langle \p[P_0]\ldots\rangle$};

  \draw[->] (T00) -- (T0);
  \draw[->, gray] (T0) -- node[lab, text=gray] {$\textsc{restart}@\p$} (T1);
  \draw[->]       (T1) -- node[lab] {$\tau@\q$} (T2);
  \draw[->, gray] (T2) -- node[lab, text=gray] {$\tau@\p$} (T4);
  \draw[->]       (T4) -- node[lab] {$\textsc{restart}@\p$} (T5);

  % --------- Bottom (restart-free) ----------
  \node[state, below=6.5 of T0] (B0) {$\langle \p[P_2]\ldots\rangle$};
  \node[state, left=0.5 of B0] (B00) {};
  \node[state, right=2.0 of B0] (B1) {$\langle \p[P_2]\ldots\rangle$};
  \node[state, right=3.4 of B1] (B2) {$\langle \p[P_0]\ldots\rangle$};

  \draw[->] (B00) -- (B0);
  \draw[->] (B0) -- node[lab] {$\tau@\q$} (B1);
  \draw[->] (B1) -- node[lab] {$\textsc{restart}@\p$} (B2);

  % --------- Correspondence (dashed blue) ----------
  \draw[map] (T0.south) .. controls +(0,-5.0) and +(0,5.0) .. (B0.north);
  \draw[map] (T2.south) .. controls +(0,-5.6) and +(0,5.6) .. (B1.north);
  \draw[map] (T5.south) .. controls +(0,-5.0) and +(0,5.0) .. (B2.north);

\end{tikzpicture}

%% file: evaluation.tex
\section{Evaluation}\label{sec:evaluation}

So far, we have argued that \Accompanist choreographies can achieve better end-to-end latency than orchestrators by deploying sidecars alongside worker services. In this section, we evaluate our approach with the following research questions:

\begin{enumerate}
  \item How does the runtime overhead of \Accompanist's sidecars compare against the overhead of orchestration as network latency increases? (\Cref{sec:microbenchmarks})
  \item How do \Accompanist choreographies compare against orchestrators within a single organization and a single data center? (\Cref{sec:evaluation-online-butique})
  \item How do \Accompanist choreographies compare against orchestrators across different organizations and data centers? (\Cref{sec:evaluation-hotel})
  \item How do \Accompanist sagas compare against orchestrated sagas in terms of end-to-end latency? (\Cref{sec:saga-evaluation})
\end{enumerate}

% COMMENTS FROM MEETING WITH FABRIZIO, 11 MARCH 2026
% The \Accompanist framework opens up to exploration of further use cases in microservices.
% We also present promising preliminary results against Temporal, a mature industry grade workflow engine and achieve up to $6.7\times$ improvements in end-to-end latency.

% However, a thorough evaluation requires several axes of deployment and features.
% Threats to validity

% - Database engine.
% - Replication of the database engine for the orchestrator vs the database engines per node (which can be asymetric).
% - Replication of the orchestrator versus the sidecars.
% - The different node failure policies (mention how many Temporal has).

% In the end of sec. 6.4 we leave this as future work.

\subsection{Microbenchmarks}\label{sec:microbenchmarks}

\begin{figure*}
  \centering
  \usefont{T1}{cmss}{m}{n}
  \begin{subfigure}[t]{0.3\textwidth}
    \centering
    \includegraphics[width=\textwidth]{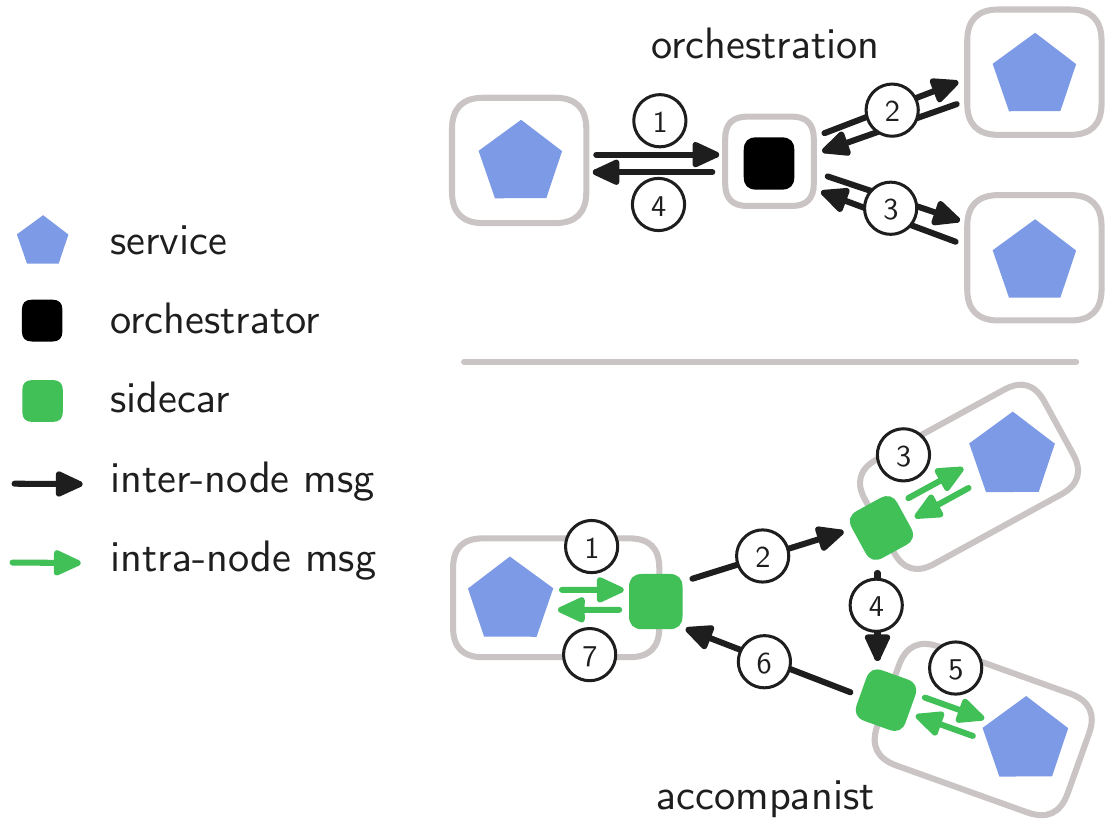}
    \subcaption{Communication diagram}%
    \label{fig:chain-latency-diagram}
  \end{subfigure}%
  \begin{subfigure}[t]{0.45\textwidth}
    \centering
    \resizebox{!}{3.2cm}{\input{graphics/equidistant_latency_2_workers.tex}}
    \caption{Two workers equidistant}%
    \label{fig:microbenchmark-equidistant}
  \end{subfigure}%
  \begin{subfigure}[t]{0.25\textwidth}
    \centering
    \resizebox{!}{3.2cm}{\input{graphics/asymmetric_latency_1_ms.tex}}
    \subcaption{$n$ workers isolated, 1ms~latency}%
    \label{fig:microbenchmark-asymmetric}
  \end{subfigure}%
  \caption{Microbenchmark evaluation results.}%
  \label{fig:chain-microbenchmark}
\end{figure*}

To develop a principled understanding of how \Accompanist compares to orchestration, we adapted the simple workflow from \Cref{fig:code-comparison} into microbenchmarks. The communication patterns from \Cref{fig:orchestrated} and \Cref{fig:choreographic} are shown at the top and bottom of \Cref{fig:chain-latency-diagram}, respectively. The diagram shows how \Accompanist \emph{decentralizes} the orchestration service into sidecars running next to each worker service. Although sidecars introduce overhead because services do not communicate with one another directly, we expect our framework to perform well when network latency is high because Choral can tightly control the most expensive communications.

Assuming intra-node messages all take time $t_{1}$, inter-node messages take time $t_{2}$, and computation time is negligible, \Cref{fig:chain-latency-diagram} indicates the orchestrated solution should have end-to-end latency $6t_{2}$ and the \Accompanist solution should have end-to-end latency $6t_{1} + 3t_{2}$; hence our framework should scale better than orchestration when $t_{1}$ is constant and $t_{2}$ increases.
We evaluated this prediction by running the microbenchmark on a single machine with Toxiproxy~\cite{toxiproxy} to simulate latency between nodes. For the orchestrated benchmark, \Cref{fig:microbenchmark-equidistant} shows end-to-end latency as a function of network latency. For the \Accompanist benchmark, we measured both the end-to-end latency and the latency between services and sidecars; by subtracting the latter from the former, we approximate how much time the benchmark spends performing expensive inter-node communications versus inexpensive sidecar RPCs. The results confirm our expectation: once network latency is high enough, \Accompanist scales better than orchestration because the overhead of sidecar communication remains small no matter the speed of the network.

In practice, network latency between services is not equal. For example, the worker services in \Cref{fig:chain-latency-diagram} could be located in a different zone or data center from the orchestrator. In such cases, the performance gap between orchestration and choreography grows with the number of worker services. Let $t_{1}$ be the latency of intra-node messages, let $t_{2}$ be the latency of messages between nodes within the same zone, and let $t_{3}$ be the latency of messages across zones. We then expect the orchestrated solution to have end-to-end latency $2t_{2} + 2nt_{3}$, whereas \Accompanist has end-to-end latency $2(n+1)t_{1} + (n-1)t_{2} + 2t_{2}$. This is indeed what we observe in \Cref{fig:microbenchmark-asymmetric}: even with only one millisecond of latency between the orchestrator and the workers, the \emph{orchestration tax} grows with the number of services, because it is more efficient for workers to communicate point-to-point than through an intermediary.

\subsection{Online Boutique} \label{sec:evaluation-online-butique}

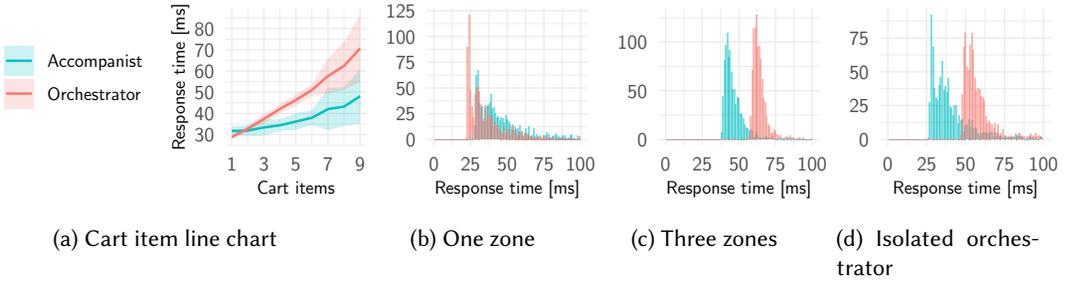
\begin{figure*}
  \usefont{T1}{cmss}{m}{n}
  \begin{subfigure}[t]{0.335\textwidth}
    \centering
    \resizebox{!}{2.8cm}{\input{benchmarks/25-01-14_AWS_plots/cart_size_latency/cart_size_latency.tex}}
    \caption{Cart item line chart}
    \label{fig:eval-cart-size}
  \end{subfigure}
  \hfill
  \begin{subfigure}[t]{0.19\textwidth}
    \centering
    \resizebox{!}{2.8cm}{\input{benchmarks/25-01-14_AWS_plots/2_nodes_1_zone_fcom/2_AWS_2_nodes_1_zone.tex}}
    \caption{One zone}
    \label{fig:eval-1-zone}
  \end{subfigure}
  \hfill
  \begin{subfigure}[t]{0.19\textwidth}
    \centering
    \resizebox{!}{2.8cm}{\input{benchmarks/25-01-14_AWS_plots/4_nodes_3_zones/1_AWS_4_nodes.tex}}
    \caption{Three zones}
    \label{fig:eval-3-zones}
  \end{subfigure}
  \hfill
  \begin{subfigure}[t]{0.19\textwidth}
    \centering
    \resizebox{!}{2.8cm}{\input{benchmarks/25-01-14_AWS_plots/2_nodes_1_zone_fcom_orchestrator_separate_zone/2_AWS_2_nodes_1_zone.tex}}
    \caption{Isolated orchestrator}
    \label{fig:eval-isolated-orchestrator}
  \end{subfigure}
  \caption{Response times for a checkout request in the \emph{Online Boutique} application in various deployments.}%
  \label{fig:evaluation-diagrams}
\end{figure*}

We evaluated the implementation of the Online Boutique (\Cref{sec:case-study-webshop}) on a 4-node AWS EKS cluster spanning 3 availability zones. This multi-zone deployment reflects production-grade architectures where services are distributed across zones to improve resilience against correlated failures through physically isolated infrastructure, or because a single zone is at capacity~\cite{young2015}.

All experiments were run in the \texttt{eu-north-1} region. Each node ran a \texttt{c5a.xlarge} instance with 4 vCPUs and 8 GB memory. Before each measurement the system ran a constant load of 120 req/s for 2 minutes to warm up the JVM. Afterwards the load changed to 60 req/s and 1000 requests were captured over a 4-minute period. The same procedure was repeated for the original orchestrator and the \Accompanist version. We found tail latencies in the original application to be quite noisy, so we only consider latencies within the bottom 99th percentile.

%Each experiment has been run multiple times with the pods redeployed to ensure consistent results.

% Remind readers why cart size affects things

\Cref{fig:eval-cart-size} shows the end-to-end response time for a user checkout workflow, where services are distributed across availability zones. As noted in \Cref{sec:case-study}, response time increases with the number of items in the cart because each item requires two RPCs from the orchestrator. However, the \Accompanist version scales better because the RPCs come from sidecars \emph{co-located} with the services they access.

The remaining figures show distributions of checkout latencies in three deployment strategies, each time with 9 items in the user's cart. \Cref{fig:eval-1-zone} schedules all services on two nodes in the same availability zone.  \Cref{fig:eval-3-zones} distributes the services randomly across four nodes in three zones. \Cref{fig:eval-isolated-orchestrator} places all RPC servers on two nodes in one zone, and places the orchestrator in a separate zone. %We ran each experiment multiple times, and observed similar distributions each time.
The results show that orchestration is efficient when all participating services can be placed close to the orchestrator, but otherwise latency can be improved up to 32\% by eliminating expensive network calls.

\subsection{Hotel Reservation}\label{sec:evaluation-hotel}

\begin{figure*}
  \centering
  \usefont{T1}{cmss}{m}{n}
  \begin{subfigure}{0.35\textwidth}
    \includegraphics[width=\textwidth]{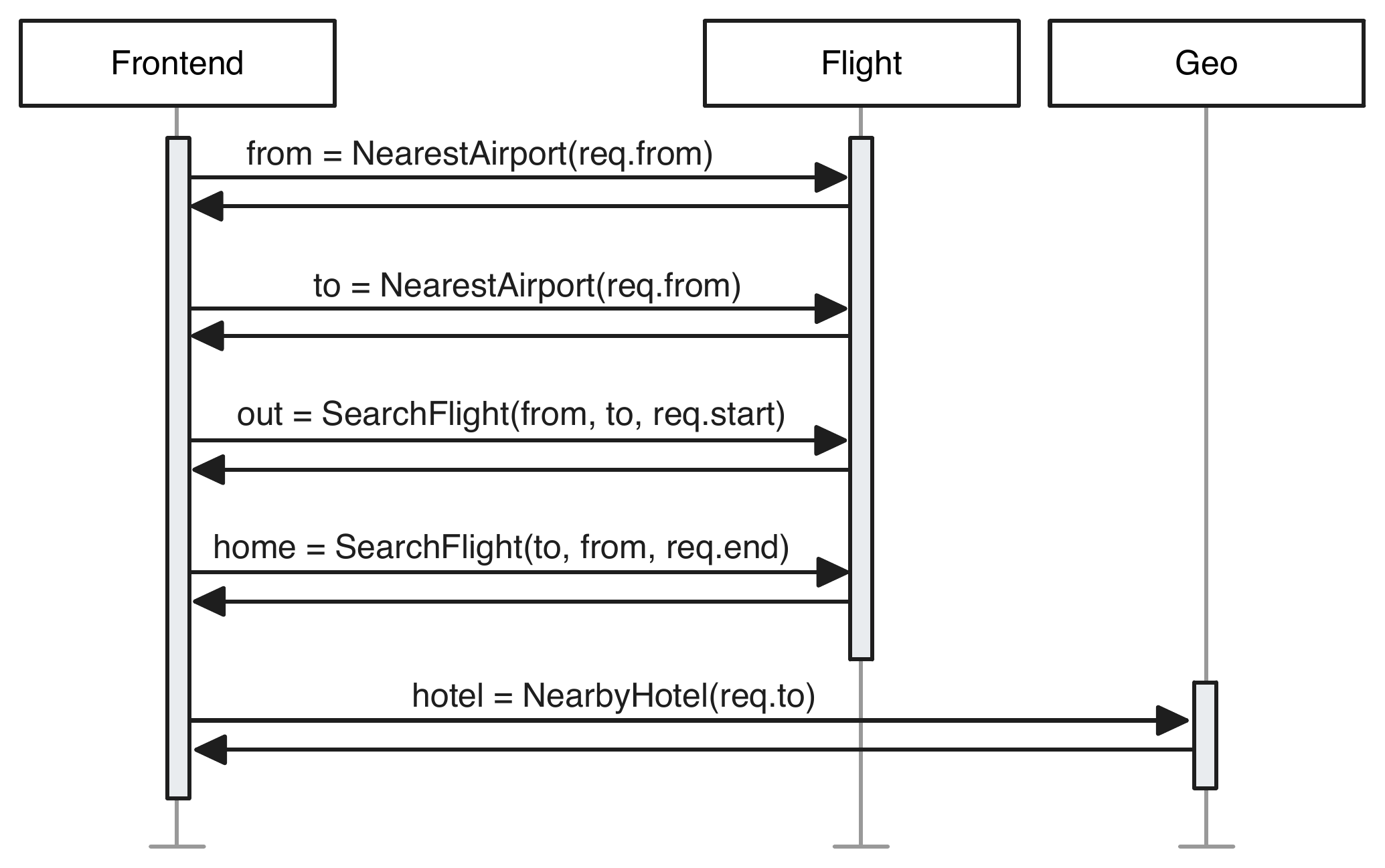}
    \subcaption{Orchestration}%
    \label{fig:hotel-orch}
  \end{subfigure}%
  \hfill
  \begin{subfigure}{0.35\textwidth}
    \includegraphics[width=\textwidth]{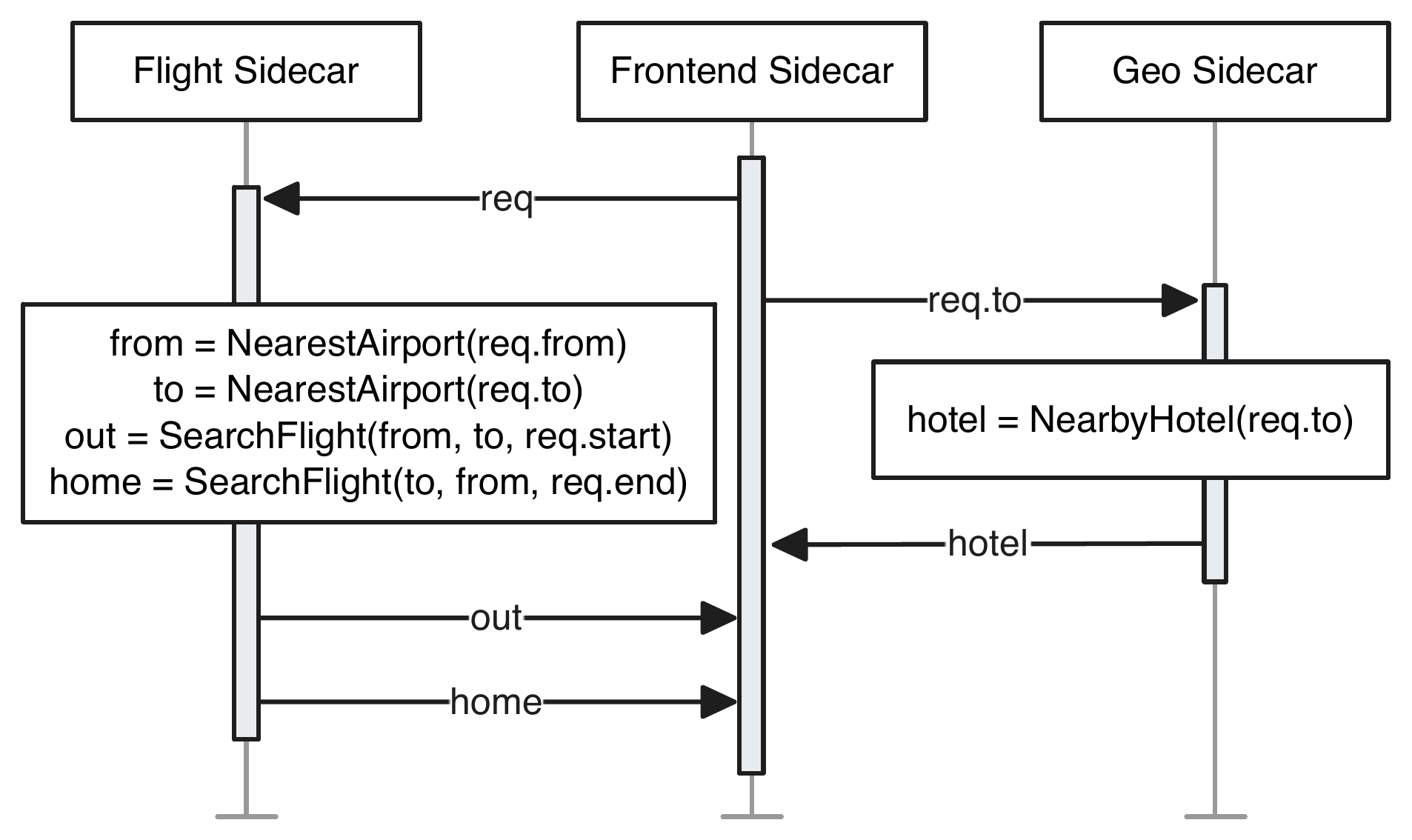}
    \subcaption{Choreography with \Accompanist}%
    \label{fig:hotel-acc}
  \end{subfigure}%
  \hfill
  \begin{subfigure}{0.28\textwidth}
    \input{benchmarks/25-04-16_AWS_multi_region/AWS_multi_region.tex}
    \subcaption{Response times}%
    \label{fig:flight-histogram}
  \end{subfigure}%
  \caption{An extension of the \emph{Hotel Reservation} app to book flights with a remote service.}%
  \label{fig:flight-benchmark}
\end{figure*}

Another common use case is cooperation between multiple parties, e.g. a distributed flight/hotel booking system. Traditionally, an orchestrator makes WAN calls to the API provided by the third party -- but the third party API might not be specialized to the orchestrator's needs, as described in \Cref{sec:case-study}. As a result, the orchestrator may need to make multiple WAN calls to complete the workflow. %It might be possible to have them deploy a sidecar to make a cross-organizational choreography.

We tested this scenario by adapting the Hotel Reservation application from \Cref{sec:case-study}. The benchmark originally consisted of a suite of microservices for searching and booking hotels; we extended the benchmark with a new flight microservice for looking up airports and flights. The flight service simulates a third party organization, deployed to a different datacenter from the rest of the microservices.
We implemented the workflow using an orchestrator (\Cref{fig:hotel-orch}) and using \Accompanist (\Cref{fig:hotel-acc}).

The flight service was deployed alone to the AWS region \texttt{eu-west-3} (Paris) on an EKS cluster consisting of a single \texttt{m5.large} node with 2 vCPUs and 8 GB memory.
All other services were deployed together in the \texttt{eu-central-1} (Frankfurt) AWS region on an EKS cluster consisting of 3 \texttt{m5.large} nodes.
In other respects, the benchmark was performed and measured in the same way as for the Online Boutique benchmark described in \Cref{sec:evaluation-online-butique}.

\Cref{fig:flight-histogram} shows the end-to-end response time for running the \Accompanist and orchestrated versions of the plan trip request.
Here, \Accompanist shows a 55\% average improvement in response time compared to the orchestrated version.

% - Motivating use case
%     - A common use case is cooperation between parties, e.g. a distributed flight booking.
%     - Typically you have to use an orchestrator that makes many WAN calls.
%     - These WAN calls have the issue the API provided by the third party might not be specialized to what we (the client) want.
%     - Maybe we can get them to deploy our code.
% - We tested this scenario by adapting the hotel reservation benchmark
%     - Explain what was already implemented before
%     - Explain what we want to add
%     - Add a diagram that compares the orchestrated vs the accompanist version
% - Explain the technical details of how you evaluated it (which data centers, how many repetitions, etc)
%     - State what the ping between the services is
% - Conclusions
%     - How big of a speedup do we see?
%     - Explain why there's a speedup

\subsection{Warehouse Saga} \label{sec:evaluation-warehouse}\label{sec:saga-evaluation}

\begin{figure*}
  \usefont{T1}{cmss}{m}{n}
  \begin{subfigure}[t]{0.32\textwidth}
    \includegraphics[width=\textwidth]{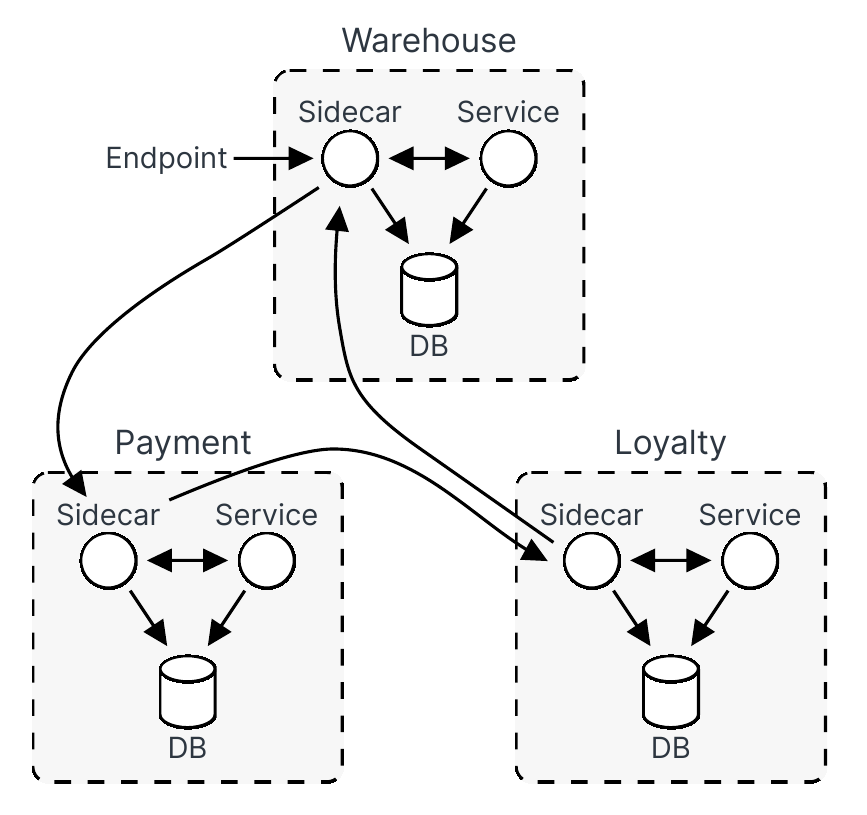}
    \caption{Warehouse deployment architecture.}
    \label{fig:warehouse-deployment}
  \end{subfigure}
  \begin{subfigure}[t]{0.32\textwidth}
    \includegraphics[width=\textwidth]{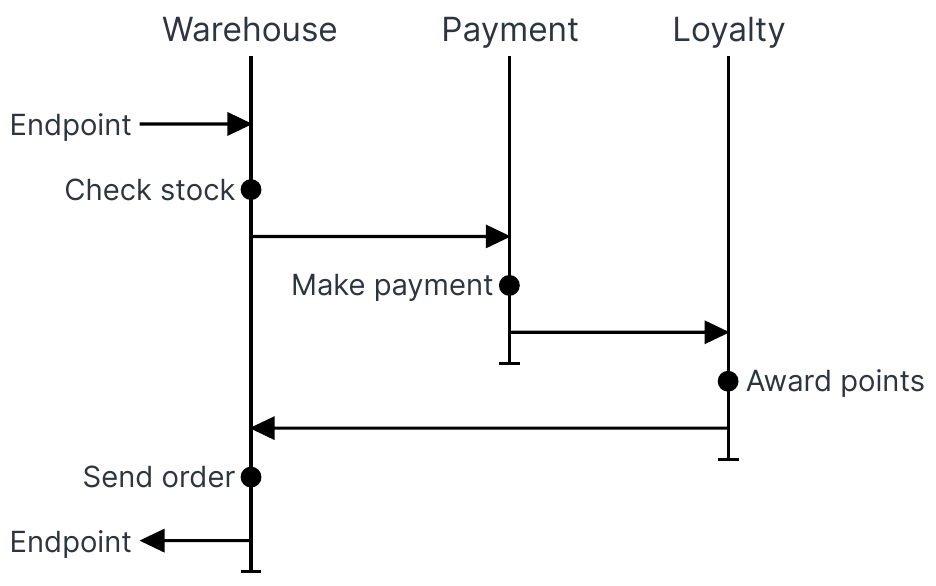}
    \caption{Warehouse choreography.}
    \label{fig:warehouse-choreography}
  \end{subfigure}
  \begin{subfigure}[t]{0.32\textwidth}
    \resizebox{!}{3.8cm}{\input{benchmarks/25-10-17_AWS_warehouse/warehouse-latencies.tex}}
    \caption{Warehouse workflow request end-to-end latency.}
    \label{fig:warehouse-histogram}
  \end{subfigure}
  \caption{A saga workflow to order products from a warehouse.}
\end{figure*}

We evaluate the performance of \Accompanist{} sagas by comparing against a mature orchestrated workflow framework, Temporal~\cite{temporal}.
Temporal uses a central orchestrator to keep track of all running workflows and to send commands to worker nodes. This design makes it simple to scale horizontally by deploying more worker nodes. This is in contrast to the \Accompanist sidecars, which can not easily be replicated. Workers in Temporal also use a pull-based model, whereas \Accompanist{} is push-based.

Because of these differences in their designs, we expect lower end-to-end latencies in \Accompanist since services communicate point-to-point instead of via a middleman. The polling design can also add unpredictable scheduling latency, so our predictive model in \Cref{sec:microbenchmarks} is only a \emph{lower bound} for Temporal's responsiveness in the worst case.

For a baseline, we used the saga example from Temporal's own documentation~\cite{temporalSagaExample} and extended it to perform real work (placing orders in the local database) and to split services across multiple Kubernetes pods. For readability, we also renamed the services using Newman's warehouse saga example~\cite[Figure 6-5]{newman2021}. The workflow consists of a warehouse, payment, and a loyalty microservice. We also implemented the same workflow using \Accompanist sagas.
\Cref{fig:warehouse-choreography} shows a sequence diagram for the choreography.

We deployed both implementations to an AWS EKS cluster consisting of 3 \texttt{m5.large} nodes in the \texttt{eu-central-1} datacenter.
Using Kubernetes affinity rules, we ensured that related services were deployed to the \emph{same} node---an ideal scenario for Temporal.
The deployment architecture for the choreographic version is shown in \Cref{fig:warehouse-deployment}.

The evaluation uses the stock production deployment configuration of Temporal\footnote{https://github.com/temporalio/helm-charts}, which uses Cassandra for its database. This comes with different characteristics than the PostgreSQL database used by \Accompanist. We also tested Temporal with PostgreSQL, but saw worse performance than Cassandra.

\Cref{fig:warehouse-histogram} shows the distribution of response times in both implementations. \Accompanist's median response time was 32ms, outperforming Temporal's 190ms median by 5.9$\times$.
\Accompanist also exhibited improved tail latency, with 99th-percentile response time of 87ms compared to Temporal's 587ms (a performance improvement of 6.7$\times$).

% We also tested the scalability of the deployment and found that \Accompanist{} could handle up to 140 requests per second, outperforming Temporal's 20 requests per second, by a $\times7$ improvement.
% The reason why Temporal handles fewer requests is that messages and control need to be aggregated at the central orchestrator and its database, whereas with \Accompanist{} this indirection is not necessary as the workers coordinates this individually.

% Durable workflow frameworks are complex which makes them difficult to compare. We highlight the following threats to validity:

% \begin{itemize}
%   \item The stock production deployment configuration of Temporal\footnote{https://github.com/temporalio/helm-charts} uses Cassandra for its database which comes with different characteristics than the Postgresql database that \Accompanist uses. We also ran the evaluation based on a custom deployment based on Postgresql and saw a worse performance reaching 12 requests per second.
%   \item Temporal supports full horizontal scalability of all aspects of the deployment, which means that Temporal can theoretically scale infinitely given enough resources. Conversely, \Accompanist{} is constrained by the individual sidecars, as it does not currently support vertical deployment of sidecars.
% \end{itemize}

%%% Local Variables:
%%% mode: latex
%%% TeX-master: "main"
%%% End:

%% file: graphics/equidistant_latency_2_workers.tex
% Created by tikzDevice version 0.12.6 on 2025-04-17 11:39:48
% !TEX encoding = UTF-8 Unicode
\begin{tikzpicture}[x=1pt,y=1pt]
\definecolor{fillColor}{RGB}{255,255,255}
\path[use as bounding box,fill=fillColor,fill opacity=0.00] (0,0) rectangle (433.62,289.08);
\begin{scope}
\path[clip] (  0.00,  0.00) rectangle (433.62,289.08);
\definecolor{drawColor}{RGB}{255,255,255}
\definecolor{fillColor}{RGB}{255,255,255}

\path[draw=drawColor,line width= 0.6pt,line join=round,line cap=round,fill=fillColor] ( -0.00,  0.00) rectangle (433.62,289.08);
\end{scope}
\begin{scope}
\path[clip] (147.51, 24.58) rectangle (199.23,237.49);
\definecolor{fillColor}{gray}{0.92}

\path[fill=fillColor] (147.51, 24.58) rectangle (199.23,237.49);
\definecolor{drawColor}{RGB}{255,255,255}

\path[draw=drawColor,line width= 0.3pt,line join=round] (147.51, 63.96) --
	(199.23, 63.96);

\path[draw=drawColor,line width= 0.3pt,line join=round] (147.51,123.37) --
	(199.23,123.37);

\path[draw=drawColor,line width= 0.3pt,line join=round] (147.51,182.78) --
	(199.23,182.78);

\path[draw=drawColor,line width= 0.6pt,line join=round] (147.51, 34.26) --
	(199.23, 34.26);

\path[draw=drawColor,line width= 0.6pt,line join=round] (147.51, 93.67) --
	(199.23, 93.67);

\path[draw=drawColor,line width= 0.6pt,line join=round] (147.51,153.08) --
	(199.23,153.08);

\path[draw=drawColor,line width= 0.6pt,line join=round] (147.51,212.49) --
	(199.23,212.49);

\path[draw=drawColor,line width= 0.6pt,line join=round] (161.62, 24.58) --
	(161.62,237.49);

\path[draw=drawColor,line width= 0.6pt,line join=round] (185.13, 24.58) --
	(185.13,237.49);
\definecolor{fillColor}{RGB}{248,118,109}

\path[fill=fillColor] (174.55, 34.26) rectangle (195.71, 42.44);
\definecolor{fillColor}{RGB}{40,191,47}

\path[fill=fillColor] (151.04, 34.26) rectangle (172.20, 41.97);
\definecolor{fillColor}{RGB}{248,118,109}

\path[fill=fillColor] (151.04, 41.97) rectangle (172.20, 56.41);
\end{scope}
\begin{scope}
\path[clip] (204.73, 24.58) rectangle (256.45,237.49);
\definecolor{fillColor}{gray}{0.92}

\path[fill=fillColor] (204.73, 24.58) rectangle (256.45,237.49);
\definecolor{drawColor}{RGB}{255,255,255}

\path[draw=drawColor,line width= 0.3pt,line join=round] (204.73, 63.96) --
	(256.45, 63.96);

\path[draw=drawColor,line width= 0.3pt,line join=round] (204.73,123.37) --
	(256.45,123.37);

\path[draw=drawColor,line width= 0.3pt,line join=round] (204.73,182.78) --
	(256.45,182.78);

\path[draw=drawColor,line width= 0.6pt,line join=round] (204.73, 34.26) --
	(256.45, 34.26);

\path[draw=drawColor,line width= 0.6pt,line join=round] (204.73, 93.67) --
	(256.45, 93.67);

\path[draw=drawColor,line width= 0.6pt,line join=round] (204.73,153.08) --
	(256.45,153.08);

\path[draw=drawColor,line width= 0.6pt,line join=round] (204.73,212.49) --
	(256.45,212.49);

\path[draw=drawColor,line width= 0.6pt,line join=round] (218.84, 24.58) --
	(218.84,237.49);

\path[draw=drawColor,line width= 0.6pt,line join=round] (242.35, 24.58) --
	(242.35,237.49);
\definecolor{fillColor}{RGB}{248,118,109}

\path[fill=fillColor] (231.77, 34.26) rectangle (252.93,112.14);
\definecolor{fillColor}{RGB}{40,191,47}

\path[fill=fillColor] (208.26, 34.26) rectangle (229.42, 42.95);
\definecolor{fillColor}{RGB}{248,118,109}

\path[fill=fillColor] (208.26, 42.95) rectangle (229.42, 91.43);
\end{scope}
\begin{scope}
\path[clip] (261.95, 24.58) rectangle (313.68,237.49);
\definecolor{fillColor}{gray}{0.92}

\path[fill=fillColor] (261.95, 24.58) rectangle (313.68,237.49);
\definecolor{drawColor}{RGB}{255,255,255}

\path[draw=drawColor,line width= 0.3pt,line join=round] (261.95, 63.96) --
	(313.68, 63.96);

\path[draw=drawColor,line width= 0.3pt,line join=round] (261.95,123.37) --
	(313.68,123.37);

\path[draw=drawColor,line width= 0.3pt,line join=round] (261.95,182.78) --
	(313.68,182.78);

\path[draw=drawColor,line width= 0.6pt,line join=round] (261.95, 34.26) --
	(313.68, 34.26);

\path[draw=drawColor,line width= 0.6pt,line join=round] (261.95, 93.67) --
	(313.68, 93.67);

\path[draw=drawColor,line width= 0.6pt,line join=round] (261.95,153.08) --
	(313.68,153.08);

\path[draw=drawColor,line width= 0.6pt,line join=round] (261.95,212.49) --
	(313.68,212.49);

\path[draw=drawColor,line width= 0.6pt,line join=round] (276.06, 24.58) --
	(276.06,237.49);

\path[draw=drawColor,line width= 0.6pt,line join=round] (299.57, 24.58) --
	(299.57,237.49);
\definecolor{fillColor}{RGB}{248,118,109}

\path[fill=fillColor] (288.99, 34.26) rectangle (310.15,172.18);
\definecolor{fillColor}{RGB}{40,191,47}

\path[fill=fillColor] (265.48, 34.26) rectangle (286.64, 42.38);
\definecolor{fillColor}{RGB}{248,118,109}

\path[fill=fillColor] (265.48, 42.38) rectangle (286.64,115.69);
\end{scope}
\begin{scope}
\path[clip] (319.18, 24.58) rectangle (370.90,237.49);
\definecolor{fillColor}{gray}{0.92}

\path[fill=fillColor] (319.18, 24.58) rectangle (370.90,237.49);
\definecolor{drawColor}{RGB}{255,255,255}

\path[draw=drawColor,line width= 0.3pt,line join=round] (319.18, 63.96) --
	(370.90, 63.96);

\path[draw=drawColor,line width= 0.3pt,line join=round] (319.18,123.37) --
	(370.90,123.37);

\path[draw=drawColor,line width= 0.3pt,line join=round] (319.18,182.78) --
	(370.90,182.78);

\path[draw=drawColor,line width= 0.6pt,line join=round] (319.18, 34.26) --
	(370.90, 34.26);

\path[draw=drawColor,line width= 0.6pt,line join=round] (319.18, 93.67) --
	(370.90, 93.67);

\path[draw=drawColor,line width= 0.6pt,line join=round] (319.18,153.08) --
	(370.90,153.08);

\path[draw=drawColor,line width= 0.6pt,line join=round] (319.18,212.49) --
	(370.90,212.49);

\path[draw=drawColor,line width= 0.6pt,line join=round] (333.28, 24.58) --
	(333.28,237.49);

\path[draw=drawColor,line width= 0.6pt,line join=round] (356.79, 24.58) --
	(356.79,237.49);
\definecolor{fillColor}{RGB}{248,118,109}

\path[fill=fillColor] (346.21, 34.26) rectangle (367.37,195.02);
\definecolor{fillColor}{RGB}{40,191,47}

\path[fill=fillColor] (322.70, 34.26) rectangle (343.86, 44.23);
\definecolor{fillColor}{RGB}{248,118,109}

\path[fill=fillColor] (322.70, 44.23) rectangle (343.86,132.63);
\end{scope}
\begin{scope}
\path[clip] (376.40, 24.58) rectangle (428.12,237.49);
\definecolor{fillColor}{gray}{0.92}

\path[fill=fillColor] (376.40, 24.58) rectangle (428.12,237.49);
\definecolor{drawColor}{RGB}{255,255,255}

\path[draw=drawColor,line width= 0.3pt,line join=round] (376.40, 63.96) --
	(428.12, 63.96);

\path[draw=drawColor,line width= 0.3pt,line join=round] (376.40,123.37) --
	(428.12,123.37);

\path[draw=drawColor,line width= 0.3pt,line join=round] (376.40,182.78) --
	(428.12,182.78);

\path[draw=drawColor,line width= 0.6pt,line join=round] (376.40, 34.26) --
	(428.12, 34.26);

\path[draw=drawColor,line width= 0.6pt,line join=round] (376.40, 93.67) --
	(428.12, 93.67);

\path[draw=drawColor,line width= 0.6pt,line join=round] (376.40,153.08) --
	(428.12,153.08);

\path[draw=drawColor,line width= 0.6pt,line join=round] (376.40,212.49) --
	(428.12,212.49);

\path[draw=drawColor,line width= 0.6pt,line join=round] (390.50, 24.58) --
	(390.50,237.49);

\path[draw=drawColor,line width= 0.6pt,line join=round] (414.01, 24.58) --
	(414.01,237.49);
\definecolor{fillColor}{RGB}{248,118,109}

\path[fill=fillColor] (403.43, 34.26) rectangle (424.59,227.81);
\definecolor{fillColor}{RGB}{40,191,47}

\path[fill=fillColor] (379.92, 34.26) rectangle (401.08, 46.11);
\definecolor{fillColor}{RGB}{248,118,109}

\path[fill=fillColor] (379.92, 46.11) rectangle (401.08,153.02);
\end{scope}
\begin{scope}
\path[clip] (147.51,237.49) rectangle (199.23,260.42);
\definecolor{drawColor}{RGB}{255,255,255}

\path[draw=drawColor,line width= 0.6pt,line join=round,line cap=round] (147.51,237.49) rectangle (199.23,260.42);
\definecolor{drawColor}{gray}{0.10}

\node[text=drawColor,anchor=base,inner sep=0pt, outer sep=0pt, scale=  1.60] at (173.37,243.44) {0 ms};
\end{scope}
\begin{scope}
\path[clip] (204.73,237.49) rectangle (256.45,260.42);
\definecolor{drawColor}{RGB}{255,255,255}

\path[draw=drawColor,line width= 0.6pt,line join=round,line cap=round] (204.73,237.49) rectangle (256.45,260.42);
\definecolor{drawColor}{gray}{0.10}

\node[text=drawColor,anchor=base,inner sep=0pt, outer sep=0pt, scale=  1.60] at (230.59,243.44) {1 ms};
\end{scope}
\begin{scope}
\path[clip] (261.95,237.49) rectangle (313.68,260.42);
\definecolor{drawColor}{RGB}{255,255,255}

\path[draw=drawColor,line width= 0.6pt,line join=round,line cap=round] (261.95,237.49) rectangle (313.68,260.42);
\definecolor{drawColor}{gray}{0.10}

\node[text=drawColor,anchor=base,inner sep=0pt, outer sep=0pt, scale=  1.60] at (287.81,243.44) {2 ms};
\end{scope}
\begin{scope}
\path[clip] (319.18,237.49) rectangle (370.90,260.42);
\definecolor{drawColor}{RGB}{255,255,255}

\path[draw=drawColor,line width= 0.6pt,line join=round,line cap=round] (319.18,237.49) rectangle (370.90,260.42);
\definecolor{drawColor}{gray}{0.10}

\node[text=drawColor,anchor=base,inner sep=0pt, outer sep=0pt, scale=  1.60] at (345.04,243.44) {3 ms};
\end{scope}
\begin{scope}
\path[clip] (376.40,237.49) rectangle (428.12,260.42);
\definecolor{drawColor}{RGB}{255,255,255}

\path[draw=drawColor,line width= 0.6pt,line join=round,line cap=round] (376.40,237.49) rectangle (428.12,260.42);
\definecolor{drawColor}{gray}{0.10}

\node[text=drawColor,anchor=base,inner sep=0pt, outer sep=0pt, scale=  1.60] at (402.26,243.44) {4 ms};
\end{scope}
\begin{scope}
\path[clip] (  0.00,  0.00) rectangle (433.62,289.08);
\definecolor{drawColor}{gray}{0.20}

\path[draw=drawColor,line width= 0.6pt,line join=round] (161.62, 21.83) --
	(161.62, 24.58);

\path[draw=drawColor,line width= 0.6pt,line join=round] (185.13, 21.83) --
	(185.13, 24.58);
\end{scope}
\begin{scope}
\path[clip] (  0.00,  0.00) rectangle (433.62,289.08);
\definecolor{drawColor}{gray}{0.30}

\node[text=drawColor,anchor=base,inner sep=0pt, outer sep=0pt, scale=  1.60] at (161.62,  8.61) {A};

\node[text=drawColor,anchor=base,inner sep=0pt, outer sep=0pt, scale=  1.60] at (185.13,  8.61) {O};
\end{scope}
\begin{scope}
\path[clip] (  0.00,  0.00) rectangle (433.62,289.08);
\definecolor{drawColor}{gray}{0.20}

\path[draw=drawColor,line width= 0.6pt,line join=round] (218.84, 21.83) --
	(218.84, 24.58);

\path[draw=drawColor,line width= 0.6pt,line join=round] (242.35, 21.83) --
	(242.35, 24.58);
\end{scope}
\begin{scope}
\path[clip] (  0.00,  0.00) rectangle (433.62,289.08);
\definecolor{drawColor}{gray}{0.30}

\node[text=drawColor,anchor=base,inner sep=0pt, outer sep=0pt, scale=  1.60] at (218.84,  8.61) {A};

\node[text=drawColor,anchor=base,inner sep=0pt, outer sep=0pt, scale=  1.60] at (242.35,  8.61) {O};
\end{scope}
\begin{scope}
\path[clip] (  0.00,  0.00) rectangle (433.62,289.08);
\definecolor{drawColor}{gray}{0.20}

\path[draw=drawColor,line width= 0.6pt,line join=round] (276.06, 21.83) --
	(276.06, 24.58);

\path[draw=drawColor,line width= 0.6pt,line join=round] (299.57, 21.83) --
	(299.57, 24.58);
\end{scope}
\begin{scope}
\path[clip] (  0.00,  0.00) rectangle (433.62,289.08);
\definecolor{drawColor}{gray}{0.30}

\node[text=drawColor,anchor=base,inner sep=0pt, outer sep=0pt, scale=  1.60] at (276.06,  8.61) {A};

\node[text=drawColor,anchor=base,inner sep=0pt, outer sep=0pt, scale=  1.60] at (299.57,  8.61) {O};
\end{scope}
\begin{scope}
\path[clip] (  0.00,  0.00) rectangle (433.62,289.08);
\definecolor{drawColor}{gray}{0.20}

\path[draw=drawColor,line width= 0.6pt,line join=round] (333.28, 21.83) --
	(333.28, 24.58);

\path[draw=drawColor,line width= 0.6pt,line join=round] (356.79, 21.83) --
	(356.79, 24.58);
\end{scope}
\begin{scope}
\path[clip] (  0.00,  0.00) rectangle (433.62,289.08);
\definecolor{drawColor}{gray}{0.30}

\node[text=drawColor,anchor=base,inner sep=0pt, outer sep=0pt, scale=  1.60] at (333.28,  8.61) {A};

\node[text=drawColor,anchor=base,inner sep=0pt, outer sep=0pt, scale=  1.60] at (356.79,  8.61) {O};
\end{scope}
\begin{scope}
\path[clip] (  0.00,  0.00) rectangle (433.62,289.08);
\definecolor{drawColor}{gray}{0.20}

\path[draw=drawColor,line width= 0.6pt,line join=round] (390.50, 21.83) --
	(390.50, 24.58);

\path[draw=drawColor,line width= 0.6pt,line join=round] (414.01, 21.83) --
	(414.01, 24.58);
\end{scope}
\begin{scope}
\path[clip] (  0.00,  0.00) rectangle (433.62,289.08);
\definecolor{drawColor}{gray}{0.30}

\node[text=drawColor,anchor=base,inner sep=0pt, outer sep=0pt, scale=  1.60] at (390.50,  8.61) {A};

\node[text=drawColor,anchor=base,inner sep=0pt, outer sep=0pt, scale=  1.60] at (414.01,  8.61) {O};
\end{scope}
\begin{scope}
\path[clip] (  0.00,  0.00) rectangle (433.62,289.08);
\definecolor{drawColor}{gray}{0.30}

\node[text=drawColor,anchor=base east,inner sep=0pt, outer sep=0pt, scale=  1.60] at (142.56, 28.75) {0};

\node[text=drawColor,anchor=base east,inner sep=0pt, outer sep=0pt, scale=  1.60] at (142.56, 88.16) {10};

\node[text=drawColor,anchor=base east,inner sep=0pt, outer sep=0pt, scale=  1.60] at (142.56,147.57) {20};

\node[text=drawColor,anchor=base east,inner sep=0pt, outer sep=0pt, scale=  1.60] at (142.56,206.98) {30};
\end{scope}
\begin{scope}
\path[clip] (  0.00,  0.00) rectangle (433.62,289.08);
\definecolor{drawColor}{gray}{0.20}

\path[draw=drawColor,line width= 0.6pt,line join=round] (144.76, 34.26) --
	(147.51, 34.26);

\path[draw=drawColor,line width= 0.6pt,line join=round] (144.76, 93.67) --
	(147.51, 93.67);

\path[draw=drawColor,line width= 0.6pt,line join=round] (144.76,153.08) --
	(147.51,153.08);

\path[draw=drawColor,line width= 0.6pt,line join=round] (144.76,212.49) --
	(147.51,212.49);
\end{scope}
\begin{scope}
\path[clip] (  0.00,  0.00) rectangle (433.62,289.08);
\definecolor{drawColor}{RGB}{0,0,0}

\node[text=drawColor,rotate= 90.00,anchor=base,inner sep=0pt, outer sep=0pt, scale=  2.00] at (119.93,131.03) {Response time [ms]};
\end{scope}
\begin{scope}
\path[clip] (  0.00,  0.00) rectangle (433.62,289.08);
\definecolor{fillColor}{RGB}{255,255,255}

\path[fill=fillColor] (  5.50,111.08) rectangle ( 95.15,150.99);
\end{scope}
\begin{scope}
\path[clip] (  0.00,  0.00) rectangle (433.62,289.08);
\definecolor{fillColor}{gray}{0.92}

\path[fill=fillColor] ( 11.00,131.03) rectangle ( 25.45,145.49);
\end{scope}
\begin{scope}
\path[clip] (  0.00,  0.00) rectangle (433.62,289.08);
\definecolor{fillColor}{RGB}{248,118,109}

\path[fill=fillColor] ( 11.71,131.75) rectangle ( 24.74,144.78);
\end{scope}
\begin{scope}
\path[clip] (  0.00,  0.00) rectangle (433.62,289.08);
\definecolor{fillColor}{gray}{0.92}

\path[fill=fillColor] ( 11.00,116.58) rectangle ( 25.45,131.03);
\end{scope}
\begin{scope}
\path[clip] (  0.00,  0.00) rectangle (433.62,289.08);
\definecolor{fillColor}{RGB}{40,191,47}

\path[fill=fillColor] ( 11.71,117.29) rectangle ( 24.74,130.32);
\end{scope}
\begin{scope}
\path[clip] (  0.00,  0.00) rectangle (433.62,289.08);
\definecolor{drawColor}{RGB}{0,0,0}

\node[text=drawColor,anchor=base west,inner sep=0pt, outer sep=0pt, scale=  1.60] at ( 30.95,132.75) {Network};
\end{scope}
\begin{scope}
\path[clip] (  0.00,  0.00) rectangle (433.62,289.08);
\definecolor{drawColor}{RGB}{0,0,0}

\node[text=drawColor,anchor=base west,inner sep=0pt, outer sep=0pt, scale=  1.60] at ( 30.95,118.30) {Sidecar};
\end{scope}
\begin{scope}
\path[clip] (  0.00,  0.00) rectangle (433.62,289.08);
\definecolor{drawColor}{RGB}{0,0,0}

\node[text=drawColor,anchor=base,inner sep=0pt, outer sep=0pt, scale=  2.00] at (287.81,269.81) {Network latency [ms]};
\end{scope}
\end{tikzpicture}

%% file: graphics/asymmetric_latency_1_ms.tex
% Created by tikzDevice version 0.12.6 on 2025-04-17 11:39:49
% !TEX encoding = UTF-8 Unicode
\begin{tikzpicture}[x=1pt,y=1pt]
\definecolor{fillColor}{RGB}{255,255,255}
\path[use as bounding box,fill=fillColor,fill opacity=0.00] (0,0) rectangle (325.21,289.08);
\begin{scope}
\path[clip] (  0.00,  0.00) rectangle (325.21,289.08);
\definecolor{drawColor}{RGB}{255,255,255}
\definecolor{fillColor}{RGB}{255,255,255}

\path[draw=drawColor,line width= 0.6pt,line join=round,line cap=round,fill=fillColor] (  0.00,  0.00) rectangle (325.21,289.08);
\end{scope}
\begin{scope}
\path[clip] ( 46.86, 24.58) rectangle ( 97.03,237.49);
\definecolor{fillColor}{gray}{0.92}

\path[fill=fillColor] ( 46.86, 24.58) rectangle ( 97.03,237.49);
\definecolor{drawColor}{RGB}{255,255,255}

\path[draw=drawColor,line width= 0.3pt,line join=round] ( 46.86, 71.18) --
	( 97.03, 71.18);

\path[draw=drawColor,line width= 0.3pt,line join=round] ( 46.86,145.04) --
	( 97.03,145.04);

\path[draw=drawColor,line width= 0.3pt,line join=round] ( 46.86,218.89) --
	( 97.03,218.89);

\path[draw=drawColor,line width= 0.6pt,line join=round] ( 46.86, 34.26) --
	( 97.03, 34.26);

\path[draw=drawColor,line width= 0.6pt,line join=round] ( 46.86,108.11) --
	( 97.03,108.11);

\path[draw=drawColor,line width= 0.6pt,line join=round] ( 46.86,181.97) --
	( 97.03,181.97);

\path[draw=drawColor,line width= 0.6pt,line join=round] ( 60.54, 24.58) --
	( 60.54,237.49);

\path[draw=drawColor,line width= 0.6pt,line join=round] ( 83.35, 24.58) --
	( 83.35,237.49);
\definecolor{fillColor}{RGB}{248,118,109}

\path[fill=fillColor] ( 73.08, 34.26) rectangle ( 93.61, 94.38);
\definecolor{fillColor}{RGB}{40,191,47}

\path[fill=fillColor] ( 50.28, 34.26) rectangle ( 70.80, 38.55);
\definecolor{fillColor}{RGB}{248,118,109}

\path[fill=fillColor] ( 50.28, 38.55) rectangle ( 70.80, 63.41);
\end{scope}
\begin{scope}
\path[clip] (102.53, 24.58) rectangle (152.70,237.49);
\definecolor{fillColor}{gray}{0.92}

\path[fill=fillColor] (102.53, 24.58) rectangle (152.70,237.49);
\definecolor{drawColor}{RGB}{255,255,255}

\path[draw=drawColor,line width= 0.3pt,line join=round] (102.53, 71.18) --
	(152.70, 71.18);

\path[draw=drawColor,line width= 0.3pt,line join=round] (102.53,145.04) --
	(152.70,145.04);

\path[draw=drawColor,line width= 0.3pt,line join=round] (102.53,218.89) --
	(152.70,218.89);

\path[draw=drawColor,line width= 0.6pt,line join=round] (102.53, 34.26) --
	(152.70, 34.26);

\path[draw=drawColor,line width= 0.6pt,line join=round] (102.53,108.11) --
	(152.70,108.11);

\path[draw=drawColor,line width= 0.6pt,line join=round] (102.53,181.97) --
	(152.70,181.97);

\path[draw=drawColor,line width= 0.6pt,line join=round] (116.21, 24.58) --
	(116.21,237.49);

\path[draw=drawColor,line width= 0.6pt,line join=round] (139.02, 24.58) --
	(139.02,237.49);
\definecolor{fillColor}{RGB}{248,118,109}

\path[fill=fillColor] (128.76, 34.26) rectangle (149.28,131.08);
\definecolor{fillColor}{RGB}{40,191,47}

\path[fill=fillColor] (105.95, 34.26) rectangle (126.48, 42.50);
\definecolor{fillColor}{RGB}{248,118,109}

\path[fill=fillColor] (105.95, 42.50) rectangle (126.48, 71.38);
\end{scope}
\begin{scope}
\path[clip] (158.20, 24.58) rectangle (208.37,237.49);
\definecolor{fillColor}{gray}{0.92}

\path[fill=fillColor] (158.20, 24.58) rectangle (208.37,237.49);
\definecolor{drawColor}{RGB}{255,255,255}

\path[draw=drawColor,line width= 0.3pt,line join=round] (158.20, 71.18) --
	(208.37, 71.18);

\path[draw=drawColor,line width= 0.3pt,line join=round] (158.20,145.04) --
	(208.37,145.04);

\path[draw=drawColor,line width= 0.3pt,line join=round] (158.20,218.89) --
	(208.37,218.89);

\path[draw=drawColor,line width= 0.6pt,line join=round] (158.20, 34.26) --
	(208.37, 34.26);

\path[draw=drawColor,line width= 0.6pt,line join=round] (158.20,108.11) --
	(208.37,108.11);

\path[draw=drawColor,line width= 0.6pt,line join=round] (158.20,181.97) --
	(208.37,181.97);

\path[draw=drawColor,line width= 0.6pt,line join=round] (171.88, 24.58) --
	(171.88,237.49);

\path[draw=drawColor,line width= 0.6pt,line join=round] (194.69, 24.58) --
	(194.69,237.49);
\definecolor{fillColor}{RGB}{248,118,109}

\path[fill=fillColor] (184.43, 34.26) rectangle (204.95,155.30);
\definecolor{fillColor}{RGB}{40,191,47}

\path[fill=fillColor] (161.62, 34.26) rectangle (182.15, 46.46);
\definecolor{fillColor}{RGB}{248,118,109}

\path[fill=fillColor] (161.62, 46.46) rectangle (182.15, 79.65);
\end{scope}
\begin{scope}
\path[clip] (213.87, 24.58) rectangle (264.04,237.49);
\definecolor{fillColor}{gray}{0.92}

\path[fill=fillColor] (213.87, 24.58) rectangle (264.04,237.49);
\definecolor{drawColor}{RGB}{255,255,255}

\path[draw=drawColor,line width= 0.3pt,line join=round] (213.87, 71.18) --
	(264.04, 71.18);

\path[draw=drawColor,line width= 0.3pt,line join=round] (213.87,145.04) --
	(264.04,145.04);

\path[draw=drawColor,line width= 0.3pt,line join=round] (213.87,218.89) --
	(264.04,218.89);

\path[draw=drawColor,line width= 0.6pt,line join=round] (213.87, 34.26) --
	(264.04, 34.26);

\path[draw=drawColor,line width= 0.6pt,line join=round] (213.87,108.11) --
	(264.04,108.11);

\path[draw=drawColor,line width= 0.6pt,line join=round] (213.87,181.97) --
	(264.04,181.97);

\path[draw=drawColor,line width= 0.6pt,line join=round] (227.56, 24.58) --
	(227.56,237.49);

\path[draw=drawColor,line width= 0.6pt,line join=round] (250.36, 24.58) --
	(250.36,237.49);
\definecolor{fillColor}{RGB}{248,118,109}

\path[fill=fillColor] (240.10, 34.26) rectangle (260.62,190.48);
\definecolor{fillColor}{RGB}{40,191,47}

\path[fill=fillColor] (217.29, 34.26) rectangle (237.82, 50.61);
\definecolor{fillColor}{RGB}{248,118,109}

\path[fill=fillColor] (217.29, 50.61) rectangle (237.82, 86.88);
\end{scope}
\begin{scope}
\path[clip] (269.54, 24.58) rectangle (319.71,237.49);
\definecolor{fillColor}{gray}{0.92}

\path[fill=fillColor] (269.54, 24.58) rectangle (319.71,237.49);
\definecolor{drawColor}{RGB}{255,255,255}

\path[draw=drawColor,line width= 0.3pt,line join=round] (269.54, 71.18) --
	(319.71, 71.18);

\path[draw=drawColor,line width= 0.3pt,line join=round] (269.54,145.04) --
	(319.71,145.04);

\path[draw=drawColor,line width= 0.3pt,line join=round] (269.54,218.89) --
	(319.71,218.89);

\path[draw=drawColor,line width= 0.6pt,line join=round] (269.54, 34.26) --
	(319.71, 34.26);

\path[draw=drawColor,line width= 0.6pt,line join=round] (269.54,108.11) --
	(319.71,108.11);

\path[draw=drawColor,line width= 0.6pt,line join=round] (269.54,181.97) --
	(319.71,181.97);

\path[draw=drawColor,line width= 0.6pt,line join=round] (283.23, 24.58) --
	(283.23,237.49);

\path[draw=drawColor,line width= 0.6pt,line join=round] (306.03, 24.58) --
	(306.03,237.49);
\definecolor{fillColor}{RGB}{248,118,109}

\path[fill=fillColor] (295.77, 34.26) rectangle (316.29,227.81);
\definecolor{fillColor}{RGB}{40,191,47}

\path[fill=fillColor] (272.96, 34.26) rectangle (293.49, 54.27);
\definecolor{fillColor}{RGB}{248,118,109}

\path[fill=fillColor] (272.96, 54.27) rectangle (293.49, 96.66);
\end{scope}
\begin{scope}
\path[clip] ( 46.86,237.49) rectangle ( 97.03,260.42);
\definecolor{drawColor}{RGB}{255,255,255}

\path[draw=drawColor,line width= 0.6pt,line join=round,line cap=round] ( 46.86,237.49) rectangle ( 97.03,260.42);
\definecolor{drawColor}{gray}{0.10}

\node[text=drawColor,anchor=base,inner sep=0pt, outer sep=0pt, scale=  1.60] at ( 71.94,243.44) {1};
\end{scope}
\begin{scope}
\path[clip] (102.53,237.49) rectangle (152.70,260.42);
\definecolor{drawColor}{RGB}{255,255,255}

\path[draw=drawColor,line width= 0.6pt,line join=round,line cap=round] (102.53,237.49) rectangle (152.70,260.42);
\definecolor{drawColor}{gray}{0.10}

\node[text=drawColor,anchor=base,inner sep=0pt, outer sep=0pt, scale=  1.60] at (127.62,243.44) {2};
\end{scope}
\begin{scope}
\path[clip] (158.20,237.49) rectangle (208.37,260.42);
\definecolor{drawColor}{RGB}{255,255,255}

\path[draw=drawColor,line width= 0.6pt,line join=round,line cap=round] (158.20,237.49) rectangle (208.37,260.42);
\definecolor{drawColor}{gray}{0.10}

\node[text=drawColor,anchor=base,inner sep=0pt, outer sep=0pt, scale=  1.60] at (183.29,243.44) {3};
\end{scope}
\begin{scope}
\path[clip] (213.87,237.49) rectangle (264.04,260.42);
\definecolor{drawColor}{RGB}{255,255,255}

\path[draw=drawColor,line width= 0.6pt,line join=round,line cap=round] (213.87,237.49) rectangle (264.04,260.42);
\definecolor{drawColor}{gray}{0.10}

\node[text=drawColor,anchor=base,inner sep=0pt, outer sep=0pt, scale=  1.60] at (238.96,243.44) {4};
\end{scope}
\begin{scope}
\path[clip] (269.54,237.49) rectangle (319.71,260.42);
\definecolor{drawColor}{RGB}{255,255,255}

\path[draw=drawColor,line width= 0.6pt,line join=round,line cap=round] (269.54,237.49) rectangle (319.71,260.42);
\definecolor{drawColor}{gray}{0.10}

\node[text=drawColor,anchor=base,inner sep=0pt, outer sep=0pt, scale=  1.60] at (294.63,243.44) {5};
\end{scope}
\begin{scope}
\path[clip] (  0.00,  0.00) rectangle (325.21,289.08);
\definecolor{drawColor}{gray}{0.20}

\path[draw=drawColor,line width= 0.6pt,line join=round] ( 60.54, 21.83) --
	( 60.54, 24.58);

\path[draw=drawColor,line width= 0.6pt,line join=round] ( 83.35, 21.83) --
	( 83.35, 24.58);
\end{scope}
\begin{scope}
\path[clip] (  0.00,  0.00) rectangle (325.21,289.08);
\definecolor{drawColor}{gray}{0.30}

\node[text=drawColor,anchor=base,inner sep=0pt, outer sep=0pt, scale=  1.60] at ( 60.54,  8.61) {A};

\node[text=drawColor,anchor=base,inner sep=0pt, outer sep=0pt, scale=  1.60] at ( 83.35,  8.61) {O};
\end{scope}
\begin{scope}
\path[clip] (  0.00,  0.00) rectangle (325.21,289.08);
\definecolor{drawColor}{gray}{0.20}

\path[draw=drawColor,line width= 0.6pt,line join=round] (116.21, 21.83) --
	(116.21, 24.58);

\path[draw=drawColor,line width= 0.6pt,line join=round] (139.02, 21.83) --
	(139.02, 24.58);
\end{scope}
\begin{scope}
\path[clip] (  0.00,  0.00) rectangle (325.21,289.08);
\definecolor{drawColor}{gray}{0.30}

\node[text=drawColor,anchor=base,inner sep=0pt, outer sep=0pt, scale=  1.60] at (116.21,  8.61) {A};

\node[text=drawColor,anchor=base,inner sep=0pt, outer sep=0pt, scale=  1.60] at (139.02,  8.61) {O};
\end{scope}
\begin{scope}
\path[clip] (  0.00,  0.00) rectangle (325.21,289.08);
\definecolor{drawColor}{gray}{0.20}

\path[draw=drawColor,line width= 0.6pt,line join=round] (171.88, 21.83) --
	(171.88, 24.58);

\path[draw=drawColor,line width= 0.6pt,line join=round] (194.69, 21.83) --
	(194.69, 24.58);
\end{scope}
\begin{scope}
\path[clip] (  0.00,  0.00) rectangle (325.21,289.08);
\definecolor{drawColor}{gray}{0.30}

\node[text=drawColor,anchor=base,inner sep=0pt, outer sep=0pt, scale=  1.60] at (171.88,  8.61) {A};

\node[text=drawColor,anchor=base,inner sep=0pt, outer sep=0pt, scale=  1.60] at (194.69,  8.61) {O};
\end{scope}
\begin{scope}
\path[clip] (  0.00,  0.00) rectangle (325.21,289.08);
\definecolor{drawColor}{gray}{0.20}

\path[draw=drawColor,line width= 0.6pt,line join=round] (227.56, 21.83) --
	(227.56, 24.58);

\path[draw=drawColor,line width= 0.6pt,line join=round] (250.36, 21.83) --
	(250.36, 24.58);
\end{scope}
\begin{scope}
\path[clip] (  0.00,  0.00) rectangle (325.21,289.08);
\definecolor{drawColor}{gray}{0.30}

\node[text=drawColor,anchor=base,inner sep=0pt, outer sep=0pt, scale=  1.60] at (227.56,  8.61) {A};

\node[text=drawColor,anchor=base,inner sep=0pt, outer sep=0pt, scale=  1.60] at (250.36,  8.61) {O};
\end{scope}
\begin{scope}
\path[clip] (  0.00,  0.00) rectangle (325.21,289.08);
\definecolor{drawColor}{gray}{0.20}

\path[draw=drawColor,line width= 0.6pt,line join=round] (283.23, 21.83) --
	(283.23, 24.58);

\path[draw=drawColor,line width= 0.6pt,line join=round] (306.03, 21.83) --
	(306.03, 24.58);
\end{scope}
\begin{scope}
\path[clip] (  0.00,  0.00) rectangle (325.21,289.08);
\definecolor{drawColor}{gray}{0.30}

\node[text=drawColor,anchor=base,inner sep=0pt, outer sep=0pt, scale=  1.60] at (283.23,  8.61) {A};

\node[text=drawColor,anchor=base,inner sep=0pt, outer sep=0pt, scale=  1.60] at (306.03,  8.61) {O};
\end{scope}
\begin{scope}
\path[clip] (  0.00,  0.00) rectangle (325.21,289.08);
\definecolor{drawColor}{gray}{0.30}

\node[text=drawColor,anchor=base east,inner sep=0pt, outer sep=0pt, scale=  1.60] at ( 41.91, 28.75) {0};

\node[text=drawColor,anchor=base east,inner sep=0pt, outer sep=0pt, scale=  1.60] at ( 41.91,102.60) {10};

\node[text=drawColor,anchor=base east,inner sep=0pt, outer sep=0pt, scale=  1.60] at ( 41.91,176.46) {20};
\end{scope}
\begin{scope}
\path[clip] (  0.00,  0.00) rectangle (325.21,289.08);
\definecolor{drawColor}{gray}{0.20}

\path[draw=drawColor,line width= 0.6pt,line join=round] ( 44.11, 34.26) --
	( 46.86, 34.26);

\path[draw=drawColor,line width= 0.6pt,line join=round] ( 44.11,108.11) --
	( 46.86,108.11);

\path[draw=drawColor,line width= 0.6pt,line join=round] ( 44.11,181.97) --
	( 46.86,181.97);
\end{scope}
\begin{scope}
\path[clip] (  0.00,  0.00) rectangle (325.21,289.08);
\definecolor{drawColor}{RGB}{0,0,0}

\node[text=drawColor,rotate= 90.00,anchor=base,inner sep=0pt, outer sep=0pt, scale=  2.00] at ( 19.27,131.03) {Response time [ms]};
\end{scope}
\begin{scope}
\path[clip] (  0.00,  0.00) rectangle (325.21,289.08);
\definecolor{drawColor}{RGB}{0,0,0}

\node[text=drawColor,anchor=base,inner sep=0pt, outer sep=0pt, scale=  2.00] at (183.29,269.81) {Number of workers};
\end{scope}
\end{tikzpicture}

%% file: benchmarks/25-01-14_AWS_plots/cart_size_latency/cart_size_latency.tex
% Created by tikzDevice version 0.12.6 on 2025-01-16 09:42:54
% !TEX encoding = UTF-8 Unicode
\begin{tikzpicture}[x=1pt,y=1pt]
\definecolor{fillColor}{RGB}{255,255,255}
\path[use as bounding box,fill=fillColor,fill opacity=0.00] (0,0) rectangle (173.45, 93.95);
\begin{scope}
\path[clip] (104.94, 28.04) rectangle (167.95, 88.45);
\definecolor{drawColor}{gray}{0.92}

\path[draw=drawColor,line width= 0.3pt,line join=round] (104.94, 28.18) --
	(167.95, 28.18);

\path[draw=drawColor,line width= 0.3pt,line join=round] (104.94, 37.54) --
	(167.95, 37.54);

\path[draw=drawColor,line width= 0.3pt,line join=round] (104.94, 46.91) --
	(167.95, 46.91);

\path[draw=drawColor,line width= 0.3pt,line join=round] (104.94, 56.27) --
	(167.95, 56.27);

\path[draw=drawColor,line width= 0.3pt,line join=round] (104.94, 65.63) --
	(167.95, 65.63);

\path[draw=drawColor,line width= 0.3pt,line join=round] (104.94, 75.00) --
	(167.95, 75.00);

\path[draw=drawColor,line width= 0.3pt,line join=round] (104.94, 84.36) --
	(167.95, 84.36);

\path[draw=drawColor,line width= 0.3pt,line join=round] (114.96, 28.04) --
	(114.96, 88.45);

\path[draw=drawColor,line width= 0.3pt,line join=round] (129.28, 28.04) --
	(129.28, 88.45);

\path[draw=drawColor,line width= 0.3pt,line join=round] (143.60, 28.04) --
	(143.60, 88.45);

\path[draw=drawColor,line width= 0.3pt,line join=round] (157.92, 28.04) --
	(157.92, 88.45);

\path[draw=drawColor,line width= 0.6pt,line join=round] (104.94, 32.86) --
	(167.95, 32.86);

\path[draw=drawColor,line width= 0.6pt,line join=round] (104.94, 42.23) --
	(167.95, 42.23);

\path[draw=drawColor,line width= 0.6pt,line join=round] (104.94, 51.59) --
	(167.95, 51.59);

\path[draw=drawColor,line width= 0.6pt,line join=round] (104.94, 60.95) --
	(167.95, 60.95);

\path[draw=drawColor,line width= 0.6pt,line join=round] (104.94, 70.32) --
	(167.95, 70.32);

\path[draw=drawColor,line width= 0.6pt,line join=round] (104.94, 79.68) --
	(167.95, 79.68);

\path[draw=drawColor,line width= 0.6pt,line join=round] (107.80, 28.04) --
	(107.80, 88.45);

\path[draw=drawColor,line width= 0.6pt,line join=round] (122.12, 28.04) --
	(122.12, 88.45);

\path[draw=drawColor,line width= 0.6pt,line join=round] (136.44, 28.04) --
	(136.44, 88.45);

\path[draw=drawColor,line width= 0.6pt,line join=round] (150.76, 28.04) --
	(150.76, 88.45);

\path[draw=drawColor,line width= 0.6pt,line join=round] (165.08, 28.04) --
	(165.08, 88.45);
\definecolor{fillColor}{RGB}{0,191,196}

\path[fill=fillColor,fill opacity=0.20] (107.80, 36.45) --
	(114.96, 37.12) --
	(122.12, 39.65) --
	(129.28, 39.63) --
	(136.44, 42.32) --
	(143.60, 43.63) --
	(150.76, 53.46) --
	(157.92, 53.83) --
	(165.08, 61.81) --
	(165.08, 37.87) --
	(157.92, 36.68) --
	(150.76, 34.86) --
	(143.60, 37.07) --
	(136.44, 35.00) --
	(129.28, 34.40) --
	(122.12, 32.46) --
	(114.96, 32.25) --
	(107.80, 32.57) --
	cycle;

\path[] (107.80, 36.45) --
	(114.96, 37.12) --
	(122.12, 39.65) --
	(129.28, 39.63) --
	(136.44, 42.32) --
	(143.60, 43.63) --
	(150.76, 53.46) --
	(157.92, 53.83) --
	(165.08, 61.81);

\path[] (165.08, 37.87) --
	(157.92, 36.68) --
	(150.76, 34.86) --
	(143.60, 37.07) --
	(136.44, 35.00) --
	(129.28, 34.40) --
	(122.12, 32.46) --
	(114.96, 32.25) --
	(107.80, 32.57);
\definecolor{fillColor}{RGB}{248,118,109}

\path[fill=fillColor,fill opacity=0.20] (107.80, 32.88) --
	(114.96, 36.97) --
	(122.12, 41.81) --
	(129.28, 46.27) --
	(136.44, 50.70) --
	(143.60, 55.25) --
	(150.76, 66.47) --
	(157.92, 73.84) --
	(165.08, 85.70) --
	(165.08, 56.10) --
	(157.92, 52.59) --
	(150.76, 51.00) --
	(143.60, 49.95) --
	(136.44, 45.70) --
	(129.28, 42.20) --
	(122.12, 37.83) --
	(114.96, 34.06) --
	(107.80, 30.78) --
	cycle;

\path[] (107.80, 32.88) --
	(114.96, 36.97) --
	(122.12, 41.81) --
	(129.28, 46.27) --
	(136.44, 50.70) --
	(143.60, 55.25) --
	(150.76, 66.47) --
	(157.92, 73.84) --
	(165.08, 85.70);

\path[] (165.08, 56.10) --
	(157.92, 52.59) --
	(150.76, 51.00) --
	(143.60, 49.95) --
	(136.44, 45.70) --
	(129.28, 42.20) --
	(122.12, 37.83) --
	(114.96, 34.06) --
	(107.80, 30.78);
\definecolor{drawColor}{RGB}{0,191,196}

\path[draw=drawColor,line width= 1.1pt,line join=round] (107.80, 34.51) --
	(114.96, 34.69) --
	(122.12, 36.05) --
	(129.28, 37.01) --
	(136.44, 38.66) --
	(143.60, 40.35) --
	(150.76, 44.16) --
	(157.92, 45.26) --
	(165.08, 49.84);
\definecolor{drawColor}{RGB}{248,118,109}

\path[draw=drawColor,line width= 1.1pt,line join=round] (107.80, 31.83) --
	(114.96, 35.52) --
	(122.12, 39.82) --
	(129.28, 44.23) --
	(136.44, 48.20) --
	(143.60, 52.60) --
	(150.76, 58.74) --
	(157.92, 63.22) --
	(165.08, 70.90);
\end{scope}
\begin{scope}
\path[clip] (  0.00,  0.00) rectangle (173.45, 93.95);
\definecolor{drawColor}{gray}{0.30}

\node[text=drawColor,anchor=base east,inner sep=0pt, outer sep=0pt, scale=  0.88] at ( 99.99, 29.83) {30};

\node[text=drawColor,anchor=base east,inner sep=0pt, outer sep=0pt, scale=  0.88] at ( 99.99, 39.20) {40};

\node[text=drawColor,anchor=base east,inner sep=0pt, outer sep=0pt, scale=  0.88] at ( 99.99, 48.56) {50};

\node[text=drawColor,anchor=base east,inner sep=0pt, outer sep=0pt, scale=  0.88] at ( 99.99, 57.92) {60};

\node[text=drawColor,anchor=base east,inner sep=0pt, outer sep=0pt, scale=  0.88] at ( 99.99, 67.29) {70};

\node[text=drawColor,anchor=base east,inner sep=0pt, outer sep=0pt, scale=  0.88] at ( 99.99, 76.65) {80};
\end{scope}
\begin{scope}
\path[clip] (  0.00,  0.00) rectangle (173.45, 93.95);
\definecolor{drawColor}{gray}{0.30}

\node[text=drawColor,anchor=base,inner sep=0pt, outer sep=0pt, scale=  0.88] at (107.80, 17.03) {1};

\node[text=drawColor,anchor=base,inner sep=0pt, outer sep=0pt, scale=  0.88] at (122.12, 17.03) {3};

\node[text=drawColor,anchor=base,inner sep=0pt, outer sep=0pt, scale=  0.88] at (136.44, 17.03) {5};

\node[text=drawColor,anchor=base,inner sep=0pt, outer sep=0pt, scale=  0.88] at (150.76, 17.03) {7};

\node[text=drawColor,anchor=base,inner sep=0pt, outer sep=0pt, scale=  0.88] at (165.08, 17.03) {9};
\end{scope}
\begin{scope}
\path[clip] (  0.00,  0.00) rectangle (173.45, 93.95);
\definecolor{drawColor}{RGB}{0,0,0}

\node[text=drawColor,anchor=base,inner sep=0pt, outer sep=0pt, scale=  0.80] at (136.44,  7.06) {Cart items};
\end{scope}
\begin{scope}
\path[clip] (  0.00,  0.00) rectangle (173.45, 93.95);
\definecolor{drawColor}{RGB}{0,0,0}

\node[text=drawColor,rotate= 90.00,anchor=base,inner sep=0pt, outer sep=0pt, scale=  0.80] at ( 86.89, 58.24) {Response time [ms]};
\end{scope}
\begin{scope}
\path[clip] (  0.00,  0.00) rectangle (173.45, 93.95);
\definecolor{fillColor}{RGB}{0,191,196}

\path[fill=fillColor,fill opacity=0.20] (  6.21, 58.96) rectangle ( 19.24, 71.99);
\end{scope}
\begin{scope}
\path[clip] (  0.00,  0.00) rectangle (173.45, 93.95);
\definecolor{drawColor}{RGB}{0,191,196}

\path[draw=drawColor,line width= 1.1pt,line join=round] (  6.95, 65.47) -- ( 18.51, 65.47);
\end{scope}
\begin{scope}
\path[clip] (  0.00,  0.00) rectangle (173.45, 93.95);
\definecolor{fillColor}{RGB}{248,118,109}

\path[fill=fillColor,fill opacity=0.20] (  6.21, 44.50) rectangle ( 19.24, 57.53);
\end{scope}
\begin{scope}
\path[clip] (  0.00,  0.00) rectangle (173.45, 93.95);
\definecolor{drawColor}{RGB}{248,118,109}

\path[draw=drawColor,line width= 1.1pt,line join=round] (  6.95, 51.02) -- ( 18.51, 51.02);
\end{scope}
\begin{scope}
\path[clip] (  0.00,  0.00) rectangle (173.45, 93.95);
\definecolor{drawColor}{RGB}{0,0,0}

\node[text=drawColor,anchor=base west,inner sep=0pt, outer sep=0pt, scale=  0.80] at ( 25.45, 62.72) {Accompanist};
\end{scope}
\begin{scope}
\path[clip] (  0.00,  0.00) rectangle (173.45, 93.95);
\definecolor{drawColor}{RGB}{0,0,0}

\node[text=drawColor,anchor=base west,inner sep=0pt, outer sep=0pt, scale=  0.80] at ( 25.45, 48.26) {Orchestrator};
\end{scope}
\end{tikzpicture}

%% file: benchmarks/25-01-14_AWS_plots/2_nodes_1_zone_fcom/2_AWS_2_nodes_1_zone.tex
% Created by tikzDevice version 0.12.6 on 2025-01-16 09:45:44
% !TEX encoding = UTF-8 Unicode
\begin{tikzpicture}[x=1pt,y=1pt]
\definecolor{fillColor}{RGB}{255,255,255}
\path[use as bounding box,fill=fillColor,fill opacity=0.00] (0,0) rectangle (101.18, 93.95);
\begin{scope}
\path[clip] ( 23.65, 28.04) rectangle ( 95.68, 88.45);
\definecolor{drawColor}{gray}{0.92}

\path[draw=drawColor,line width= 0.3pt,line join=round] ( 23.65, 36.46) --
	( 95.68, 36.46);

\path[draw=drawColor,line width= 0.3pt,line join=round] ( 23.65, 47.80) --
	( 95.68, 47.80);

\path[draw=drawColor,line width= 0.3pt,line join=round] ( 23.65, 59.15) --
	( 95.68, 59.15);

\path[draw=drawColor,line width= 0.3pt,line join=round] ( 23.65, 70.50) --
	( 95.68, 70.50);

\path[draw=drawColor,line width= 0.3pt,line join=round] ( 23.65, 81.85) --
	( 95.68, 81.85);

\path[draw=drawColor,line width= 0.3pt,line join=round] ( 35.11, 28.04) --
	( 35.11, 88.45);

\path[draw=drawColor,line width= 0.3pt,line join=round] ( 51.48, 28.04) --
	( 51.48, 88.45);

\path[draw=drawColor,line width= 0.3pt,line join=round] ( 67.85, 28.04) --
	( 67.85, 88.45);

\path[draw=drawColor,line width= 0.3pt,line join=round] ( 84.22, 28.04) --
	( 84.22, 88.45);

\path[draw=drawColor,line width= 0.6pt,line join=round] ( 23.65, 30.78) --
	( 95.68, 30.78);

\path[draw=drawColor,line width= 0.6pt,line join=round] ( 23.65, 42.13) --
	( 95.68, 42.13);

\path[draw=drawColor,line width= 0.6pt,line join=round] ( 23.65, 53.48) --
	( 95.68, 53.48);

\path[draw=drawColor,line width= 0.6pt,line join=round] ( 23.65, 64.83) --
	( 95.68, 64.83);

\path[draw=drawColor,line width= 0.6pt,line join=round] ( 23.65, 76.17) --
	( 95.68, 76.17);

\path[draw=drawColor,line width= 0.6pt,line join=round] ( 23.65, 87.52) --
	( 95.68, 87.52);

\path[draw=drawColor,line width= 0.6pt,line join=round] ( 26.92, 28.04) --
	( 26.92, 88.45);

\path[draw=drawColor,line width= 0.6pt,line join=round] ( 43.29, 28.04) --
	( 43.29, 88.45);

\path[draw=drawColor,line width= 0.6pt,line join=round] ( 59.66, 28.04) --
	( 59.66, 88.45);

\path[draw=drawColor,line width= 0.6pt,line join=round] ( 76.03, 28.04) --
	( 76.03, 88.45);

\path[draw=drawColor,line width= 0.6pt,line join=round] ( 92.40, 28.04) --
	( 92.40, 88.45);
\definecolor{fillColor}{RGB}{0,191,196}

\path[fill=fillColor,fill opacity=0.50] ( 27.25, 30.78) rectangle ( 27.90, 30.78);

\path[fill=fillColor,fill opacity=0.50] ( 27.90, 30.78) rectangle ( 28.56, 30.78);

\path[fill=fillColor,fill opacity=0.50] ( 28.56, 30.78) rectangle ( 29.21, 30.78);

\path[fill=fillColor,fill opacity=0.50] ( 29.21, 30.78) rectangle ( 29.87, 30.78);

\path[fill=fillColor,fill opacity=0.50] ( 29.87, 30.78) rectangle ( 30.52, 30.78);

\path[fill=fillColor,fill opacity=0.50] ( 30.52, 30.78) rectangle ( 31.18, 30.78);

\path[fill=fillColor,fill opacity=0.50] ( 31.18, 30.78) rectangle ( 31.83, 30.78);

\path[fill=fillColor,fill opacity=0.50] ( 31.83, 30.78) rectangle ( 32.49, 30.78);

\path[fill=fillColor,fill opacity=0.50] ( 32.49, 30.78) rectangle ( 33.14, 30.78);

\path[fill=fillColor,fill opacity=0.50] ( 33.14, 30.78) rectangle ( 33.80, 30.78);

\path[fill=fillColor,fill opacity=0.50] ( 33.80, 30.78) rectangle ( 34.45, 30.78);

\path[fill=fillColor,fill opacity=0.50] ( 34.45, 30.78) rectangle ( 35.11, 30.78);

\path[fill=fillColor,fill opacity=0.50] ( 35.11, 30.78) rectangle ( 35.76, 30.78);

\path[fill=fillColor,fill opacity=0.50] ( 35.76, 30.78) rectangle ( 36.42, 30.78);

\path[fill=fillColor,fill opacity=0.50] ( 36.42, 30.78) rectangle ( 37.07, 30.78);

\path[fill=fillColor,fill opacity=0.50] ( 37.07, 30.78) rectangle ( 37.73, 30.78);

\path[fill=fillColor,fill opacity=0.50] ( 37.73, 30.78) rectangle ( 38.38, 30.78);

\path[fill=fillColor,fill opacity=0.50] ( 38.38, 30.78) rectangle ( 39.04, 30.78);

\path[fill=fillColor,fill opacity=0.50] ( 39.04, 30.78) rectangle ( 39.69, 30.78);

\path[fill=fillColor,fill opacity=0.50] ( 39.69, 30.78) rectangle ( 40.34, 30.78);

\path[fill=fillColor,fill opacity=0.50] ( 40.34, 30.78) rectangle ( 41.00, 30.78);

\path[fill=fillColor,fill opacity=0.50] ( 41.00, 30.78) rectangle ( 41.65, 30.78);

\path[fill=fillColor,fill opacity=0.50] ( 41.65, 30.78) rectangle ( 42.31, 30.78);

\path[fill=fillColor,fill opacity=0.50] ( 42.31, 30.78) rectangle ( 42.96, 30.78);

\path[fill=fillColor,fill opacity=0.50] ( 42.96, 30.78) rectangle ( 43.62, 30.78);

\path[fill=fillColor,fill opacity=0.50] ( 43.62, 30.78) rectangle ( 44.27, 30.78);

\path[fill=fillColor,fill opacity=0.50] ( 44.27, 30.78) rectangle ( 44.93, 30.78);

\path[fill=fillColor,fill opacity=0.50] ( 44.93, 30.78) rectangle ( 45.58, 36.68);

\path[fill=fillColor,fill opacity=0.50] ( 45.58, 30.78) rectangle ( 46.24, 58.92);

\path[fill=fillColor,fill opacity=0.50] ( 46.24, 30.78) rectangle ( 46.89, 61.19);

\path[fill=fillColor,fill opacity=0.50] ( 46.89, 30.78) rectangle ( 47.55, 46.67);

\path[fill=fillColor,fill opacity=0.50] ( 47.55, 30.78) rectangle ( 48.20, 44.85);

\path[fill=fillColor,fill opacity=0.50] ( 48.20, 30.78) rectangle ( 48.86, 45.31);

\path[fill=fillColor,fill opacity=0.50] ( 48.86, 30.78) rectangle ( 49.51, 38.50);

\path[fill=fillColor,fill opacity=0.50] ( 49.51, 30.78) rectangle ( 50.17, 45.76);

\path[fill=fillColor,fill opacity=0.50] ( 50.17, 30.78) rectangle ( 50.82, 45.31);

\path[fill=fillColor,fill opacity=0.50] ( 50.82, 30.78) rectangle ( 51.48, 45.31);

\path[fill=fillColor,fill opacity=0.50] ( 51.48, 30.78) rectangle ( 52.13, 47.12);

\path[fill=fillColor,fill opacity=0.50] ( 52.13, 30.78) rectangle ( 52.79, 50.75);

\path[fill=fillColor,fill opacity=0.50] ( 52.79, 30.78) rectangle ( 53.44, 44.40);

\path[fill=fillColor,fill opacity=0.50] ( 53.44, 30.78) rectangle ( 54.10, 44.85);

\path[fill=fillColor,fill opacity=0.50] ( 54.10, 30.78) rectangle ( 54.75, 42.13);

\path[fill=fillColor,fill opacity=0.50] ( 54.75, 30.78) rectangle ( 55.41, 40.31);

\path[fill=fillColor,fill opacity=0.50] ( 55.41, 30.78) rectangle ( 56.06, 40.77);

\path[fill=fillColor,fill opacity=0.50] ( 56.06, 30.78) rectangle ( 56.72, 38.95);

\path[fill=fillColor,fill opacity=0.50] ( 56.72, 30.78) rectangle ( 57.37, 40.31);

\path[fill=fillColor,fill opacity=0.50] ( 57.37, 30.78) rectangle ( 58.03, 39.86);

\path[fill=fillColor,fill opacity=0.50] ( 58.03, 30.78) rectangle ( 58.68, 38.05);

\path[fill=fillColor,fill opacity=0.50] ( 58.68, 30.78) rectangle ( 59.34, 41.22);

\path[fill=fillColor,fill opacity=0.50] ( 59.34, 30.78) rectangle ( 59.99, 39.41);

\path[fill=fillColor,fill opacity=0.50] ( 59.99, 30.78) rectangle ( 60.64, 38.50);

\path[fill=fillColor,fill opacity=0.50] ( 60.64, 30.78) rectangle ( 61.30, 37.14);

\path[fill=fillColor,fill opacity=0.50] ( 61.30, 30.78) rectangle ( 61.95, 35.32);

\path[fill=fillColor,fill opacity=0.50] ( 61.95, 30.78) rectangle ( 62.61, 37.14);

\path[fill=fillColor,fill opacity=0.50] ( 62.61, 30.78) rectangle ( 63.26, 34.87);

\path[fill=fillColor,fill opacity=0.50] ( 63.26, 30.78) rectangle ( 63.92, 35.78);

\path[fill=fillColor,fill opacity=0.50] ( 63.92, 30.78) rectangle ( 64.57, 35.78);

\path[fill=fillColor,fill opacity=0.50] ( 64.57, 30.78) rectangle ( 65.23, 34.41);

\path[fill=fillColor,fill opacity=0.50] ( 65.23, 30.78) rectangle ( 65.88, 36.68);

\path[fill=fillColor,fill opacity=0.50] ( 65.88, 30.78) rectangle ( 66.54, 34.41);

\path[fill=fillColor,fill opacity=0.50] ( 66.54, 30.78) rectangle ( 67.19, 32.60);

\path[fill=fillColor,fill opacity=0.50] ( 67.19, 30.78) rectangle ( 67.85, 33.51);

\path[fill=fillColor,fill opacity=0.50] ( 67.85, 30.78) rectangle ( 68.50, 35.78);

\path[fill=fillColor,fill opacity=0.50] ( 68.50, 30.78) rectangle ( 69.16, 33.51);

\path[fill=fillColor,fill opacity=0.50] ( 69.16, 30.78) rectangle ( 69.81, 36.23);

\path[fill=fillColor,fill opacity=0.50] ( 69.81, 30.78) rectangle ( 70.47, 32.14);

\path[fill=fillColor,fill opacity=0.50] ( 70.47, 30.78) rectangle ( 71.12, 32.60);

\path[fill=fillColor,fill opacity=0.50] ( 71.12, 30.78) rectangle ( 71.78, 35.78);

\path[fill=fillColor,fill opacity=0.50] ( 71.78, 30.78) rectangle ( 72.43, 32.14);

\path[fill=fillColor,fill opacity=0.50] ( 72.43, 30.78) rectangle ( 73.09, 33.05);

\path[fill=fillColor,fill opacity=0.50] ( 73.09, 30.78) rectangle ( 73.74, 32.14);

\path[fill=fillColor,fill opacity=0.50] ( 73.74, 30.78) rectangle ( 74.40, 32.60);

\path[fill=fillColor,fill opacity=0.50] ( 74.40, 30.78) rectangle ( 75.05, 31.24);

\path[fill=fillColor,fill opacity=0.50] ( 75.05, 30.78) rectangle ( 75.71, 30.78);

\path[fill=fillColor,fill opacity=0.50] ( 75.71, 30.78) rectangle ( 76.36, 32.60);

\path[fill=fillColor,fill opacity=0.50] ( 76.36, 30.78) rectangle ( 77.02, 32.60);

\path[fill=fillColor,fill opacity=0.50] ( 77.02, 30.78) rectangle ( 77.67, 33.96);

\path[fill=fillColor,fill opacity=0.50] ( 77.67, 30.78) rectangle ( 78.33, 33.05);

\path[fill=fillColor,fill opacity=0.50] ( 78.33, 30.78) rectangle ( 78.98, 31.24);

\path[fill=fillColor,fill opacity=0.50] ( 78.98, 30.78) rectangle ( 79.63, 31.69);

\path[fill=fillColor,fill opacity=0.50] ( 79.63, 30.78) rectangle ( 80.29, 32.14);

\path[fill=fillColor,fill opacity=0.50] ( 80.29, 30.78) rectangle ( 80.94, 32.60);

\path[fill=fillColor,fill opacity=0.50] ( 80.94, 30.78) rectangle ( 81.60, 33.51);

\path[fill=fillColor,fill opacity=0.50] ( 81.60, 30.78) rectangle ( 82.25, 32.60);

\path[fill=fillColor,fill opacity=0.50] ( 82.25, 30.78) rectangle ( 82.91, 32.14);

\path[fill=fillColor,fill opacity=0.50] ( 82.91, 30.78) rectangle ( 83.56, 31.69);

\path[fill=fillColor,fill opacity=0.50] ( 83.56, 30.78) rectangle ( 84.22, 31.69);

\path[fill=fillColor,fill opacity=0.50] ( 84.22, 30.78) rectangle ( 84.87, 32.60);

\path[fill=fillColor,fill opacity=0.50] ( 84.87, 30.78) rectangle ( 85.53, 32.60);

\path[fill=fillColor,fill opacity=0.50] ( 85.53, 30.78) rectangle ( 86.18, 31.69);

\path[fill=fillColor,fill opacity=0.50] ( 86.18, 30.78) rectangle ( 86.84, 31.24);

\path[fill=fillColor,fill opacity=0.50] ( 86.84, 30.78) rectangle ( 87.49, 31.69);

\path[fill=fillColor,fill opacity=0.50] ( 87.49, 30.78) rectangle ( 88.15, 32.60);

\path[fill=fillColor,fill opacity=0.50] ( 88.15, 30.78) rectangle ( 88.80, 33.05);

\path[fill=fillColor,fill opacity=0.50] ( 88.80, 30.78) rectangle ( 89.46, 31.24);

\path[fill=fillColor,fill opacity=0.50] ( 89.46, 30.78) rectangle ( 90.11, 31.24);

\path[fill=fillColor,fill opacity=0.50] ( 90.11, 30.78) rectangle ( 90.77, 30.78);

\path[fill=fillColor,fill opacity=0.50] ( 90.77, 30.78) rectangle ( 91.42, 31.24);

\path[fill=fillColor,fill opacity=0.50] ( 91.42, 30.78) rectangle ( 92.08, 32.14);
\definecolor{fillColor}{RGB}{248,118,109}

\path[fill=fillColor,fill opacity=0.50] ( 27.25, 30.78) rectangle ( 27.90, 30.78);

\path[fill=fillColor,fill opacity=0.50] ( 27.90, 30.78) rectangle ( 28.56, 30.78);

\path[fill=fillColor,fill opacity=0.50] ( 28.56, 30.78) rectangle ( 29.21, 30.78);

\path[fill=fillColor,fill opacity=0.50] ( 29.21, 30.78) rectangle ( 29.87, 30.78);

\path[fill=fillColor,fill opacity=0.50] ( 29.87, 30.78) rectangle ( 30.52, 30.78);

\path[fill=fillColor,fill opacity=0.50] ( 30.52, 30.78) rectangle ( 31.18, 30.78);

\path[fill=fillColor,fill opacity=0.50] ( 31.18, 30.78) rectangle ( 31.83, 30.78);

\path[fill=fillColor,fill opacity=0.50] ( 31.83, 30.78) rectangle ( 32.49, 30.78);

\path[fill=fillColor,fill opacity=0.50] ( 32.49, 30.78) rectangle ( 33.14, 30.78);

\path[fill=fillColor,fill opacity=0.50] ( 33.14, 30.78) rectangle ( 33.80, 30.78);

\path[fill=fillColor,fill opacity=0.50] ( 33.80, 30.78) rectangle ( 34.45, 30.78);

\path[fill=fillColor,fill opacity=0.50] ( 34.45, 30.78) rectangle ( 35.11, 30.78);

\path[fill=fillColor,fill opacity=0.50] ( 35.11, 30.78) rectangle ( 35.76, 30.78);

\path[fill=fillColor,fill opacity=0.50] ( 35.76, 30.78) rectangle ( 36.42, 30.78);

\path[fill=fillColor,fill opacity=0.50] ( 36.42, 30.78) rectangle ( 37.07, 30.78);

\path[fill=fillColor,fill opacity=0.50] ( 37.07, 30.78) rectangle ( 37.73, 30.78);

\path[fill=fillColor,fill opacity=0.50] ( 37.73, 30.78) rectangle ( 38.38, 30.78);

\path[fill=fillColor,fill opacity=0.50] ( 38.38, 30.78) rectangle ( 39.04, 30.78);

\path[fill=fillColor,fill opacity=0.50] ( 39.04, 30.78) rectangle ( 39.69, 30.78);

\path[fill=fillColor,fill opacity=0.50] ( 39.69, 30.78) rectangle ( 40.34, 30.78);

\path[fill=fillColor,fill opacity=0.50] ( 40.34, 30.78) rectangle ( 41.00, 30.78);

\path[fill=fillColor,fill opacity=0.50] ( 41.00, 30.78) rectangle ( 41.65, 31.24);

\path[fill=fillColor,fill opacity=0.50] ( 41.65, 30.78) rectangle ( 42.31, 71.63);

\path[fill=fillColor,fill opacity=0.50] ( 42.31, 30.78) rectangle ( 42.96, 85.70);

\path[fill=fillColor,fill opacity=0.50] ( 42.96, 30.78) rectangle ( 43.62, 52.57);

\path[fill=fillColor,fill opacity=0.50] ( 43.62, 30.78) rectangle ( 44.27, 45.31);

\path[fill=fillColor,fill opacity=0.50] ( 44.27, 30.78) rectangle ( 44.93, 40.31);

\path[fill=fillColor,fill opacity=0.50] ( 44.93, 30.78) rectangle ( 45.58, 50.75);

\path[fill=fillColor,fill opacity=0.50] ( 45.58, 30.78) rectangle ( 46.24, 53.93);

\path[fill=fillColor,fill opacity=0.50] ( 46.24, 30.78) rectangle ( 46.89, 53.93);

\path[fill=fillColor,fill opacity=0.50] ( 46.89, 30.78) rectangle ( 47.55, 52.12);

\path[fill=fillColor,fill opacity=0.50] ( 47.55, 30.78) rectangle ( 48.20, 46.22);

\path[fill=fillColor,fill opacity=0.50] ( 48.20, 30.78) rectangle ( 48.86, 38.95);

\path[fill=fillColor,fill opacity=0.50] ( 48.86, 30.78) rectangle ( 49.51, 39.86);

\path[fill=fillColor,fill opacity=0.50] ( 49.51, 30.78) rectangle ( 50.17, 44.40);

\path[fill=fillColor,fill opacity=0.50] ( 50.17, 30.78) rectangle ( 50.82, 39.41);

\path[fill=fillColor,fill opacity=0.50] ( 50.82, 30.78) rectangle ( 51.48, 46.22);

\path[fill=fillColor,fill opacity=0.50] ( 51.48, 30.78) rectangle ( 52.13, 41.22);

\path[fill=fillColor,fill opacity=0.50] ( 52.13, 30.78) rectangle ( 52.79, 38.95);

\path[fill=fillColor,fill opacity=0.50] ( 52.79, 30.78) rectangle ( 53.44, 41.22);

\path[fill=fillColor,fill opacity=0.50] ( 53.44, 30.78) rectangle ( 54.10, 36.23);

\path[fill=fillColor,fill opacity=0.50] ( 54.10, 30.78) rectangle ( 54.75, 38.05);

\path[fill=fillColor,fill opacity=0.50] ( 54.75, 30.78) rectangle ( 55.41, 36.68);

\path[fill=fillColor,fill opacity=0.50] ( 55.41, 30.78) rectangle ( 56.06, 36.68);

\path[fill=fillColor,fill opacity=0.50] ( 56.06, 30.78) rectangle ( 56.72, 35.32);

\path[fill=fillColor,fill opacity=0.50] ( 56.72, 30.78) rectangle ( 57.37, 33.51);

\path[fill=fillColor,fill opacity=0.50] ( 57.37, 30.78) rectangle ( 58.03, 33.51);

\path[fill=fillColor,fill opacity=0.50] ( 58.03, 30.78) rectangle ( 58.68, 36.23);

\path[fill=fillColor,fill opacity=0.50] ( 58.68, 30.78) rectangle ( 59.34, 33.05);

\path[fill=fillColor,fill opacity=0.50] ( 59.34, 30.78) rectangle ( 59.99, 35.32);

\path[fill=fillColor,fill opacity=0.50] ( 59.99, 30.78) rectangle ( 60.64, 35.32);

\path[fill=fillColor,fill opacity=0.50] ( 60.64, 30.78) rectangle ( 61.30, 32.60);

\path[fill=fillColor,fill opacity=0.50] ( 61.30, 30.78) rectangle ( 61.95, 36.23);

\path[fill=fillColor,fill opacity=0.50] ( 61.95, 30.78) rectangle ( 62.61, 34.87);

\path[fill=fillColor,fill opacity=0.50] ( 62.61, 30.78) rectangle ( 63.26, 36.23);

\path[fill=fillColor,fill opacity=0.50] ( 63.26, 30.78) rectangle ( 63.92, 33.51);

\path[fill=fillColor,fill opacity=0.50] ( 63.92, 30.78) rectangle ( 64.57, 32.60);

\path[fill=fillColor,fill opacity=0.50] ( 64.57, 30.78) rectangle ( 65.23, 33.05);

\path[fill=fillColor,fill opacity=0.50] ( 65.23, 30.78) rectangle ( 65.88, 33.96);

\path[fill=fillColor,fill opacity=0.50] ( 65.88, 30.78) rectangle ( 66.54, 34.41);

\path[fill=fillColor,fill opacity=0.50] ( 66.54, 30.78) rectangle ( 67.19, 32.14);

\path[fill=fillColor,fill opacity=0.50] ( 67.19, 30.78) rectangle ( 67.85, 33.96);

\path[fill=fillColor,fill opacity=0.50] ( 67.85, 30.78) rectangle ( 68.50, 33.51);

\path[fill=fillColor,fill opacity=0.50] ( 68.50, 30.78) rectangle ( 69.16, 32.60);

\path[fill=fillColor,fill opacity=0.50] ( 69.16, 30.78) rectangle ( 69.81, 32.60);

\path[fill=fillColor,fill opacity=0.50] ( 69.81, 30.78) rectangle ( 70.47, 32.14);

\path[fill=fillColor,fill opacity=0.50] ( 70.47, 30.78) rectangle ( 71.12, 32.60);

\path[fill=fillColor,fill opacity=0.50] ( 71.12, 30.78) rectangle ( 71.78, 31.69);

\path[fill=fillColor,fill opacity=0.50] ( 71.78, 30.78) rectangle ( 72.43, 31.69);

\path[fill=fillColor,fill opacity=0.50] ( 72.43, 30.78) rectangle ( 73.09, 31.69);

\path[fill=fillColor,fill opacity=0.50] ( 73.09, 30.78) rectangle ( 73.74, 31.69);

\path[fill=fillColor,fill opacity=0.50] ( 73.74, 30.78) rectangle ( 74.40, 30.78);

\path[fill=fillColor,fill opacity=0.50] ( 74.40, 30.78) rectangle ( 75.05, 32.14);

\path[fill=fillColor,fill opacity=0.50] ( 75.05, 30.78) rectangle ( 75.71, 31.69);

\path[fill=fillColor,fill opacity=0.50] ( 75.71, 30.78) rectangle ( 76.36, 32.60);

\path[fill=fillColor,fill opacity=0.50] ( 76.36, 30.78) rectangle ( 77.02, 31.24);

\path[fill=fillColor,fill opacity=0.50] ( 77.02, 30.78) rectangle ( 77.67, 31.69);

\path[fill=fillColor,fill opacity=0.50] ( 77.67, 30.78) rectangle ( 78.33, 31.24);

\path[fill=fillColor,fill opacity=0.50] ( 78.33, 30.78) rectangle ( 78.98, 31.69);

\path[fill=fillColor,fill opacity=0.50] ( 78.98, 30.78) rectangle ( 79.63, 31.24);

\path[fill=fillColor,fill opacity=0.50] ( 79.63, 30.78) rectangle ( 80.29, 32.14);

\path[fill=fillColor,fill opacity=0.50] ( 80.29, 30.78) rectangle ( 80.94, 30.78);

\path[fill=fillColor,fill opacity=0.50] ( 80.94, 30.78) rectangle ( 81.60, 31.69);

\path[fill=fillColor,fill opacity=0.50] ( 81.60, 30.78) rectangle ( 82.25, 31.24);

\path[fill=fillColor,fill opacity=0.50] ( 82.25, 30.78) rectangle ( 82.91, 31.24);

\path[fill=fillColor,fill opacity=0.50] ( 82.91, 30.78) rectangle ( 83.56, 31.24);

\path[fill=fillColor,fill opacity=0.50] ( 83.56, 30.78) rectangle ( 84.22, 31.24);

\path[fill=fillColor,fill opacity=0.50] ( 84.22, 30.78) rectangle ( 84.87, 31.69);

\path[fill=fillColor,fill opacity=0.50] ( 84.87, 30.78) rectangle ( 85.53, 30.78);

\path[fill=fillColor,fill opacity=0.50] ( 85.53, 30.78) rectangle ( 86.18, 31.24);

\path[fill=fillColor,fill opacity=0.50] ( 86.18, 30.78) rectangle ( 86.84, 30.78);

\path[fill=fillColor,fill opacity=0.50] ( 86.84, 30.78) rectangle ( 87.49, 31.24);

\path[fill=fillColor,fill opacity=0.50] ( 87.49, 30.78) rectangle ( 88.15, 31.24);

\path[fill=fillColor,fill opacity=0.50] ( 88.15, 30.78) rectangle ( 88.80, 31.69);

\path[fill=fillColor,fill opacity=0.50] ( 88.80, 30.78) rectangle ( 89.46, 30.78);

\path[fill=fillColor,fill opacity=0.50] ( 89.46, 30.78) rectangle ( 90.11, 31.24);

\path[fill=fillColor,fill opacity=0.50] ( 90.11, 30.78) rectangle ( 90.77, 30.78);

\path[fill=fillColor,fill opacity=0.50] ( 90.77, 30.78) rectangle ( 91.42, 32.60);

\path[fill=fillColor,fill opacity=0.50] ( 91.42, 30.78) rectangle ( 92.08, 30.78);
\end{scope}
\begin{scope}
\path[clip] (  0.00,  0.00) rectangle (101.18, 93.95);
\definecolor{drawColor}{gray}{0.30}

\node[text=drawColor,anchor=base east,inner sep=0pt, outer sep=0pt, scale=  0.88] at ( 18.70, 27.75) {0};

\node[text=drawColor,anchor=base east,inner sep=0pt, outer sep=0pt, scale=  0.88] at ( 18.70, 39.10) {25};

\node[text=drawColor,anchor=base east,inner sep=0pt, outer sep=0pt, scale=  0.88] at ( 18.70, 50.45) {50};

\node[text=drawColor,anchor=base east,inner sep=0pt, outer sep=0pt, scale=  0.88] at ( 18.70, 61.79) {75};

\node[text=drawColor,anchor=base east,inner sep=0pt, outer sep=0pt, scale=  0.88] at ( 18.70, 73.14) {100};

\node[text=drawColor,anchor=base east,inner sep=0pt, outer sep=0pt, scale=  0.88] at ( 18.70, 84.49) {125};
\end{scope}
\begin{scope}
\path[clip] (  0.00,  0.00) rectangle (101.18, 93.95);
\definecolor{drawColor}{gray}{0.30}

\node[text=drawColor,anchor=base,inner sep=0pt, outer sep=0pt, scale=  0.88] at ( 26.92, 17.03) {0};

\node[text=drawColor,anchor=base,inner sep=0pt, outer sep=0pt, scale=  0.88] at ( 43.29, 17.03) {25};

\node[text=drawColor,anchor=base,inner sep=0pt, outer sep=0pt, scale=  0.88] at ( 59.66, 17.03) {50};

\node[text=drawColor,anchor=base,inner sep=0pt, outer sep=0pt, scale=  0.88] at ( 76.03, 17.03) {75};

\node[text=drawColor,anchor=base,inner sep=0pt, outer sep=0pt, scale=  0.88] at ( 92.40, 17.03) {100};
\end{scope}
\begin{scope}
\path[clip] (  0.00,  0.00) rectangle (101.18, 93.95);
\definecolor{drawColor}{RGB}{0,0,0}

\node[text=drawColor,anchor=base,inner sep=0pt, outer sep=0pt, scale=  0.80] at ( 59.66,  7.06) {Response time [ms]};
\end{scope}
\end{tikzpicture}

%% file: benchmarks/25-01-14_AWS_plots/4_nodes_3_zones/1_AWS_4_nodes.tex
% Created by tikzDevice version 0.12.6 on 2025-01-16 10:10:08
% !TEX encoding = UTF-8 Unicode
\begin{tikzpicture}[x=1pt,y=1pt]
\definecolor{fillColor}{RGB}{255,255,255}
\path[use as bounding box,fill=fillColor,fill opacity=0.00] (0,0) rectangle (101.18, 93.95);
\begin{scope}
\path[clip] ( 23.65, 28.04) rectangle ( 95.68, 88.45);
\definecolor{drawColor}{gray}{0.92}

\path[draw=drawColor,line width= 0.3pt,line join=round] ( 23.65, 41.51) --
	( 95.68, 41.51);

\path[draw=drawColor,line width= 0.3pt,line join=round] ( 23.65, 62.96) --
	( 95.68, 62.96);

\path[draw=drawColor,line width= 0.3pt,line join=round] ( 23.65, 84.42) --
	( 95.68, 84.42);

\path[draw=drawColor,line width= 0.3pt,line join=round] ( 35.11, 28.04) --
	( 35.11, 88.45);

\path[draw=drawColor,line width= 0.3pt,line join=round] ( 51.48, 28.04) --
	( 51.48, 88.45);

\path[draw=drawColor,line width= 0.3pt,line join=round] ( 67.85, 28.04) --
	( 67.85, 88.45);

\path[draw=drawColor,line width= 0.3pt,line join=round] ( 84.22, 28.04) --
	( 84.22, 88.45);

\path[draw=drawColor,line width= 0.6pt,line join=round] ( 23.65, 30.78) --
	( 95.68, 30.78);

\path[draw=drawColor,line width= 0.6pt,line join=round] ( 23.65, 52.24) --
	( 95.68, 52.24);

\path[draw=drawColor,line width= 0.6pt,line join=round] ( 23.65, 73.69) --
	( 95.68, 73.69);

\path[draw=drawColor,line width= 0.6pt,line join=round] ( 26.92, 28.04) --
	( 26.92, 88.45);

\path[draw=drawColor,line width= 0.6pt,line join=round] ( 43.29, 28.04) --
	( 43.29, 88.45);

\path[draw=drawColor,line width= 0.6pt,line join=round] ( 59.66, 28.04) --
	( 59.66, 88.45);

\path[draw=drawColor,line width= 0.6pt,line join=round] ( 76.03, 28.04) --
	( 76.03, 88.45);

\path[draw=drawColor,line width= 0.6pt,line join=round] ( 92.40, 28.04) --
	( 92.40, 88.45);
\definecolor{fillColor}{RGB}{0,191,196}

\path[fill=fillColor,fill opacity=0.50] ( 27.25, 30.78) rectangle ( 27.90, 30.78);

\path[fill=fillColor,fill opacity=0.50] ( 27.90, 30.78) rectangle ( 28.56, 30.78);

\path[fill=fillColor,fill opacity=0.50] ( 28.56, 30.78) rectangle ( 29.21, 30.78);

\path[fill=fillColor,fill opacity=0.50] ( 29.21, 30.78) rectangle ( 29.87, 30.78);

\path[fill=fillColor,fill opacity=0.50] ( 29.87, 30.78) rectangle ( 30.52, 30.78);

\path[fill=fillColor,fill opacity=0.50] ( 30.52, 30.78) rectangle ( 31.18, 30.78);

\path[fill=fillColor,fill opacity=0.50] ( 31.18, 30.78) rectangle ( 31.83, 30.78);

\path[fill=fillColor,fill opacity=0.50] ( 31.83, 30.78) rectangle ( 32.49, 30.78);

\path[fill=fillColor,fill opacity=0.50] ( 32.49, 30.78) rectangle ( 33.14, 30.78);

\path[fill=fillColor,fill opacity=0.50] ( 33.14, 30.78) rectangle ( 33.80, 30.78);

\path[fill=fillColor,fill opacity=0.50] ( 33.80, 30.78) rectangle ( 34.45, 30.78);

\path[fill=fillColor,fill opacity=0.50] ( 34.45, 30.78) rectangle ( 35.11, 30.78);

\path[fill=fillColor,fill opacity=0.50] ( 35.11, 30.78) rectangle ( 35.76, 30.78);

\path[fill=fillColor,fill opacity=0.50] ( 35.76, 30.78) rectangle ( 36.42, 30.78);

\path[fill=fillColor,fill opacity=0.50] ( 36.42, 30.78) rectangle ( 37.07, 30.78);

\path[fill=fillColor,fill opacity=0.50] ( 37.07, 30.78) rectangle ( 37.73, 30.78);

\path[fill=fillColor,fill opacity=0.50] ( 37.73, 30.78) rectangle ( 38.38, 30.78);

\path[fill=fillColor,fill opacity=0.50] ( 38.38, 30.78) rectangle ( 39.04, 30.78);

\path[fill=fillColor,fill opacity=0.50] ( 39.04, 30.78) rectangle ( 39.69, 30.78);

\path[fill=fillColor,fill opacity=0.50] ( 39.69, 30.78) rectangle ( 40.34, 30.78);

\path[fill=fillColor,fill opacity=0.50] ( 40.34, 30.78) rectangle ( 41.00, 30.78);

\path[fill=fillColor,fill opacity=0.50] ( 41.00, 30.78) rectangle ( 41.65, 30.78);

\path[fill=fillColor,fill opacity=0.50] ( 41.65, 30.78) rectangle ( 42.31, 30.78);

\path[fill=fillColor,fill opacity=0.50] ( 42.31, 30.78) rectangle ( 42.96, 30.78);

\path[fill=fillColor,fill opacity=0.50] ( 42.96, 30.78) rectangle ( 43.62, 30.78);

\path[fill=fillColor,fill opacity=0.50] ( 43.62, 30.78) rectangle ( 44.27, 30.78);

\path[fill=fillColor,fill opacity=0.50] ( 44.27, 30.78) rectangle ( 44.93, 30.78);

\path[fill=fillColor,fill opacity=0.50] ( 44.93, 30.78) rectangle ( 45.58, 30.78);

\path[fill=fillColor,fill opacity=0.50] ( 45.58, 30.78) rectangle ( 46.24, 30.78);

\path[fill=fillColor,fill opacity=0.50] ( 46.24, 30.78) rectangle ( 46.89, 30.78);

\path[fill=fillColor,fill opacity=0.50] ( 46.89, 30.78) rectangle ( 47.55, 30.78);

\path[fill=fillColor,fill opacity=0.50] ( 47.55, 30.78) rectangle ( 48.20, 30.78);

\path[fill=fillColor,fill opacity=0.50] ( 48.20, 30.78) rectangle ( 48.86, 30.78);

\path[fill=fillColor,fill opacity=0.50] ( 48.86, 30.78) rectangle ( 49.51, 30.78);

\path[fill=fillColor,fill opacity=0.50] ( 49.51, 30.78) rectangle ( 50.17, 30.78);

\path[fill=fillColor,fill opacity=0.50] ( 50.17, 30.78) rectangle ( 50.82, 30.78);

\path[fill=fillColor,fill opacity=0.50] ( 50.82, 30.78) rectangle ( 51.48, 30.78);

\path[fill=fillColor,fill opacity=0.50] ( 51.48, 30.78) rectangle ( 52.13, 33.79);

\path[fill=fillColor,fill opacity=0.50] ( 52.13, 30.78) rectangle ( 52.79, 49.66);

\path[fill=fillColor,fill opacity=0.50] ( 52.79, 30.78) rectangle ( 53.44, 65.54);

\path[fill=fillColor,fill opacity=0.50] ( 53.44, 30.78) rectangle ( 54.10, 72.40);

\path[fill=fillColor,fill opacity=0.50] ( 54.10, 30.78) rectangle ( 54.75, 77.55);

\path[fill=fillColor,fill opacity=0.50] ( 54.75, 30.78) rectangle ( 55.41, 67.25);

\path[fill=fillColor,fill opacity=0.50] ( 55.41, 30.78) rectangle ( 56.06, 70.26);

\path[fill=fillColor,fill opacity=0.50] ( 56.06, 30.78) rectangle ( 56.72, 60.39);

\path[fill=fillColor,fill opacity=0.50] ( 56.72, 30.78) rectangle ( 57.37, 52.67);

\path[fill=fillColor,fill opacity=0.50] ( 57.37, 30.78) rectangle ( 58.03, 53.09);

\path[fill=fillColor,fill opacity=0.50] ( 58.03, 30.78) rectangle ( 58.68, 46.23);

\path[fill=fillColor,fill opacity=0.50] ( 58.68, 30.78) rectangle ( 59.34, 49.23);

\path[fill=fillColor,fill opacity=0.50] ( 59.34, 30.78) rectangle ( 59.99, 48.80);

\path[fill=fillColor,fill opacity=0.50] ( 59.99, 30.78) rectangle ( 60.64, 42.37);

\path[fill=fillColor,fill opacity=0.50] ( 60.64, 30.78) rectangle ( 61.30, 41.94);

\path[fill=fillColor,fill opacity=0.50] ( 61.30, 30.78) rectangle ( 61.95, 40.22);

\path[fill=fillColor,fill opacity=0.50] ( 61.95, 30.78) rectangle ( 62.61, 37.22);

\path[fill=fillColor,fill opacity=0.50] ( 62.61, 30.78) rectangle ( 63.26, 35.93);

\path[fill=fillColor,fill opacity=0.50] ( 63.26, 30.78) rectangle ( 63.92, 35.50);

\path[fill=fillColor,fill opacity=0.50] ( 63.92, 30.78) rectangle ( 64.57, 34.64);

\path[fill=fillColor,fill opacity=0.50] ( 64.57, 30.78) rectangle ( 65.23, 35.07);

\path[fill=fillColor,fill opacity=0.50] ( 65.23, 30.78) rectangle ( 65.88, 34.64);

\path[fill=fillColor,fill opacity=0.50] ( 65.88, 30.78) rectangle ( 66.54, 31.64);

\path[fill=fillColor,fill opacity=0.50] ( 66.54, 30.78) rectangle ( 67.19, 32.50);

\path[fill=fillColor,fill opacity=0.50] ( 67.19, 30.78) rectangle ( 67.85, 34.22);

\path[fill=fillColor,fill opacity=0.50] ( 67.85, 30.78) rectangle ( 68.50, 32.50);

\path[fill=fillColor,fill opacity=0.50] ( 68.50, 30.78) rectangle ( 69.16, 31.64);

\path[fill=fillColor,fill opacity=0.50] ( 69.16, 30.78) rectangle ( 69.81, 31.64);

\path[fill=fillColor,fill opacity=0.50] ( 69.81, 30.78) rectangle ( 70.47, 31.21);

\path[fill=fillColor,fill opacity=0.50] ( 70.47, 30.78) rectangle ( 71.12, 31.64);

\path[fill=fillColor,fill opacity=0.50] ( 71.12, 30.78) rectangle ( 71.78, 32.07);

\path[fill=fillColor,fill opacity=0.50] ( 71.78, 30.78) rectangle ( 72.43, 32.07);

\path[fill=fillColor,fill opacity=0.50] ( 72.43, 30.78) rectangle ( 73.09, 31.64);

\path[fill=fillColor,fill opacity=0.50] ( 73.09, 30.78) rectangle ( 73.74, 31.21);

\path[fill=fillColor,fill opacity=0.50] ( 73.74, 30.78) rectangle ( 74.40, 31.64);

\path[fill=fillColor,fill opacity=0.50] ( 74.40, 30.78) rectangle ( 75.05, 32.07);

\path[fill=fillColor,fill opacity=0.50] ( 75.05, 30.78) rectangle ( 75.71, 31.21);

\path[fill=fillColor,fill opacity=0.50] ( 75.71, 30.78) rectangle ( 76.36, 30.78);

\path[fill=fillColor,fill opacity=0.50] ( 76.36, 30.78) rectangle ( 77.02, 30.78);

\path[fill=fillColor,fill opacity=0.50] ( 77.02, 30.78) rectangle ( 77.67, 30.78);

\path[fill=fillColor,fill opacity=0.50] ( 77.67, 30.78) rectangle ( 78.33, 31.21);

\path[fill=fillColor,fill opacity=0.50] ( 78.33, 30.78) rectangle ( 78.98, 30.78);

\path[fill=fillColor,fill opacity=0.50] ( 78.98, 30.78) rectangle ( 79.63, 30.78);

\path[fill=fillColor,fill opacity=0.50] ( 79.63, 30.78) rectangle ( 80.29, 30.78);

\path[fill=fillColor,fill opacity=0.50] ( 80.29, 30.78) rectangle ( 80.94, 31.21);

\path[fill=fillColor,fill opacity=0.50] ( 80.94, 30.78) rectangle ( 81.60, 30.78);

\path[fill=fillColor,fill opacity=0.50] ( 81.60, 30.78) rectangle ( 82.25, 31.64);

\path[fill=fillColor,fill opacity=0.50] ( 82.25, 30.78) rectangle ( 82.91, 30.78);

\path[fill=fillColor,fill opacity=0.50] ( 82.91, 30.78) rectangle ( 83.56, 30.78);

\path[fill=fillColor,fill opacity=0.50] ( 83.56, 30.78) rectangle ( 84.22, 31.21);

\path[fill=fillColor,fill opacity=0.50] ( 84.22, 30.78) rectangle ( 84.87, 30.78);

\path[fill=fillColor,fill opacity=0.50] ( 84.87, 30.78) rectangle ( 85.53, 30.78);

\path[fill=fillColor,fill opacity=0.50] ( 85.53, 30.78) rectangle ( 86.18, 30.78);

\path[fill=fillColor,fill opacity=0.50] ( 86.18, 30.78) rectangle ( 86.84, 30.78);

\path[fill=fillColor,fill opacity=0.50] ( 86.84, 30.78) rectangle ( 87.49, 30.78);

\path[fill=fillColor,fill opacity=0.50] ( 87.49, 30.78) rectangle ( 88.15, 30.78);

\path[fill=fillColor,fill opacity=0.50] ( 88.15, 30.78) rectangle ( 88.80, 30.78);

\path[fill=fillColor,fill opacity=0.50] ( 88.80, 30.78) rectangle ( 89.46, 30.78);

\path[fill=fillColor,fill opacity=0.50] ( 89.46, 30.78) rectangle ( 90.11, 30.78);

\path[fill=fillColor,fill opacity=0.50] ( 90.11, 30.78) rectangle ( 90.77, 30.78);

\path[fill=fillColor,fill opacity=0.50] ( 90.77, 30.78) rectangle ( 91.42, 30.78);

\path[fill=fillColor,fill opacity=0.50] ( 91.42, 30.78) rectangle ( 92.08, 30.78);
\definecolor{fillColor}{RGB}{248,118,109}

\path[fill=fillColor,fill opacity=0.50] ( 27.25, 30.78) rectangle ( 27.90, 30.78);

\path[fill=fillColor,fill opacity=0.50] ( 27.90, 30.78) rectangle ( 28.56, 30.78);

\path[fill=fillColor,fill opacity=0.50] ( 28.56, 30.78) rectangle ( 29.21, 30.78);

\path[fill=fillColor,fill opacity=0.50] ( 29.21, 30.78) rectangle ( 29.87, 30.78);

\path[fill=fillColor,fill opacity=0.50] ( 29.87, 30.78) rectangle ( 30.52, 30.78);

\path[fill=fillColor,fill opacity=0.50] ( 30.52, 30.78) rectangle ( 31.18, 30.78);

\path[fill=fillColor,fill opacity=0.50] ( 31.18, 30.78) rectangle ( 31.83, 30.78);

\path[fill=fillColor,fill opacity=0.50] ( 31.83, 30.78) rectangle ( 32.49, 30.78);

\path[fill=fillColor,fill opacity=0.50] ( 32.49, 30.78) rectangle ( 33.14, 30.78);

\path[fill=fillColor,fill opacity=0.50] ( 33.14, 30.78) rectangle ( 33.80, 30.78);

\path[fill=fillColor,fill opacity=0.50] ( 33.80, 30.78) rectangle ( 34.45, 30.78);

\path[fill=fillColor,fill opacity=0.50] ( 34.45, 30.78) rectangle ( 35.11, 30.78);

\path[fill=fillColor,fill opacity=0.50] ( 35.11, 30.78) rectangle ( 35.76, 30.78);

\path[fill=fillColor,fill opacity=0.50] ( 35.76, 30.78) rectangle ( 36.42, 30.78);

\path[fill=fillColor,fill opacity=0.50] ( 36.42, 30.78) rectangle ( 37.07, 30.78);

\path[fill=fillColor,fill opacity=0.50] ( 37.07, 30.78) rectangle ( 37.73, 30.78);

\path[fill=fillColor,fill opacity=0.50] ( 37.73, 30.78) rectangle ( 38.38, 30.78);

\path[fill=fillColor,fill opacity=0.50] ( 38.38, 30.78) rectangle ( 39.04, 30.78);

\path[fill=fillColor,fill opacity=0.50] ( 39.04, 30.78) rectangle ( 39.69, 30.78);

\path[fill=fillColor,fill opacity=0.50] ( 39.69, 30.78) rectangle ( 40.34, 30.78);

\path[fill=fillColor,fill opacity=0.50] ( 40.34, 30.78) rectangle ( 41.00, 30.78);

\path[fill=fillColor,fill opacity=0.50] ( 41.00, 30.78) rectangle ( 41.65, 30.78);

\path[fill=fillColor,fill opacity=0.50] ( 41.65, 30.78) rectangle ( 42.31, 30.78);

\path[fill=fillColor,fill opacity=0.50] ( 42.31, 30.78) rectangle ( 42.96, 30.78);

\path[fill=fillColor,fill opacity=0.50] ( 42.96, 30.78) rectangle ( 43.62, 30.78);

\path[fill=fillColor,fill opacity=0.50] ( 43.62, 30.78) rectangle ( 44.27, 30.78);

\path[fill=fillColor,fill opacity=0.50] ( 44.27, 30.78) rectangle ( 44.93, 30.78);

\path[fill=fillColor,fill opacity=0.50] ( 44.93, 30.78) rectangle ( 45.58, 30.78);

\path[fill=fillColor,fill opacity=0.50] ( 45.58, 30.78) rectangle ( 46.24, 30.78);

\path[fill=fillColor,fill opacity=0.50] ( 46.24, 30.78) rectangle ( 46.89, 30.78);

\path[fill=fillColor,fill opacity=0.50] ( 46.89, 30.78) rectangle ( 47.55, 30.78);

\path[fill=fillColor,fill opacity=0.50] ( 47.55, 30.78) rectangle ( 48.20, 30.78);

\path[fill=fillColor,fill opacity=0.50] ( 48.20, 30.78) rectangle ( 48.86, 30.78);

\path[fill=fillColor,fill opacity=0.50] ( 48.86, 30.78) rectangle ( 49.51, 30.78);

\path[fill=fillColor,fill opacity=0.50] ( 49.51, 30.78) rectangle ( 50.17, 30.78);

\path[fill=fillColor,fill opacity=0.50] ( 50.17, 30.78) rectangle ( 50.82, 30.78);

\path[fill=fillColor,fill opacity=0.50] ( 50.82, 30.78) rectangle ( 51.48, 30.78);

\path[fill=fillColor,fill opacity=0.50] ( 51.48, 30.78) rectangle ( 52.13, 30.78);

\path[fill=fillColor,fill opacity=0.50] ( 52.13, 30.78) rectangle ( 52.79, 30.78);

\path[fill=fillColor,fill opacity=0.50] ( 52.79, 30.78) rectangle ( 53.44, 30.78);

\path[fill=fillColor,fill opacity=0.50] ( 53.44, 30.78) rectangle ( 54.10, 30.78);

\path[fill=fillColor,fill opacity=0.50] ( 54.10, 30.78) rectangle ( 54.75, 30.78);

\path[fill=fillColor,fill opacity=0.50] ( 54.75, 30.78) rectangle ( 55.41, 30.78);

\path[fill=fillColor,fill opacity=0.50] ( 55.41, 30.78) rectangle ( 56.06, 30.78);

\path[fill=fillColor,fill opacity=0.50] ( 56.06, 30.78) rectangle ( 56.72, 30.78);

\path[fill=fillColor,fill opacity=0.50] ( 56.72, 30.78) rectangle ( 57.37, 30.78);

\path[fill=fillColor,fill opacity=0.50] ( 57.37, 30.78) rectangle ( 58.03, 30.78);

\path[fill=fillColor,fill opacity=0.50] ( 58.03, 30.78) rectangle ( 58.68, 30.78);

\path[fill=fillColor,fill opacity=0.50] ( 58.68, 30.78) rectangle ( 59.34, 30.78);

\path[fill=fillColor,fill opacity=0.50] ( 59.34, 30.78) rectangle ( 59.99, 30.78);

\path[fill=fillColor,fill opacity=0.50] ( 59.99, 30.78) rectangle ( 60.64, 30.78);

\path[fill=fillColor,fill opacity=0.50] ( 60.64, 30.78) rectangle ( 61.30, 30.78);

\path[fill=fillColor,fill opacity=0.50] ( 61.30, 30.78) rectangle ( 61.95, 30.78);

\path[fill=fillColor,fill opacity=0.50] ( 61.95, 30.78) rectangle ( 62.61, 30.78);

\path[fill=fillColor,fill opacity=0.50] ( 62.61, 30.78) rectangle ( 63.26, 30.78);

\path[fill=fillColor,fill opacity=0.50] ( 63.26, 30.78) rectangle ( 63.92, 30.78);

\path[fill=fillColor,fill opacity=0.50] ( 63.92, 30.78) rectangle ( 64.57, 32.07);

\path[fill=fillColor,fill opacity=0.50] ( 64.57, 30.78) rectangle ( 65.23, 41.08);

\path[fill=fillColor,fill opacity=0.50] ( 65.23, 30.78) rectangle ( 65.88, 61.25);

\path[fill=fillColor,fill opacity=0.50] ( 65.88, 30.78) rectangle ( 66.54, 79.70);

\path[fill=fillColor,fill opacity=0.50] ( 66.54, 30.78) rectangle ( 67.19, 80.13);

\path[fill=fillColor,fill opacity=0.50] ( 67.19, 30.78) rectangle ( 67.85, 85.70);

\path[fill=fillColor,fill opacity=0.50] ( 67.85, 30.78) rectangle ( 68.50, 71.55);

\path[fill=fillColor,fill opacity=0.50] ( 68.50, 30.78) rectangle ( 69.16, 71.97);

\path[fill=fillColor,fill opacity=0.50] ( 69.16, 30.78) rectangle ( 69.81, 66.40);

\path[fill=fillColor,fill opacity=0.50] ( 69.81, 30.78) rectangle ( 70.47, 57.39);

\path[fill=fillColor,fill opacity=0.50] ( 70.47, 30.78) rectangle ( 71.12, 48.80);

\path[fill=fillColor,fill opacity=0.50] ( 71.12, 30.78) rectangle ( 71.78, 47.09);

\path[fill=fillColor,fill opacity=0.50] ( 71.78, 30.78) rectangle ( 72.43, 41.08);

\path[fill=fillColor,fill opacity=0.50] ( 72.43, 30.78) rectangle ( 73.09, 37.22);

\path[fill=fillColor,fill opacity=0.50] ( 73.09, 30.78) rectangle ( 73.74, 36.36);

\path[fill=fillColor,fill opacity=0.50] ( 73.74, 30.78) rectangle ( 74.40, 33.36);

\path[fill=fillColor,fill opacity=0.50] ( 74.40, 30.78) rectangle ( 75.05, 33.79);

\path[fill=fillColor,fill opacity=0.50] ( 75.05, 30.78) rectangle ( 75.71, 35.07);

\path[fill=fillColor,fill opacity=0.50] ( 75.71, 30.78) rectangle ( 76.36, 34.22);

\path[fill=fillColor,fill opacity=0.50] ( 76.36, 30.78) rectangle ( 77.02, 32.93);

\path[fill=fillColor,fill opacity=0.50] ( 77.02, 30.78) rectangle ( 77.67, 32.07);

\path[fill=fillColor,fill opacity=0.50] ( 77.67, 30.78) rectangle ( 78.33, 32.50);

\path[fill=fillColor,fill opacity=0.50] ( 78.33, 30.78) rectangle ( 78.98, 31.64);

\path[fill=fillColor,fill opacity=0.50] ( 78.98, 30.78) rectangle ( 79.63, 31.21);

\path[fill=fillColor,fill opacity=0.50] ( 79.63, 30.78) rectangle ( 80.29, 31.64);

\path[fill=fillColor,fill opacity=0.50] ( 80.29, 30.78) rectangle ( 80.94, 32.07);

\path[fill=fillColor,fill opacity=0.50] ( 80.94, 30.78) rectangle ( 81.60, 30.78);

\path[fill=fillColor,fill opacity=0.50] ( 81.60, 30.78) rectangle ( 82.25, 31.64);

\path[fill=fillColor,fill opacity=0.50] ( 82.25, 30.78) rectangle ( 82.91, 32.50);

\path[fill=fillColor,fill opacity=0.50] ( 82.91, 30.78) rectangle ( 83.56, 31.64);

\path[fill=fillColor,fill opacity=0.50] ( 83.56, 30.78) rectangle ( 84.22, 31.64);

\path[fill=fillColor,fill opacity=0.50] ( 84.22, 30.78) rectangle ( 84.87, 31.21);

\path[fill=fillColor,fill opacity=0.50] ( 84.87, 30.78) rectangle ( 85.53, 30.78);

\path[fill=fillColor,fill opacity=0.50] ( 85.53, 30.78) rectangle ( 86.18, 31.21);

\path[fill=fillColor,fill opacity=0.50] ( 86.18, 30.78) rectangle ( 86.84, 31.64);

\path[fill=fillColor,fill opacity=0.50] ( 86.84, 30.78) rectangle ( 87.49, 31.21);

\path[fill=fillColor,fill opacity=0.50] ( 87.49, 30.78) rectangle ( 88.15, 30.78);

\path[fill=fillColor,fill opacity=0.50] ( 88.15, 30.78) rectangle ( 88.80, 31.64);

\path[fill=fillColor,fill opacity=0.50] ( 88.80, 30.78) rectangle ( 89.46, 30.78);

\path[fill=fillColor,fill opacity=0.50] ( 89.46, 30.78) rectangle ( 90.11, 30.78);

\path[fill=fillColor,fill opacity=0.50] ( 90.11, 30.78) rectangle ( 90.77, 30.78);

\path[fill=fillColor,fill opacity=0.50] ( 90.77, 30.78) rectangle ( 91.42, 30.78);

\path[fill=fillColor,fill opacity=0.50] ( 91.42, 30.78) rectangle ( 92.08, 31.21);
\end{scope}
\begin{scope}
\path[clip] (  0.00,  0.00) rectangle (101.18, 93.95);
\definecolor{drawColor}{gray}{0.30}

\node[text=drawColor,anchor=base east,inner sep=0pt, outer sep=0pt, scale=  0.88] at ( 18.70, 27.75) {0};

\node[text=drawColor,anchor=base east,inner sep=0pt, outer sep=0pt, scale=  0.88] at ( 18.70, 49.21) {50};

\node[text=drawColor,anchor=base east,inner sep=0pt, outer sep=0pt, scale=  0.88] at ( 18.70, 70.66) {100};
\end{scope}
\begin{scope}
\path[clip] (  0.00,  0.00) rectangle (101.18, 93.95);
\definecolor{drawColor}{gray}{0.30}

\node[text=drawColor,anchor=base,inner sep=0pt, outer sep=0pt, scale=  0.88] at ( 26.92, 17.03) {0};

\node[text=drawColor,anchor=base,inner sep=0pt, outer sep=0pt, scale=  0.88] at ( 43.29, 17.03) {25};

\node[text=drawColor,anchor=base,inner sep=0pt, outer sep=0pt, scale=  0.88] at ( 59.66, 17.03) {50};

\node[text=drawColor,anchor=base,inner sep=0pt, outer sep=0pt, scale=  0.88] at ( 76.03, 17.03) {75};

\node[text=drawColor,anchor=base,inner sep=0pt, outer sep=0pt, scale=  0.88] at ( 92.40, 17.03) {100};
\end{scope}
\begin{scope}
\path[clip] (  0.00,  0.00) rectangle (101.18, 93.95);
\definecolor{drawColor}{RGB}{0,0,0}

\node[text=drawColor,anchor=base,inner sep=0pt, outer sep=0pt, scale=  0.80] at ( 59.66,  7.06) {Response time [ms]};
\end{scope}
\end{tikzpicture}

%% file: benchmarks/25-01-14_AWS_plots/2_nodes_1_zone_fcom_orchestrator_separate_zone/2_AWS_2_nodes_1_zone.tex
% Created by tikzDevice version 0.12.6 on 2025-01-16 09:46:37
% !TEX encoding = UTF-8 Unicode
\begin{tikzpicture}[x=1pt,y=1pt]
\definecolor{fillColor}{RGB}{255,255,255}
\path[use as bounding box,fill=fillColor,fill opacity=0.00] (0,0) rectangle (101.18, 93.95);
\begin{scope}
\path[clip] ( 19.25, 28.04) rectangle ( 95.68, 88.45);
\definecolor{drawColor}{gray}{0.92}

\path[draw=drawColor,line width= 0.3pt,line join=round] ( 19.25, 38.24) --
	( 95.68, 38.24);

\path[draw=drawColor,line width= 0.3pt,line join=round] ( 19.25, 53.17) --
	( 95.68, 53.17);

\path[draw=drawColor,line width= 0.3pt,line join=round] ( 19.25, 68.09) --
	( 95.68, 68.09);

\path[draw=drawColor,line width= 0.3pt,line join=round] ( 19.25, 83.02) --
	( 95.68, 83.02);

\path[draw=drawColor,line width= 0.3pt,line join=round] ( 31.41, 28.04) --
	( 31.41, 88.45);

\path[draw=drawColor,line width= 0.3pt,line join=round] ( 48.78, 28.04) --
	( 48.78, 88.45);

\path[draw=drawColor,line width= 0.3pt,line join=round] ( 66.15, 28.04) --
	( 66.15, 88.45);

\path[draw=drawColor,line width= 0.3pt,line join=round] ( 83.52, 28.04) --
	( 83.52, 88.45);

\path[draw=drawColor,line width= 0.6pt,line join=round] ( 19.25, 30.78) --
	( 95.68, 30.78);

\path[draw=drawColor,line width= 0.6pt,line join=round] ( 19.25, 45.71) --
	( 95.68, 45.71);

\path[draw=drawColor,line width= 0.6pt,line join=round] ( 19.25, 60.63) --
	( 95.68, 60.63);

\path[draw=drawColor,line width= 0.6pt,line join=round] ( 19.25, 75.56) --
	( 95.68, 75.56);

\path[draw=drawColor,line width= 0.6pt,line join=round] ( 22.72, 28.04) --
	( 22.72, 88.45);

\path[draw=drawColor,line width= 0.6pt,line join=round] ( 40.09, 28.04) --
	( 40.09, 88.45);

\path[draw=drawColor,line width= 0.6pt,line join=round] ( 57.46, 28.04) --
	( 57.46, 88.45);

\path[draw=drawColor,line width= 0.6pt,line join=round] ( 74.83, 28.04) --
	( 74.83, 88.45);

\path[draw=drawColor,line width= 0.6pt,line join=round] ( 92.20, 28.04) --
	( 92.20, 88.45);
\definecolor{fillColor}{RGB}{0,191,196}

\path[fill=fillColor,fill opacity=0.50] ( 23.07, 30.78) rectangle ( 23.76, 30.78);

\path[fill=fillColor,fill opacity=0.50] ( 23.76, 30.78) rectangle ( 24.46, 30.78);

\path[fill=fillColor,fill opacity=0.50] ( 24.46, 30.78) rectangle ( 25.15, 30.78);

\path[fill=fillColor,fill opacity=0.50] ( 25.15, 30.78) rectangle ( 25.85, 30.78);

\path[fill=fillColor,fill opacity=0.50] ( 25.85, 30.78) rectangle ( 26.54, 30.78);

\path[fill=fillColor,fill opacity=0.50] ( 26.54, 30.78) rectangle ( 27.24, 30.78);

\path[fill=fillColor,fill opacity=0.50] ( 27.24, 30.78) rectangle ( 27.93, 30.78);

\path[fill=fillColor,fill opacity=0.50] ( 27.93, 30.78) rectangle ( 28.63, 30.78);

\path[fill=fillColor,fill opacity=0.50] ( 28.63, 30.78) rectangle ( 29.32, 30.78);

\path[fill=fillColor,fill opacity=0.50] ( 29.32, 30.78) rectangle ( 30.02, 30.78);

\path[fill=fillColor,fill opacity=0.50] ( 30.02, 30.78) rectangle ( 30.71, 30.78);

\path[fill=fillColor,fill opacity=0.50] ( 30.71, 30.78) rectangle ( 31.41, 30.78);

\path[fill=fillColor,fill opacity=0.50] ( 31.41, 30.78) rectangle ( 32.10, 30.78);

\path[fill=fillColor,fill opacity=0.50] ( 32.10, 30.78) rectangle ( 32.80, 30.78);

\path[fill=fillColor,fill opacity=0.50] ( 32.80, 30.78) rectangle ( 33.49, 30.78);

\path[fill=fillColor,fill opacity=0.50] ( 33.49, 30.78) rectangle ( 34.19, 30.78);

\path[fill=fillColor,fill opacity=0.50] ( 34.19, 30.78) rectangle ( 34.88, 30.78);

\path[fill=fillColor,fill opacity=0.50] ( 34.88, 30.78) rectangle ( 35.58, 30.78);

\path[fill=fillColor,fill opacity=0.50] ( 35.58, 30.78) rectangle ( 36.27, 30.78);

\path[fill=fillColor,fill opacity=0.50] ( 36.27, 30.78) rectangle ( 36.97, 30.78);

\path[fill=fillColor,fill opacity=0.50] ( 36.97, 30.78) rectangle ( 37.66, 30.78);

\path[fill=fillColor,fill opacity=0.50] ( 37.66, 30.78) rectangle ( 38.36, 30.78);

\path[fill=fillColor,fill opacity=0.50] ( 38.36, 30.78) rectangle ( 39.05, 30.78);

\path[fill=fillColor,fill opacity=0.50] ( 39.05, 30.78) rectangle ( 39.75, 30.78);

\path[fill=fillColor,fill opacity=0.50] ( 39.75, 30.78) rectangle ( 40.44, 30.78);

\path[fill=fillColor,fill opacity=0.50] ( 40.44, 30.78) rectangle ( 41.13, 31.38);

\path[fill=fillColor,fill opacity=0.50] ( 41.13, 30.78) rectangle ( 41.83, 46.90);

\path[fill=fillColor,fill opacity=0.50] ( 41.83, 30.78) rectangle ( 42.52, 85.70);

\path[fill=fillColor,fill opacity=0.50] ( 42.52, 30.78) rectangle ( 43.22, 71.38);

\path[fill=fillColor,fill opacity=0.50] ( 43.22, 30.78) rectangle ( 43.91, 53.47);

\path[fill=fillColor,fill opacity=0.50] ( 43.91, 30.78) rectangle ( 44.61, 49.29);

\path[fill=fillColor,fill opacity=0.50] ( 44.61, 30.78) rectangle ( 45.30, 48.10);

\path[fill=fillColor,fill opacity=0.50] ( 45.30, 30.78) rectangle ( 46.00, 54.66);

\path[fill=fillColor,fill opacity=0.50] ( 46.00, 30.78) rectangle ( 46.69, 58.24);

\path[fill=fillColor,fill opacity=0.50] ( 46.69, 30.78) rectangle ( 47.39, 65.41);

\path[fill=fillColor,fill opacity=0.50] ( 47.39, 30.78) rectangle ( 48.08, 52.27);

\path[fill=fillColor,fill opacity=0.50] ( 48.08, 30.78) rectangle ( 48.78, 54.06);

\path[fill=fillColor,fill opacity=0.50] ( 48.78, 30.78) rectangle ( 49.47, 51.68);

\path[fill=fillColor,fill opacity=0.50] ( 49.47, 30.78) rectangle ( 50.17, 57.65);

\path[fill=fillColor,fill opacity=0.50] ( 50.17, 30.78) rectangle ( 50.86, 45.11);

\path[fill=fillColor,fill opacity=0.50] ( 50.86, 30.78) rectangle ( 51.56, 44.51);

\path[fill=fillColor,fill opacity=0.50] ( 51.56, 30.78) rectangle ( 52.25, 41.53);

\path[fill=fillColor,fill opacity=0.50] ( 52.25, 30.78) rectangle ( 52.95, 45.71);

\path[fill=fillColor,fill opacity=0.50] ( 52.95, 30.78) rectangle ( 53.64, 45.71);

\path[fill=fillColor,fill opacity=0.50] ( 53.64, 30.78) rectangle ( 54.34, 37.95);

\path[fill=fillColor,fill opacity=0.50] ( 54.34, 30.78) rectangle ( 55.03, 41.53);

\path[fill=fillColor,fill opacity=0.50] ( 55.03, 30.78) rectangle ( 55.73, 39.14);

\path[fill=fillColor,fill opacity=0.50] ( 55.73, 30.78) rectangle ( 56.42, 39.74);

\path[fill=fillColor,fill opacity=0.50] ( 56.42, 30.78) rectangle ( 57.12, 34.36);

\path[fill=fillColor,fill opacity=0.50] ( 57.12, 30.78) rectangle ( 57.81, 36.75);

\path[fill=fillColor,fill opacity=0.50] ( 57.81, 30.78) rectangle ( 58.51, 37.35);

\path[fill=fillColor,fill opacity=0.50] ( 58.51, 30.78) rectangle ( 59.20, 39.74);

\path[fill=fillColor,fill opacity=0.50] ( 59.20, 30.78) rectangle ( 59.89, 36.16);

\path[fill=fillColor,fill opacity=0.50] ( 59.89, 30.78) rectangle ( 60.59, 39.14);

\path[fill=fillColor,fill opacity=0.50] ( 60.59, 30.78) rectangle ( 61.28, 35.56);

\path[fill=fillColor,fill opacity=0.50] ( 61.28, 30.78) rectangle ( 61.98, 36.16);

\path[fill=fillColor,fill opacity=0.50] ( 61.98, 30.78) rectangle ( 62.67, 35.56);

\path[fill=fillColor,fill opacity=0.50] ( 62.67, 30.78) rectangle ( 63.37, 33.17);

\path[fill=fillColor,fill opacity=0.50] ( 63.37, 30.78) rectangle ( 64.06, 32.57);

\path[fill=fillColor,fill opacity=0.50] ( 64.06, 30.78) rectangle ( 64.76, 33.17);

\path[fill=fillColor,fill opacity=0.50] ( 64.76, 30.78) rectangle ( 65.45, 33.77);

\path[fill=fillColor,fill opacity=0.50] ( 65.45, 30.78) rectangle ( 66.15, 33.77);

\path[fill=fillColor,fill opacity=0.50] ( 66.15, 30.78) rectangle ( 66.84, 33.77);

\path[fill=fillColor,fill opacity=0.50] ( 66.84, 30.78) rectangle ( 67.54, 33.17);

\path[fill=fillColor,fill opacity=0.50] ( 67.54, 30.78) rectangle ( 68.23, 33.17);

\path[fill=fillColor,fill opacity=0.50] ( 68.23, 30.78) rectangle ( 68.93, 34.36);

\path[fill=fillColor,fill opacity=0.50] ( 68.93, 30.78) rectangle ( 69.62, 33.77);

\path[fill=fillColor,fill opacity=0.50] ( 69.62, 30.78) rectangle ( 70.32, 31.98);

\path[fill=fillColor,fill opacity=0.50] ( 70.32, 30.78) rectangle ( 71.01, 33.77);

\path[fill=fillColor,fill opacity=0.50] ( 71.01, 30.78) rectangle ( 71.71, 31.98);

\path[fill=fillColor,fill opacity=0.50] ( 71.71, 30.78) rectangle ( 72.40, 33.17);

\path[fill=fillColor,fill opacity=0.50] ( 72.40, 30.78) rectangle ( 73.10, 32.57);

\path[fill=fillColor,fill opacity=0.50] ( 73.10, 30.78) rectangle ( 73.79, 31.98);

\path[fill=fillColor,fill opacity=0.50] ( 73.79, 30.78) rectangle ( 74.49, 31.38);

\path[fill=fillColor,fill opacity=0.50] ( 74.49, 30.78) rectangle ( 75.18, 32.57);

\path[fill=fillColor,fill opacity=0.50] ( 75.18, 30.78) rectangle ( 75.88, 31.38);

\path[fill=fillColor,fill opacity=0.50] ( 75.88, 30.78) rectangle ( 76.57, 31.38);

\path[fill=fillColor,fill opacity=0.50] ( 76.57, 30.78) rectangle ( 77.27, 31.38);

\path[fill=fillColor,fill opacity=0.50] ( 77.27, 30.78) rectangle ( 77.96, 31.38);

\path[fill=fillColor,fill opacity=0.50] ( 77.96, 30.78) rectangle ( 78.65, 31.38);

\path[fill=fillColor,fill opacity=0.50] ( 78.65, 30.78) rectangle ( 79.35, 31.98);

\path[fill=fillColor,fill opacity=0.50] ( 79.35, 30.78) rectangle ( 80.04, 33.17);

\path[fill=fillColor,fill opacity=0.50] ( 80.04, 30.78) rectangle ( 80.74, 32.57);

\path[fill=fillColor,fill opacity=0.50] ( 80.74, 30.78) rectangle ( 81.43, 31.98);

\path[fill=fillColor,fill opacity=0.50] ( 81.43, 30.78) rectangle ( 82.13, 31.98);

\path[fill=fillColor,fill opacity=0.50] ( 82.13, 30.78) rectangle ( 82.82, 32.57);

\path[fill=fillColor,fill opacity=0.50] ( 82.82, 30.78) rectangle ( 83.52, 31.38);

\path[fill=fillColor,fill opacity=0.50] ( 83.52, 30.78) rectangle ( 84.21, 30.78);

\path[fill=fillColor,fill opacity=0.50] ( 84.21, 30.78) rectangle ( 84.91, 30.78);

\path[fill=fillColor,fill opacity=0.50] ( 84.91, 30.78) rectangle ( 85.60, 30.78);

\path[fill=fillColor,fill opacity=0.50] ( 85.60, 30.78) rectangle ( 86.30, 31.98);

\path[fill=fillColor,fill opacity=0.50] ( 86.30, 30.78) rectangle ( 86.99, 31.38);

\path[fill=fillColor,fill opacity=0.50] ( 86.99, 30.78) rectangle ( 87.69, 31.38);

\path[fill=fillColor,fill opacity=0.50] ( 87.69, 30.78) rectangle ( 88.38, 30.78);

\path[fill=fillColor,fill opacity=0.50] ( 88.38, 30.78) rectangle ( 89.08, 31.38);

\path[fill=fillColor,fill opacity=0.50] ( 89.08, 30.78) rectangle ( 89.77, 30.78);

\path[fill=fillColor,fill opacity=0.50] ( 89.77, 30.78) rectangle ( 90.47, 30.78);

\path[fill=fillColor,fill opacity=0.50] ( 90.47, 30.78) rectangle ( 91.16, 31.98);

\path[fill=fillColor,fill opacity=0.50] ( 91.16, 30.78) rectangle ( 91.86, 31.38);
\definecolor{fillColor}{RGB}{248,118,109}

\path[fill=fillColor,fill opacity=0.50] ( 23.07, 30.78) rectangle ( 23.76, 30.78);

\path[fill=fillColor,fill opacity=0.50] ( 23.76, 30.78) rectangle ( 24.46, 30.78);

\path[fill=fillColor,fill opacity=0.50] ( 24.46, 30.78) rectangle ( 25.15, 30.78);

\path[fill=fillColor,fill opacity=0.50] ( 25.15, 30.78) rectangle ( 25.85, 30.78);

\path[fill=fillColor,fill opacity=0.50] ( 25.85, 30.78) rectangle ( 26.54, 30.78);

\path[fill=fillColor,fill opacity=0.50] ( 26.54, 30.78) rectangle ( 27.24, 30.78);

\path[fill=fillColor,fill opacity=0.50] ( 27.24, 30.78) rectangle ( 27.93, 30.78);

\path[fill=fillColor,fill opacity=0.50] ( 27.93, 30.78) rectangle ( 28.63, 30.78);

\path[fill=fillColor,fill opacity=0.50] ( 28.63, 30.78) rectangle ( 29.32, 30.78);

\path[fill=fillColor,fill opacity=0.50] ( 29.32, 30.78) rectangle ( 30.02, 30.78);

\path[fill=fillColor,fill opacity=0.50] ( 30.02, 30.78) rectangle ( 30.71, 30.78);

\path[fill=fillColor,fill opacity=0.50] ( 30.71, 30.78) rectangle ( 31.41, 30.78);

\path[fill=fillColor,fill opacity=0.50] ( 31.41, 30.78) rectangle ( 32.10, 30.78);

\path[fill=fillColor,fill opacity=0.50] ( 32.10, 30.78) rectangle ( 32.80, 30.78);

\path[fill=fillColor,fill opacity=0.50] ( 32.80, 30.78) rectangle ( 33.49, 30.78);

\path[fill=fillColor,fill opacity=0.50] ( 33.49, 30.78) rectangle ( 34.19, 30.78);

\path[fill=fillColor,fill opacity=0.50] ( 34.19, 30.78) rectangle ( 34.88, 30.78);

\path[fill=fillColor,fill opacity=0.50] ( 34.88, 30.78) rectangle ( 35.58, 30.78);

\path[fill=fillColor,fill opacity=0.50] ( 35.58, 30.78) rectangle ( 36.27, 30.78);

\path[fill=fillColor,fill opacity=0.50] ( 36.27, 30.78) rectangle ( 36.97, 30.78);

\path[fill=fillColor,fill opacity=0.50] ( 36.97, 30.78) rectangle ( 37.66, 30.78);

\path[fill=fillColor,fill opacity=0.50] ( 37.66, 30.78) rectangle ( 38.36, 30.78);

\path[fill=fillColor,fill opacity=0.50] ( 38.36, 30.78) rectangle ( 39.05, 30.78);

\path[fill=fillColor,fill opacity=0.50] ( 39.05, 30.78) rectangle ( 39.75, 30.78);

\path[fill=fillColor,fill opacity=0.50] ( 39.75, 30.78) rectangle ( 40.44, 30.78);

\path[fill=fillColor,fill opacity=0.50] ( 40.44, 30.78) rectangle ( 41.13, 30.78);

\path[fill=fillColor,fill opacity=0.50] ( 41.13, 30.78) rectangle ( 41.83, 30.78);

\path[fill=fillColor,fill opacity=0.50] ( 41.83, 30.78) rectangle ( 42.52, 30.78);

\path[fill=fillColor,fill opacity=0.50] ( 42.52, 30.78) rectangle ( 43.22, 30.78);

\path[fill=fillColor,fill opacity=0.50] ( 43.22, 30.78) rectangle ( 43.91, 30.78);

\path[fill=fillColor,fill opacity=0.50] ( 43.91, 30.78) rectangle ( 44.61, 30.78);

\path[fill=fillColor,fill opacity=0.50] ( 44.61, 30.78) rectangle ( 45.30, 30.78);

\path[fill=fillColor,fill opacity=0.50] ( 45.30, 30.78) rectangle ( 46.00, 30.78);

\path[fill=fillColor,fill opacity=0.50] ( 46.00, 30.78) rectangle ( 46.69, 30.78);

\path[fill=fillColor,fill opacity=0.50] ( 46.69, 30.78) rectangle ( 47.39, 30.78);

\path[fill=fillColor,fill opacity=0.50] ( 47.39, 30.78) rectangle ( 48.08, 30.78);

\path[fill=fillColor,fill opacity=0.50] ( 48.08, 30.78) rectangle ( 48.78, 30.78);

\path[fill=fillColor,fill opacity=0.50] ( 48.78, 30.78) rectangle ( 49.47, 30.78);

\path[fill=fillColor,fill opacity=0.50] ( 49.47, 30.78) rectangle ( 50.17, 30.78);

\path[fill=fillColor,fill opacity=0.50] ( 50.17, 30.78) rectangle ( 50.86, 30.78);

\path[fill=fillColor,fill opacity=0.50] ( 50.86, 30.78) rectangle ( 51.56, 30.78);

\path[fill=fillColor,fill opacity=0.50] ( 51.56, 30.78) rectangle ( 52.25, 30.78);

\path[fill=fillColor,fill opacity=0.50] ( 52.25, 30.78) rectangle ( 52.95, 30.78);

\path[fill=fillColor,fill opacity=0.50] ( 52.95, 30.78) rectangle ( 53.64, 30.78);

\path[fill=fillColor,fill opacity=0.50] ( 53.64, 30.78) rectangle ( 54.34, 30.78);

\path[fill=fillColor,fill opacity=0.50] ( 54.34, 30.78) rectangle ( 55.03, 30.78);

\path[fill=fillColor,fill opacity=0.50] ( 55.03, 30.78) rectangle ( 55.73, 32.57);

\path[fill=fillColor,fill opacity=0.50] ( 55.73, 30.78) rectangle ( 56.42, 50.48);

\path[fill=fillColor,fill opacity=0.50] ( 56.42, 30.78) rectangle ( 57.12, 71.38);

\path[fill=fillColor,fill opacity=0.50] ( 57.12, 30.78) rectangle ( 57.81, 77.94);

\path[fill=fillColor,fill opacity=0.50] ( 57.81, 30.78) rectangle ( 58.51, 62.42);

\path[fill=fillColor,fill opacity=0.50] ( 58.51, 30.78) rectangle ( 59.20, 61.23);

\path[fill=fillColor,fill opacity=0.50] ( 59.20, 30.78) rectangle ( 59.89, 72.57);

\path[fill=fillColor,fill opacity=0.50] ( 59.89, 30.78) rectangle ( 60.59, 77.94);

\path[fill=fillColor,fill opacity=0.50] ( 60.59, 30.78) rectangle ( 61.28, 70.18);

\path[fill=fillColor,fill opacity=0.50] ( 61.28, 30.78) rectangle ( 61.98, 57.65);

\path[fill=fillColor,fill opacity=0.50] ( 61.98, 30.78) rectangle ( 62.67, 52.27);

\path[fill=fillColor,fill opacity=0.50] ( 62.67, 30.78) rectangle ( 63.37, 53.47);

\path[fill=fillColor,fill opacity=0.50] ( 63.37, 30.78) rectangle ( 64.06, 52.87);

\path[fill=fillColor,fill opacity=0.50] ( 64.06, 30.78) rectangle ( 64.76, 49.89);

\path[fill=fillColor,fill opacity=0.50] ( 64.76, 30.78) rectangle ( 65.45, 49.29);

\path[fill=fillColor,fill opacity=0.50] ( 65.45, 30.78) rectangle ( 66.15, 48.10);

\path[fill=fillColor,fill opacity=0.50] ( 66.15, 30.78) rectangle ( 66.84, 40.93);

\path[fill=fillColor,fill opacity=0.50] ( 66.84, 30.78) rectangle ( 67.54, 37.95);

\path[fill=fillColor,fill opacity=0.50] ( 67.54, 30.78) rectangle ( 68.23, 39.14);

\path[fill=fillColor,fill opacity=0.50] ( 68.23, 30.78) rectangle ( 68.93, 37.95);

\path[fill=fillColor,fill opacity=0.50] ( 68.93, 30.78) rectangle ( 69.62, 37.35);

\path[fill=fillColor,fill opacity=0.50] ( 69.62, 30.78) rectangle ( 70.32, 35.56);

\path[fill=fillColor,fill opacity=0.50] ( 70.32, 30.78) rectangle ( 71.01, 37.95);

\path[fill=fillColor,fill opacity=0.50] ( 71.01, 30.78) rectangle ( 71.71, 33.77);

\path[fill=fillColor,fill opacity=0.50] ( 71.71, 30.78) rectangle ( 72.40, 37.35);

\path[fill=fillColor,fill opacity=0.50] ( 72.40, 30.78) rectangle ( 73.10, 32.57);

\path[fill=fillColor,fill opacity=0.50] ( 73.10, 30.78) rectangle ( 73.79, 33.17);

\path[fill=fillColor,fill opacity=0.50] ( 73.79, 30.78) rectangle ( 74.49, 34.96);

\path[fill=fillColor,fill opacity=0.50] ( 74.49, 30.78) rectangle ( 75.18, 31.98);

\path[fill=fillColor,fill opacity=0.50] ( 75.18, 30.78) rectangle ( 75.88, 31.98);

\path[fill=fillColor,fill opacity=0.50] ( 75.88, 30.78) rectangle ( 76.57, 31.98);

\path[fill=fillColor,fill opacity=0.50] ( 76.57, 30.78) rectangle ( 77.27, 31.98);

\path[fill=fillColor,fill opacity=0.50] ( 77.27, 30.78) rectangle ( 77.96, 32.57);

\path[fill=fillColor,fill opacity=0.50] ( 77.96, 30.78) rectangle ( 78.65, 31.98);

\path[fill=fillColor,fill opacity=0.50] ( 78.65, 30.78) rectangle ( 79.35, 31.38);

\path[fill=fillColor,fill opacity=0.50] ( 79.35, 30.78) rectangle ( 80.04, 31.38);

\path[fill=fillColor,fill opacity=0.50] ( 80.04, 30.78) rectangle ( 80.74, 32.57);

\path[fill=fillColor,fill opacity=0.50] ( 80.74, 30.78) rectangle ( 81.43, 33.17);

\path[fill=fillColor,fill opacity=0.50] ( 81.43, 30.78) rectangle ( 82.13, 33.17);

\path[fill=fillColor,fill opacity=0.50] ( 82.13, 30.78) rectangle ( 82.82, 31.98);

\path[fill=fillColor,fill opacity=0.50] ( 82.82, 30.78) rectangle ( 83.52, 31.98);

\path[fill=fillColor,fill opacity=0.50] ( 83.52, 30.78) rectangle ( 84.21, 31.98);

\path[fill=fillColor,fill opacity=0.50] ( 84.21, 30.78) rectangle ( 84.91, 31.38);

\path[fill=fillColor,fill opacity=0.50] ( 84.91, 30.78) rectangle ( 85.60, 31.98);

\path[fill=fillColor,fill opacity=0.50] ( 85.60, 30.78) rectangle ( 86.30, 32.57);

\path[fill=fillColor,fill opacity=0.50] ( 86.30, 30.78) rectangle ( 86.99, 33.17);

\path[fill=fillColor,fill opacity=0.50] ( 86.99, 30.78) rectangle ( 87.69, 31.98);

\path[fill=fillColor,fill opacity=0.50] ( 87.69, 30.78) rectangle ( 88.38, 32.57);

\path[fill=fillColor,fill opacity=0.50] ( 88.38, 30.78) rectangle ( 89.08, 33.17);

\path[fill=fillColor,fill opacity=0.50] ( 89.08, 30.78) rectangle ( 89.77, 31.98);

\path[fill=fillColor,fill opacity=0.50] ( 89.77, 30.78) rectangle ( 90.47, 32.57);

\path[fill=fillColor,fill opacity=0.50] ( 90.47, 30.78) rectangle ( 91.16, 31.98);

\path[fill=fillColor,fill opacity=0.50] ( 91.16, 30.78) rectangle ( 91.86, 31.38);
\end{scope}
\begin{scope}
\path[clip] (  0.00,  0.00) rectangle (101.18, 93.95);
\definecolor{drawColor}{gray}{0.30}

\node[text=drawColor,anchor=base east,inner sep=0pt, outer sep=0pt, scale=  0.88] at ( 14.30, 27.75) {0};

\node[text=drawColor,anchor=base east,inner sep=0pt, outer sep=0pt, scale=  0.88] at ( 14.30, 42.68) {25};

\node[text=drawColor,anchor=base east,inner sep=0pt, outer sep=0pt, scale=  0.88] at ( 14.30, 57.60) {50};

\node[text=drawColor,anchor=base east,inner sep=0pt, outer sep=0pt, scale=  0.88] at ( 14.30, 72.53) {75};
\end{scope}
\begin{scope}
\path[clip] (  0.00,  0.00) rectangle (101.18, 93.95);
\definecolor{drawColor}{gray}{0.30}

\node[text=drawColor,anchor=base,inner sep=0pt, outer sep=0pt, scale=  0.88] at ( 22.72, 17.03) {0};

\node[text=drawColor,anchor=base,inner sep=0pt, outer sep=0pt, scale=  0.88] at ( 40.09, 17.03) {25};

\node[text=drawColor,anchor=base,inner sep=0pt, outer sep=0pt, scale=  0.88] at ( 57.46, 17.03) {50};

\node[text=drawColor,anchor=base,inner sep=0pt, outer sep=0pt, scale=  0.88] at ( 74.83, 17.03) {75};

\node[text=drawColor,anchor=base,inner sep=0pt, outer sep=0pt, scale=  0.88] at ( 92.20, 17.03) {100};
\end{scope}
\begin{scope}
\path[clip] (  0.00,  0.00) rectangle (101.18, 93.95);
\definecolor{drawColor}{RGB}{0,0,0}

\node[text=drawColor,anchor=base,inner sep=0pt, outer sep=0pt, scale=  0.80] at ( 57.46,  7.06) {Response time [ms]};
\end{scope}
\end{tikzpicture}

%% file: benchmarks/25-04-16_AWS_multi_region/AWS_multi_region.tex
% Created by tikzDevice version 0.12.6 on 2025-04-18 01:42:55
% !TEX encoding = UTF-8 Unicode
\begin{tikzpicture}[x=1pt,y=1pt]
\definecolor{fillColor}{RGB}{255,255,255}
\path[use as bounding box,fill=fillColor,fill opacity=0.00] (0,0) rectangle (108.41,108.41);
\begin{scope}
\path[clip] ( 23.65, 28.04) rectangle (102.90,102.91);
\definecolor{drawColor}{gray}{0.92}

\path[draw=drawColor,line width= 0.3pt,line join=round] ( 23.65, 39.56) --
	(102.90, 39.56);

\path[draw=drawColor,line width= 0.3pt,line join=round] ( 23.65, 55.81) --
	(102.90, 55.81);

\path[draw=drawColor,line width= 0.3pt,line join=round] ( 23.65, 72.05) --
	(102.90, 72.05);

\path[draw=drawColor,line width= 0.3pt,line join=round] ( 23.65, 88.29) --
	(102.90, 88.29);

\path[draw=drawColor,line width= 0.3pt,line join=round] ( 38.33, 28.04) --
	( 38.33,102.91);

\path[draw=drawColor,line width= 0.3pt,line join=round] ( 60.50, 28.04) --
	( 60.50,102.91);

\path[draw=drawColor,line width= 0.3pt,line join=round] ( 82.67, 28.04) --
	( 82.67,102.91);

\path[draw=drawColor,line width= 0.6pt,line join=round] ( 23.65, 31.44) --
	(102.90, 31.44);

\path[draw=drawColor,line width= 0.6pt,line join=round] ( 23.65, 47.68) --
	(102.90, 47.68);

\path[draw=drawColor,line width= 0.6pt,line join=round] ( 23.65, 63.93) --
	(102.90, 63.93);

\path[draw=drawColor,line width= 0.6pt,line join=round] ( 23.65, 80.17) --
	(102.90, 80.17);

\path[draw=drawColor,line width= 0.6pt,line join=round] ( 23.65, 96.42) --
	(102.90, 96.42);

\path[draw=drawColor,line width= 0.6pt,line join=round] ( 27.25, 28.04) --
	( 27.25,102.91);

\path[draw=drawColor,line width= 0.6pt,line join=round] ( 49.42, 28.04) --
	( 49.42,102.91);

\path[draw=drawColor,line width= 0.6pt,line join=round] ( 71.59, 28.04) --
	( 71.59,102.91);

\path[draw=drawColor,line width= 0.6pt,line join=round] ( 93.76, 28.04) --
	( 93.76,102.91);
\definecolor{fillColor}{RGB}{0,191,196}

\path[fill=fillColor,fill opacity=0.50] ( 27.80, 31.44) rectangle ( 28.91, 31.44);

\path[fill=fillColor,fill opacity=0.50] ( 28.91, 31.44) rectangle ( 30.02, 31.44);

\path[fill=fillColor,fill opacity=0.50] ( 30.02, 31.44) rectangle ( 31.13, 31.44);

\path[fill=fillColor,fill opacity=0.50] ( 31.13, 31.44) rectangle ( 32.24, 31.44);

\path[fill=fillColor,fill opacity=0.50] ( 32.24, 31.44) rectangle ( 33.35, 31.44);

\path[fill=fillColor,fill opacity=0.50] ( 33.35, 31.44) rectangle ( 34.45, 31.44);

\path[fill=fillColor,fill opacity=0.50] ( 34.45, 31.44) rectangle ( 35.56, 31.44);

\path[fill=fillColor,fill opacity=0.50] ( 35.56, 31.44) rectangle ( 36.67, 31.44);

\path[fill=fillColor,fill opacity=0.50] ( 36.67, 31.44) rectangle ( 37.78, 31.44);

\path[fill=fillColor,fill opacity=0.50] ( 37.78, 31.44) rectangle ( 38.89, 31.44);

\path[fill=fillColor,fill opacity=0.50] ( 38.89, 31.44) rectangle ( 40.00, 31.44);

\path[fill=fillColor,fill opacity=0.50] ( 40.00, 31.44) rectangle ( 41.11, 31.44);

\path[fill=fillColor,fill opacity=0.50] ( 41.11, 31.44) rectangle ( 42.21, 31.44);

\path[fill=fillColor,fill opacity=0.50] ( 42.21, 31.44) rectangle ( 43.32, 31.44);

\path[fill=fillColor,fill opacity=0.50] ( 43.32, 31.44) rectangle ( 44.43, 31.44);

\path[fill=fillColor,fill opacity=0.50] ( 44.43, 31.44) rectangle ( 45.54, 31.44);

\path[fill=fillColor,fill opacity=0.50] ( 45.54, 31.44) rectangle ( 46.65, 31.44);

\path[fill=fillColor,fill opacity=0.50] ( 46.65, 31.44) rectangle ( 47.76, 31.44);

\path[fill=fillColor,fill opacity=0.50] ( 47.76, 31.44) rectangle ( 48.87, 31.44);

\path[fill=fillColor,fill opacity=0.50] ( 48.87, 31.44) rectangle ( 49.97, 31.44);

\path[fill=fillColor,fill opacity=0.50] ( 49.97, 31.44) rectangle ( 51.08, 41.67);

\path[fill=fillColor,fill opacity=0.50] ( 51.08, 31.44) rectangle ( 52.19, 84.39);

\path[fill=fillColor,fill opacity=0.50] ( 52.19, 31.44) rectangle ( 53.30, 91.22);

\path[fill=fillColor,fill opacity=0.50] ( 53.30, 31.44) rectangle ( 54.41, 55.32);

\path[fill=fillColor,fill opacity=0.50] ( 54.41, 31.44) rectangle ( 55.52, 37.61);

\path[fill=fillColor,fill opacity=0.50] ( 55.52, 31.44) rectangle ( 56.62, 34.36);

\path[fill=fillColor,fill opacity=0.50] ( 56.62, 31.44) rectangle ( 57.73, 33.23);

\path[fill=fillColor,fill opacity=0.50] ( 57.73, 31.44) rectangle ( 58.84, 32.41);

\path[fill=fillColor,fill opacity=0.50] ( 58.84, 31.44) rectangle ( 59.95, 31.93);

\path[fill=fillColor,fill opacity=0.50] ( 59.95, 31.44) rectangle ( 61.06, 31.93);

\path[fill=fillColor,fill opacity=0.50] ( 61.06, 31.44) rectangle ( 62.17, 32.25);

\path[fill=fillColor,fill opacity=0.50] ( 62.17, 31.44) rectangle ( 63.28, 32.09);

\path[fill=fillColor,fill opacity=0.50] ( 63.28, 31.44) rectangle ( 64.38, 31.60);

\path[fill=fillColor,fill opacity=0.50] ( 64.38, 31.44) rectangle ( 65.49, 31.93);

\path[fill=fillColor,fill opacity=0.50] ( 65.49, 31.44) rectangle ( 66.60, 31.60);

\path[fill=fillColor,fill opacity=0.50] ( 66.60, 31.44) rectangle ( 67.71, 31.44);

\path[fill=fillColor,fill opacity=0.50] ( 67.71, 31.44) rectangle ( 68.82, 31.44);

\path[fill=fillColor,fill opacity=0.50] ( 68.82, 31.44) rectangle ( 69.93, 31.44);

\path[fill=fillColor,fill opacity=0.50] ( 69.93, 31.44) rectangle ( 71.04, 31.44);

\path[fill=fillColor,fill opacity=0.50] ( 71.04, 31.44) rectangle ( 72.14, 31.60);

\path[fill=fillColor,fill opacity=0.50] ( 72.14, 31.44) rectangle ( 73.25, 31.60);

\path[fill=fillColor,fill opacity=0.50] ( 73.25, 31.44) rectangle ( 74.36, 31.44);

\path[fill=fillColor,fill opacity=0.50] ( 74.36, 31.44) rectangle ( 75.47, 31.60);

\path[fill=fillColor,fill opacity=0.50] ( 75.47, 31.44) rectangle ( 76.58, 31.44);

\path[fill=fillColor,fill opacity=0.50] ( 76.58, 31.44) rectangle ( 77.69, 31.44);

\path[fill=fillColor,fill opacity=0.50] ( 77.69, 31.44) rectangle ( 78.80, 31.44);

\path[fill=fillColor,fill opacity=0.50] ( 78.80, 31.44) rectangle ( 79.90, 31.44);

\path[fill=fillColor,fill opacity=0.50] ( 79.90, 31.44) rectangle ( 81.01, 31.44);

\path[fill=fillColor,fill opacity=0.50] ( 81.01, 31.44) rectangle ( 82.12, 31.44);

\path[fill=fillColor,fill opacity=0.50] ( 82.12, 31.44) rectangle ( 83.23, 31.44);

\path[fill=fillColor,fill opacity=0.50] ( 83.23, 31.44) rectangle ( 84.34, 31.44);

\path[fill=fillColor,fill opacity=0.50] ( 84.34, 31.44) rectangle ( 85.45, 31.44);

\path[fill=fillColor,fill opacity=0.50] ( 85.45, 31.44) rectangle ( 86.55, 31.44);

\path[fill=fillColor,fill opacity=0.50] ( 86.55, 31.44) rectangle ( 87.66, 31.44);

\path[fill=fillColor,fill opacity=0.50] ( 87.66, 31.44) rectangle ( 88.77, 31.44);

\path[fill=fillColor,fill opacity=0.50] ( 88.77, 31.44) rectangle ( 89.88, 31.44);

\path[fill=fillColor,fill opacity=0.50] ( 89.88, 31.44) rectangle ( 90.99, 31.44);

\path[fill=fillColor,fill opacity=0.50] ( 90.99, 31.44) rectangle ( 92.10, 31.44);

\path[fill=fillColor,fill opacity=0.50] ( 92.10, 31.44) rectangle ( 93.21, 31.44);

\path[fill=fillColor,fill opacity=0.50] ( 93.21, 31.44) rectangle ( 94.31, 31.44);

\path[fill=fillColor,fill opacity=0.50] ( 94.31, 31.44) rectangle ( 95.42, 31.44);

\path[fill=fillColor,fill opacity=0.50] ( 95.42, 31.44) rectangle ( 96.53, 31.44);

\path[fill=fillColor,fill opacity=0.50] ( 96.53, 31.44) rectangle ( 97.64, 31.44);

\path[fill=fillColor,fill opacity=0.50] ( 97.64, 31.44) rectangle ( 98.75, 31.44);
\definecolor{fillColor}{RGB}{248,118,109}

\path[fill=fillColor,fill opacity=0.50] ( 27.80, 31.44) rectangle ( 28.91, 31.44);

\path[fill=fillColor,fill opacity=0.50] ( 28.91, 31.44) rectangle ( 30.02, 31.44);

\path[fill=fillColor,fill opacity=0.50] ( 30.02, 31.44) rectangle ( 31.13, 31.44);

\path[fill=fillColor,fill opacity=0.50] ( 31.13, 31.44) rectangle ( 32.24, 31.44);

\path[fill=fillColor,fill opacity=0.50] ( 32.24, 31.44) rectangle ( 33.35, 31.44);

\path[fill=fillColor,fill opacity=0.50] ( 33.35, 31.44) rectangle ( 34.45, 31.44);

\path[fill=fillColor,fill opacity=0.50] ( 34.45, 31.44) rectangle ( 35.56, 31.44);

\path[fill=fillColor,fill opacity=0.50] ( 35.56, 31.44) rectangle ( 36.67, 31.44);

\path[fill=fillColor,fill opacity=0.50] ( 36.67, 31.44) rectangle ( 37.78, 31.44);

\path[fill=fillColor,fill opacity=0.50] ( 37.78, 31.44) rectangle ( 38.89, 31.44);

\path[fill=fillColor,fill opacity=0.50] ( 38.89, 31.44) rectangle ( 40.00, 31.44);

\path[fill=fillColor,fill opacity=0.50] ( 40.00, 31.44) rectangle ( 41.11, 31.44);

\path[fill=fillColor,fill opacity=0.50] ( 41.11, 31.44) rectangle ( 42.21, 31.44);

\path[fill=fillColor,fill opacity=0.50] ( 42.21, 31.44) rectangle ( 43.32, 31.44);

\path[fill=fillColor,fill opacity=0.50] ( 43.32, 31.44) rectangle ( 44.43, 31.44);

\path[fill=fillColor,fill opacity=0.50] ( 44.43, 31.44) rectangle ( 45.54, 31.44);

\path[fill=fillColor,fill opacity=0.50] ( 45.54, 31.44) rectangle ( 46.65, 31.44);

\path[fill=fillColor,fill opacity=0.50] ( 46.65, 31.44) rectangle ( 47.76, 31.44);

\path[fill=fillColor,fill opacity=0.50] ( 47.76, 31.44) rectangle ( 48.87, 31.44);

\path[fill=fillColor,fill opacity=0.50] ( 48.87, 31.44) rectangle ( 49.97, 31.44);

\path[fill=fillColor,fill opacity=0.50] ( 49.97, 31.44) rectangle ( 51.08, 31.44);

\path[fill=fillColor,fill opacity=0.50] ( 51.08, 31.44) rectangle ( 52.19, 31.44);

\path[fill=fillColor,fill opacity=0.50] ( 52.19, 31.44) rectangle ( 53.30, 31.44);

\path[fill=fillColor,fill opacity=0.50] ( 53.30, 31.44) rectangle ( 54.41, 31.44);

\path[fill=fillColor,fill opacity=0.50] ( 54.41, 31.44) rectangle ( 55.52, 31.44);

\path[fill=fillColor,fill opacity=0.50] ( 55.52, 31.44) rectangle ( 56.62, 31.44);

\path[fill=fillColor,fill opacity=0.50] ( 56.62, 31.44) rectangle ( 57.73, 31.44);

\path[fill=fillColor,fill opacity=0.50] ( 57.73, 31.44) rectangle ( 58.84, 31.44);

\path[fill=fillColor,fill opacity=0.50] ( 58.84, 31.44) rectangle ( 59.95, 31.44);

\path[fill=fillColor,fill opacity=0.50] ( 59.95, 31.44) rectangle ( 61.06, 31.44);

\path[fill=fillColor,fill opacity=0.50] ( 61.06, 31.44) rectangle ( 62.17, 31.44);

\path[fill=fillColor,fill opacity=0.50] ( 62.17, 31.44) rectangle ( 63.28, 31.44);

\path[fill=fillColor,fill opacity=0.50] ( 63.28, 31.44) rectangle ( 64.38, 31.44);

\path[fill=fillColor,fill opacity=0.50] ( 64.38, 31.44) rectangle ( 65.49, 31.44);

\path[fill=fillColor,fill opacity=0.50] ( 65.49, 31.44) rectangle ( 66.60, 31.44);

\path[fill=fillColor,fill opacity=0.50] ( 66.60, 31.44) rectangle ( 67.71, 31.44);

\path[fill=fillColor,fill opacity=0.50] ( 67.71, 31.44) rectangle ( 68.82, 31.44);

\path[fill=fillColor,fill opacity=0.50] ( 68.82, 31.44) rectangle ( 69.93, 31.44);

\path[fill=fillColor,fill opacity=0.50] ( 69.93, 31.44) rectangle ( 71.04, 31.44);

\path[fill=fillColor,fill opacity=0.50] ( 71.04, 31.44) rectangle ( 72.14, 31.44);

\path[fill=fillColor,fill opacity=0.50] ( 72.14, 31.44) rectangle ( 73.25, 31.44);

\path[fill=fillColor,fill opacity=0.50] ( 73.25, 31.44) rectangle ( 74.36, 31.44);

\path[fill=fillColor,fill opacity=0.50] ( 74.36, 31.44) rectangle ( 75.47, 31.44);

\path[fill=fillColor,fill opacity=0.50] ( 75.47, 31.44) rectangle ( 76.58, 31.44);

\path[fill=fillColor,fill opacity=0.50] ( 76.58, 31.44) rectangle ( 77.69, 31.44);

\path[fill=fillColor,fill opacity=0.50] ( 77.69, 31.44) rectangle ( 78.80, 31.44);

\path[fill=fillColor,fill opacity=0.50] ( 78.80, 31.44) rectangle ( 79.90, 31.44);

\path[fill=fillColor,fill opacity=0.50] ( 79.90, 31.44) rectangle ( 81.01, 31.44);

\path[fill=fillColor,fill opacity=0.50] ( 81.01, 31.44) rectangle ( 82.12, 31.93);

\path[fill=fillColor,fill opacity=0.50] ( 82.12, 31.44) rectangle ( 83.23, 56.62);

\path[fill=fillColor,fill opacity=0.50] ( 83.23, 31.44) rectangle ( 84.34, 99.50);

\path[fill=fillColor,fill opacity=0.50] ( 84.34, 31.44) rectangle ( 85.45, 77.57);

\path[fill=fillColor,fill opacity=0.50] ( 85.45, 31.44) rectangle ( 86.55, 43.95);

\path[fill=fillColor,fill opacity=0.50] ( 86.55, 31.44) rectangle ( 87.66, 35.66);

\path[fill=fillColor,fill opacity=0.50] ( 87.66, 31.44) rectangle ( 88.77, 33.55);

\path[fill=fillColor,fill opacity=0.50] ( 88.77, 31.44) rectangle ( 89.88, 32.25);

\path[fill=fillColor,fill opacity=0.50] ( 89.88, 31.44) rectangle ( 90.99, 32.25);

\path[fill=fillColor,fill opacity=0.50] ( 90.99, 31.44) rectangle ( 92.10, 31.76);

\path[fill=fillColor,fill opacity=0.50] ( 92.10, 31.44) rectangle ( 93.21, 32.09);

\path[fill=fillColor,fill opacity=0.50] ( 93.21, 31.44) rectangle ( 94.31, 31.76);

\path[fill=fillColor,fill opacity=0.50] ( 94.31, 31.44) rectangle ( 95.42, 31.44);

\path[fill=fillColor,fill opacity=0.50] ( 95.42, 31.44) rectangle ( 96.53, 31.60);

\path[fill=fillColor,fill opacity=0.50] ( 96.53, 31.44) rectangle ( 97.64, 31.60);

\path[fill=fillColor,fill opacity=0.50] ( 97.64, 31.44) rectangle ( 98.75, 31.44);
\end{scope}
\begin{scope}
\path[clip] (  0.00,  0.00) rectangle (108.41,108.41);
\definecolor{drawColor}{gray}{0.30}

\node[text=drawColor,anchor=base east,inner sep=0pt, outer sep=0pt, scale=  0.88] at ( 18.70, 28.41) {0};

\node[text=drawColor,anchor=base east,inner sep=0pt, outer sep=0pt, scale=  0.88] at ( 18.70, 44.65) {100};

\node[text=drawColor,anchor=base east,inner sep=0pt, outer sep=0pt, scale=  0.88] at ( 18.70, 60.90) {200};

\node[text=drawColor,anchor=base east,inner sep=0pt, outer sep=0pt, scale=  0.88] at ( 18.70, 77.14) {300};

\node[text=drawColor,anchor=base east,inner sep=0pt, outer sep=0pt, scale=  0.88] at ( 18.70, 93.39) {400};
\end{scope}
\begin{scope}
\path[clip] (  0.00,  0.00) rectangle (108.41,108.41);
\definecolor{drawColor}{gray}{0.30}

\node[text=drawColor,anchor=base,inner sep=0pt, outer sep=0pt, scale=  0.88] at ( 27.25, 17.03) {0};

\node[text=drawColor,anchor=base,inner sep=0pt, outer sep=0pt, scale=  0.88] at ( 49.42, 17.03) {20};

\node[text=drawColor,anchor=base,inner sep=0pt, outer sep=0pt, scale=  0.88] at ( 71.59, 17.03) {40};

\node[text=drawColor,anchor=base,inner sep=0pt, outer sep=0pt, scale=  0.88] at ( 93.76, 17.03) {60};
\end{scope}
\begin{scope}
\path[clip] (  0.00,  0.00) rectangle (108.41,108.41);
\definecolor{drawColor}{RGB}{0,0,0}

\node[text=drawColor,anchor=base,inner sep=0pt, outer sep=0pt, scale=  0.80] at ( 63.28,  7.06) {Response time [ms]};
\end{scope}
\end{tikzpicture}

%% file: conclusion.tex
\section{Related Work}\label{sec:related-work}

Existing approaches to distributed programming address correctness and fault tolerance at different levels of abstraction. We classify them into 
\begin{enumerate*}[label=(\roman*), itemjoin={{; }}, itemjoin*={{; and }}] 
\item
language-level approaches, which extend programming models with communication and failure-handling constructs
\item
runtime-based systems, which provide fault tolerance through infrastructure
\item
orchestration and workflow frameworks, which manage long-running distributed processes
\end{enumerate*}.
This classification highlights the distinctive position of \Accompanist as a runtime-level solution for resilient choreographic programs, enabling decentralized sagas and replay-based fault-tolerance, without extending the source language.

\paragraph{Language-Level Approaches}

A large body of work addresses correctness and fault handling by extending programming languages for distributed systems.

Wiersdorf and Greenman \cite{chorex} study fail-stop failures\footnote{
  Under the fail-stop failure mode, failures occur as nodes crashing and the rest of the system can reliably detect this failure. 
} in choreographic programming and introduce a distributed try-catch mechanism for handling with their Elixir DSL Chorex.
When a try-catch block is executed, Chorex creates a checkpoint and when a participant to the try-block crashes it is restored to this checkpoint and all participants move to the catch-block. 
This mechanism relies on the reliable detection of node crashes and on global synchronization to ensure all participants agree on whether the block succeeded or failed whereas \Accompanist's solution avoids global synchronization.
While Chorex does not come with a formal semantics and a formal correctness result for its recovery mechanism, its try-catch mechanism shares some similarities with models explored in the adjacent context of multiparty session types \cite{mezzina2025,VHEZ21}.

Graversen et al.~\cite{graversen2025} study omission failures\footnote{
  Omission failures occur when a component fails to send or receive a message or when the transport fails to transmit it.
} in choreographic programming and introduce linguistic primitives for programming custom, fine-grained failure-handling and recovery strategies.
Their approach demonstrates that certain classes of non-malicious communication failures can be handled at the language level while preserving key safety guarantees of choreographic programs.
These approaches provide strong correctness guarantees and expressive abstractions, but they require enriching the programming model with additional constructs and place a burden on programmers to explicitly structure code around failure-handling logic (e.g., retries, acknowledgments, or timeouts). 
In contrast, \Accompanist shifts fault tolerance to the runtime layer, enabling existing choreographic programs to tolerate failures without modifying the source language or compiler, while retaining the clarity and readability characteristic of choreographic programming.

Beyond choreographic languages, several interaction and multi-tier programming languages adopt the paradigm of writing a single program that compiles into distributed code~\cite{casadei2023,wael2015,weisenburger2020,tagueu_service_2025}. Frameworks such as ScalaLoci~\cite{weisenburger2018a} extend this idea with reactive abstractions, allowing streams to be shared across nodes of the system and propagating failures along dataflow dependencies while supporting the programming of manual recovery strategies at each participant.
In RPC Chains~\cite{DBLP:conf/nsdi/SongAKM09} sequential RPCs are chained together before returned to the initiator to reduce latency while sending periodic heartbeats along the chain to detect failures that break it. 
These approaches have seen limited application in polyglot microservice development. 
Our framework is particularly well-suited to this domain: the communication medium can be customized via Choral's channel API; communications and data locality are statically enforced by the type system; execution is deterministic under reasonable assumptions; and opportunities for parallelism are exploited automatically by virtue of the compilation strategy.

\paragraph{Runtime-Based Approaches}

Runtime-based approaches provide fault tolerance and scalability through infrastructure rather than language design.

Actor runtimes and frameworks such as Erlang~\cite{armstrong2003}, Orleans~\cite{bykov2011}, and Akka~\cite{akka} provide flexible abstractions for building distributed systems, including supervision hierarchies and failure recovery mechanisms. 
Systems like Ray~\cite{moritz2018} further support distributed futures, decoupling control flow from data movement. However, these abstractions rely on runtime mechanisms such as weighted reference counting~\cite{wang2021} and can be nontrivial to use correctly in practice~\cite{ray-get-order,ray-get-unnecessary,ray-get-loop}.
While these systems offer significant flexibility, they lack a global specification mechanism and do not guarantee properties such as communication safety or (global) recovery.
As a result, developers must manually ensure protocol correctness and recovery behaviour. 
In contrast, choreographic programming provides global specifications with strong correctness guarantees, and \Accompanist complements these guarantees with runtime fault tolerance.
%Moreover, fault tolerance in actor systems is typically based on local supervision and restart strategies, whereas \Accompanist provides coordinated recovery via replay of the entire distributed computation.

%(While the ScalaLoci runtime is also important but the linguistic approach it supports is the crucial innovation, in our opinion.)

\paragraph{Orchestration, Workflows, and Sagas}

%Programmers typically implement service choreographies with message brokers or RPC libraries, possibly with the aid of an external data store~\cite{hohpe2004}. These systems can be hard to maintain at scale due to poor programming support~\cite{netflixtechnologyblog2016}. Many organizations have developed orchestration systems as an alternative~\cite{Camunda,LogicApps,airflow,burckhardt_durable_2021}, but orchestration scales poorly as latency increases.

In practice, service choreographies are often implemented using message brokers or RPC libraries, possibly with the aid of an external data store~\cite{hohpe2004}. 
While widely adopted, these approaches provide limited high-level abstractions, making large-scale systems difficult to reason about and maintain~\cite{netflixtechnologyblog2016}.

To address these challenges, many organizations have developed orchestration and workflow systems~\cite{Camunda,LogicApps,airflow,burckhardt_durable_2021,temporal}, which centralize control over distributed processes and provide built-in support for fault tolerance and recovery. 
These systems simplify reasoning about long-running workflows and often implement saga-style execution models, but rely on a centralized orchestrator that can introduce latency, limit scalability, and constrain (or prevent) data locality.
%Furthermore, while these systems support saga-style execution, they do not provide formal guarantees such as deadlock-freedom or correctness-by-construction, nor do they support decentralized execution of sagas.

%\Accompanist's architecture -- which improves end-to-end latency by `decentralizing' the orchestrator service into a collection of sidecars -- is not inherently tied to choreographic programming. For example, we could have replaced Choral with a conventional language like Java and an actor framework. But choreographic programming is particularly well-suited to the task: migrating an orchestrator to Choral preserves the `shape' of the original program; the type system guarantees rewritten code will not introduce communication mismatches; and the compiler guarantees generated code is deterministic and deadlock-free.

\Accompanist's architecture can be seen as ``decentralizing the orchestrator'' into a collection of reactive sidecars, improving end-to-end latency while preserving fault tolerance. 
Unlike traditional orchestration systems, this approach avoids central coordination during execution.
Although this architecture is not inherently tied to choreographic programming, choreographies provide a natural fit: migrating an orchestrator to Choral preserves the structure of the original program, the type system guarantees the absence of communication mismatches, and the compiler ensures deterministic and deadlock-free execution. In contrast to centralized workflow engines, \Accompanist achieves similar fault-tolerance goals while maintaining decentralized execution.

\section{Conclusion}\label{sec:conclusion}

Our results show \Accompanist adds modest overhead and scales better than orchestration when network costs are high. %We expect overhead to decrease as we optimize our prototype; for instance, many of the optimizations in ServiceWeaver~\cite{ghemawat2023} at the serialization and transport layer apply equally well to our approach. We also plan to extend our evaluation to a wider range of applications, particularly edge computing~\cite{wolski2019} and cross-organizational workflows~\cite{lai2020}.
Perhaps our most important contribution is to put forward \emph{choreographic programming as a better interface} than RPC for application developers to express preferences to the computational substrate. If application developers can be convinced to express workflows on the hot path as choreographic programs, systems researchers could harness the extra information to improve performance. One promising opportunity is to integrate choreographic programs into lower layers of the network stack, taking advantage of features like in-network computing and eliminating the overhead needed to support general-purpose RPCs~\cite{zhu2023,saxena2023,giallorenzo2025}. Since choreographic programs can specify how data is routed from one location to another, analysis tools could also identify how choreographic programs could be re-routed to reduce congestion and prioritize flows~\cite{zhu2015}. These questions invite opportunities for deeper collaboration between systems researchers and specialists in compilers and programming languages.

%Although our prototype already supports failure recovery via traditional means like SAGAs, we expect choreographic programming could offer a more idiomatic and fool-proof API for such transactions.

%We believe that horizontal scalability could be achieved by replicating the sidecars together with the replica. And using sticky sessions such that messages for a particular choreography session are handled by the same sidecar replica.

% How does Kubernetes do it? We just need session stickiness for sidecars.
% - Kubernetes would automatically replicate the sidecar as well as it's apart of the same pod
% - There are multiple loadbalancers that can support sticky sessions based on eg. HTTP headers that we could use.

%\todo[inline]{We don't guarantee that interactions \emph{between} choreographies are safe. That would require something like linearizability checking!}

\section{Data-Availability Statement}

The full source code for the \Accompanist framework and instructions on how to carry out the evaluations for reproducibility will be submitted for artifact submission.